\documentclass{aa} \newcommand{\twice}{2} \newcommand{\zerosix}{1} \newcommand{\oneseven}{1.7}

\usepackage[varg]{txfonts}
\usepackage{pifont} 
\usepackage{lscape}
\usepackage{graphicx}
\usepackage[]{natbib}
\usepackage[usenames,dvipsnames,svgnames,table]{xcolor}
\usepackage{array,longtable}
\bibpunct{(}{)}{;}{a}{}{,} % to follow the A&A style

\newcommand{\izw}{I~Zw\,18\,}
\newcommand{\mso}{~\rm M_{\odot}}
\definecolor{gold}{HTML}{FFD700}
\definecolor{crimson}{HTML}{DC143C}
\definecolor{bluee}{HTML}{00008B}
\definecolor{WRblue}{HTML}{1E90FF}

\begin{document}
\title{
Low-metallicity massive single stars with rotation
}
\subtitle{Evolutionary models applicable to I~Zwicky\,18}
\titlerunning{Low-metallicity massive single stars}
\authorrunning{Dorottya Sz\'ecsi et al.}
\author{Dorottya Sz\'ecsi$^1$
\and Norbert Langer$^1$
\and Sung-Chul Yoon$^2$
\and Debashis Sanyal$^1$
\and Selma de Mink$^3$
\and Christopher J. Evans$^4$
\and Tyl Dermine$^1$
}
\institute{$^1$~Argelander-Institut f\"ur Astronomie der Universit\"at Bonn, Auf dem H\"ugel 71, 53121 Bonn, Germany\\
$^2$~Department of Physics \& Astronomy, Seoul National University, Gwanak-ro 1, Gwanak-gu, 151-742, Seoul, South Korea\\
$^3$~Astronomical Institute Anton Pannekoek, University of Amsterdam, 1098 XH Amsterdam, The Netherlands\\
$^4$~UK Astronomy Technology Centre, Royal Observatory, Blackford Hill, Edinburgh, EH9 3HJ, UK
}
\date{Received \today / Accepted ...}

\abstract 
%---%Context.
{Low-metallicity environments such as the early Universe and compact star-forming dwarf galaxies contain many massive stars. These stars influence their surroundings through intense UV radiation, strong winds and explosive deaths. A good understanding of low-metallicity environments requires a detailed theoretical comprehension of the evolution of their massive stars.} %Context.
%---%Aims.
{We aim to investigate the role of metallicity and rotation in shaping the evolutionary paths of massive stars and to provide theoretical predictions that can be tested by observations of metal-poor environments.} %Aims.
%---%Method.
{Massive rotating single stars with an initial metal composition appropriate for the dwarf galaxy I~Zw\,18 ([Fe/H]=$-$1.7) are modelled during hydrogen burning for initial masses of 9-300~M$_{\odot}$ and rotational velocities of 0-900~km~s$^{-1}$. Internal mixing processes in these models were calibrated based on an observed sample of OB-type stars in the Magellanic Clouds.} %Method.
%---%Results.
{Even moderately fast rotators, which may be abundant at this metallicity, are found to undergo efficient mixing induced by rotation resulting in quasi chemically-homogeneous evolution. These homogeneously-evolving models reach effective temperatures of up to 90~kK during core hydrogen burning. This, together with their moderate mass-loss rates, make them Transparent Wind Ultraviolet INtense stars (TWUIN star), and their expected numbers might explain the observed He\,II ionizing photon flux in I~Zw\,18 and other low-metallicity He\,II galaxies. Our slowly rotating stars above $\sim$80~M$_{\odot}$ evolve into late B- to M-type supergiants during core hydrogen burning, with visual magnitudes up to 19$^{\mathrm{m}}$ at the distance of I~Zw\,18. Both types of stars, TWUIN stars and luminous late-type supergiants, are only predicted at low metallicity.} %Results.
%---%Conclusions.
{Massive star evolution at low metallicity is shown to differ qualitatively from that in metal-rich environments. Our grid can be used to interpret observations of local star-forming dwarf galaxies and high-redshift galaxies, as well as the metal-poor components of our Milky Way and its globular clusters.} %Conclusions.

 %Max 6 keys
\maketitle

%---------------------------------------------------------------------------
% Introduction

%\tableofcontents

\section{Introduction}\label{sec:Introduction}

Many of the first stars in the Universe are thought to have started out very massive and almost metal-free \citep{Abel:2002,Bromm:2004,Frebel:2005}. Direct observations of these stars are not possible with current telescopes. However, low-metallicity massive stars can also be found in the local Universe: some of the nearby dwarf galaxies form massive stars at a high rate \citep{Tolstoy:2009,Weisz:2014}. As these galaxies can be directly observed and as their metallicity happens to be close to that of the first stars, they can be used as laboratories to study massive stellar evolution at low (i.e. substantially subsolar) metallicity. Such studies may lead us to a better understanding of the metallicity dependence of stellar evolution, including the first stars in the Universe.

Apart from the cosmological implications of stars at high redshift, there are another reasons to study stellar evolution at low metallicity. The initial chemical composition of a star influences the whole evolutionary path, internal structure, circumstellar surroundings and even the final fate of the star \citep{Meynet:2002,Hirschi:2005,Meynet:2005,Brott:2011a,Yoon:2012,Yusof:2013}. There is observational evidence that long-duration gamma-ray bursts tend to prefer low-metallicity environments \citep{Levesque:2010,Modjaz:2011,Graham:2013} and high redshifts \citep{Horvath:2014}. Theoretical studies have shown that fast rotating stars at low metallicity may evolve quasi chemically-homogeneously \citep{Yoon:2006,Brott:2011a}. 
These homogeneously-evolving stellar models are predicted to become fast rotating Wolf--Rayet (WR) type objects during the post main-sequence phase. They are, therefore, candidates of long-duration gamma-ray burst progenitors within the collapsar scenario
\citep{MacFadyen:1999,Yoon:2005,Woosley:2006}. 
Moreover, broad line type Ic supernovae \citep{Arcavi:2010,Sanders:2012} that are associated with gamma-ray bursts \citep{Modjaz:2011,Graham:2013} as well as the recently identified superluminous supernovae \citep{Quimby:2011,Lunnan:2013} occur preferentially in low-metallicity dwarf galaxies. This may corroborate the idea that reduced wind mass-loss at low metallicity \citep{Vink:2001,Mokiem:2007} may allow for rapid rotation rates \citep{Yoon:2006,Georgy:2009} and very massive \citep{Langer:2007,Yusof:2013,Kozyreva:2014} supernova progenitors. 
A good understanding of the evolution of metal-poor massive stars is, therefore, important to probe the origin of these extremely energetic explosions.

The first stars are thought to have consisted of mostly hydrogen and helium with a $^7$Li mass fraction of about 10$^{-9}$ \citep{Mathews:2005}. This first generation synthesized heavy elements via nuclear fusion, either in hydrostatic equilibrium or during an explosion. Stars that have formed from material processed by the first stars therefore also have non-zero metallicity.
This second generation of stars may also be important in the re-ionisation history and chemical evolution of the early Universe \citep{Yoshida:2007,Greif:2010,Hosokawa:2012}.
Additionally, the imprint of the first nucleosynthesis events is thought to be present in extremely metal-poor Galactic halo stars \citep{Beers:2005,Keller:2014}, for which our understanding is still incomplete \citep{Heger:2010,Lee:2014}. 

Galactic globular clusters are also observed to have a low metal content (\mbox{[Fe/H]=$-$2.2...$-$0.2}) \citep{Gratton:2001,Yong:2003,Caretta:2005,DAntona:2010,Caretta:2010}. Although we observe only low-mass stars in globular clusters today, there was probably a generation of massive stars during their early epoch \citep{PortegiesZwart:2010,Denissenkov:2014,Longmore:2014}. A theoretical understanding of massive stars at this metallicity might help to explain some of the most intriguing phenomena concerning globular clusters, e.g. the abundance anomalies and multiple populations observed in these objects \citep{Decressin:2007,deMink:2009,Bastian:2013}.

We can observe environments at very low but finite metallicity, if we turn to nearby blue compact dwarf galaxies (BCDG) \citep{Searle:1972,Zhao:2013}. BCDGs are typically small, high surface-brightness galaxies of low metallicity, with blue colours and intense emission lines \citep{Hunter:1995,Vaduvescu:2007}. Additionally, some of them contain WR stars, e.g. \izw \citep{Legrand:1997,Aloisi:1999,Schaerer:1999a,Shirazi:2012,Kehrig:2013}. 
Moreover, nearby BCDGs form massive stars at a high rate of up to 1~M$_{\odot}$~yr$^{-1}$ \citep{Annibali:2013}. 
Given that their metallicity is observed to be low, they are laboratories to study the evolution of metal-poor massive stars \citep{Izotov:2002, Izotov:2004, Annibali:2013}. As mentioned above, modelling stellar evolution with a composition suitable for these dwarf galaxies can be an important step towards a deeper understanding of low-metallicity environments. 

Recent studies theorized about the presence of metal-free Population~III (Pop~III) stars in finite-metallicity environments to explain various observational phenomena such as unusually high He~II and Lyman-$\alpha$ emission in local dwarf galaxies or high-redshift galaxies \citep{Heap:2015,Kehrig:2015,Sobral:2015}. However, the detailed evolutionary behaviour of low- but finite-metallicity massive stars has not been investigated comprehensively. With this study, we aim to shed new light on this issue.

We computed stellar evolutionary sequences of single stars in the mass range 9-300~M$_{\odot}$ with rotational velocities between 0-900~km~s$^{-1}$ and with an initial composition of Z=0.0002. Here we present the core-hydrogen-burning phase of these models.
We emphasize therefore that the present study applies only to the main-sequence evolution of low-metallicity massive stars. The post-main-sequence evolution and final fates of our models will be discussed in a following study.

We include rotation into our models because massive stars are generally rapid rotators \citep{Penny:2009,Huang:2010,RamirezAgudelo:2013,Dufton:2013}. Rotation may influence the life of massive stars in many ways \citep{HegerI:2000,Meynet:2000,Hirschi:2005,Yoon:2006,Ekstroem:2008,Georgy:2012}. at low metallicity, rotation may be particularly important because the stellar wind induced spin-down is much weaker \citep[cf.][]{Brott:2011a}, and the stars remain rapidly rotating such that rotational mixing is facilitated \citep{Maeder:2000,Langer:2012}. 

We consider the evolution of isolated single stars. The majority of massive stars may form in binary systems that lead to interaction during their lives \citep{Chini:2012,Sana:2012}, often already during their main-sequence evolution. This can drastically affect their evolution \citep{Eldridge:2008,Eldridge:2011} and binary products may be abundantly present among the brightest stars in dwarf galaxies \citep{deMink:2014,Schneider:2014}. However, in many cases stars are spun up early during their evolution \citep{deMink:2013}. This means that our models provide a fair approximation to the evolution of stars spun up in binary systems. 

Our paper is organised as follows. 
First we summarise the physical assumptions made for calculating the stellar evolutionary models in Sect.~\ref{sec:input}. %2
Then we give an overview of the grid of stellar model sequences and the classification system that describes the different types of evolution at low metallicity in Sect.~\ref{sec:gridclass}. %3
We explain the behaviour of individual stellar tracks in the Hertzsprung--Russell (HR) diagram in Sect.~\ref{sec:HR}. %4
In Sect.~\ref{sec:hbrs}, we analyse the models that evolve into core-hydrogen-burning cool supergiants. %5
In Sect.~\ref{sec:wrHR}, we present the models that evolve into transparent wind UV-intense stars. %6 
An analysis of the helium abundance at the surface and in the core is given in Sect.~\ref{sec:YcYs}. %7
A closer look into the mass-loss history is taken in Sect.~\ref{sec:masshist}. %8
The evolution of the rotational velocity is presented in Sect.~\ref{sec:rotation}. %9
In Sect.~\ref{sec:flux}, we provide information on the ionizing fluxes predicted by our models. %10
In Sect.~\ref{sec:geneva}, we discusse the results in context of previous publications of massive-star evolution at low metallicity. %11
A summary of the results is given in Sect.~\ref{sec:Summary}. %12
The appendices (available only in the online version) provide a summary of the models, a table of the ionizing fluxes, as well as isochrones.
%The evolutionary model sequences and isochrones are available via the CDS archive.

%---------------------------------------------------------------------------
% Summary of input physics

\section{Physical assumptions}\label{sec:input}

We use a one-dimensional hydrodynamic binary evolutionary code (BEC) to compute rotating and non-rotating single stellar evolutionary sequences \citep[see][and references therein]{HegerI:2000,HegerII:2000,Brott:2011a, Yoon:2012}. BEC solves the five stellar structure equations using the implicit Henyey method. It contains detailed state-of-the-art treatment of rotation, magnetic fields, angular momentum transport and mass-loss.

Stellar model sequences are computed under the physical assumptions described in this section. The time between two consecutive models in the sequence is chosen adaptively, resolving the structural changes in detail. 
We typically resolve the core-hydrogen-burning evolution with $\sim$2000 time steps, for which each stellar model is resolved into a similar number of mass shells. The whole set of evolutionary sequences with different initial masses and rotational velocities (but the same initial composition) is referred to as our grid of models.

The calculations were stopped when the central helium abundance reaches Y$_{\mathrm{C}}$=0.98. We choose this as the terminal age main-sequence (TAMS). After this point, as the hydrogen fraction in the core becomes very small, the central temperature increases substantially due to an overall contraction, and the star falls out of thermal equilibrium. For this reason, we exclude this short contracting phase from the analysis of the main-sequence evolution of our stellar models. 

\subsection{Initial chemical composition}

Stellar models with the same initial mass M$_{\rm ini}$ and same initial rotational velocity v$_{\rm ini}$ but different initial composition Z$_{\rm ini}$ may evolve differently for at least two reasons. First, the metallicity has a fundamental impact on the mass-loss rate of a star: the higher the total metal abundance at the surface, the stronger the stellar wind \citep{Kudritzki:1987,Vink:2001,Mokiem:2007,Puls:2008}. Second, due to the reduced radiative opacity and the low amount of CNO nuclei as initial catalysts, metal-poor stars are more compact than corresponding metal-rich ones \citep{Ekstroem:2011,Yoon:2006}.

\begin{figure}[h!]
\resizebox{\hsize}{!}{\includegraphics[angle=270,width=1.\columnwidth]{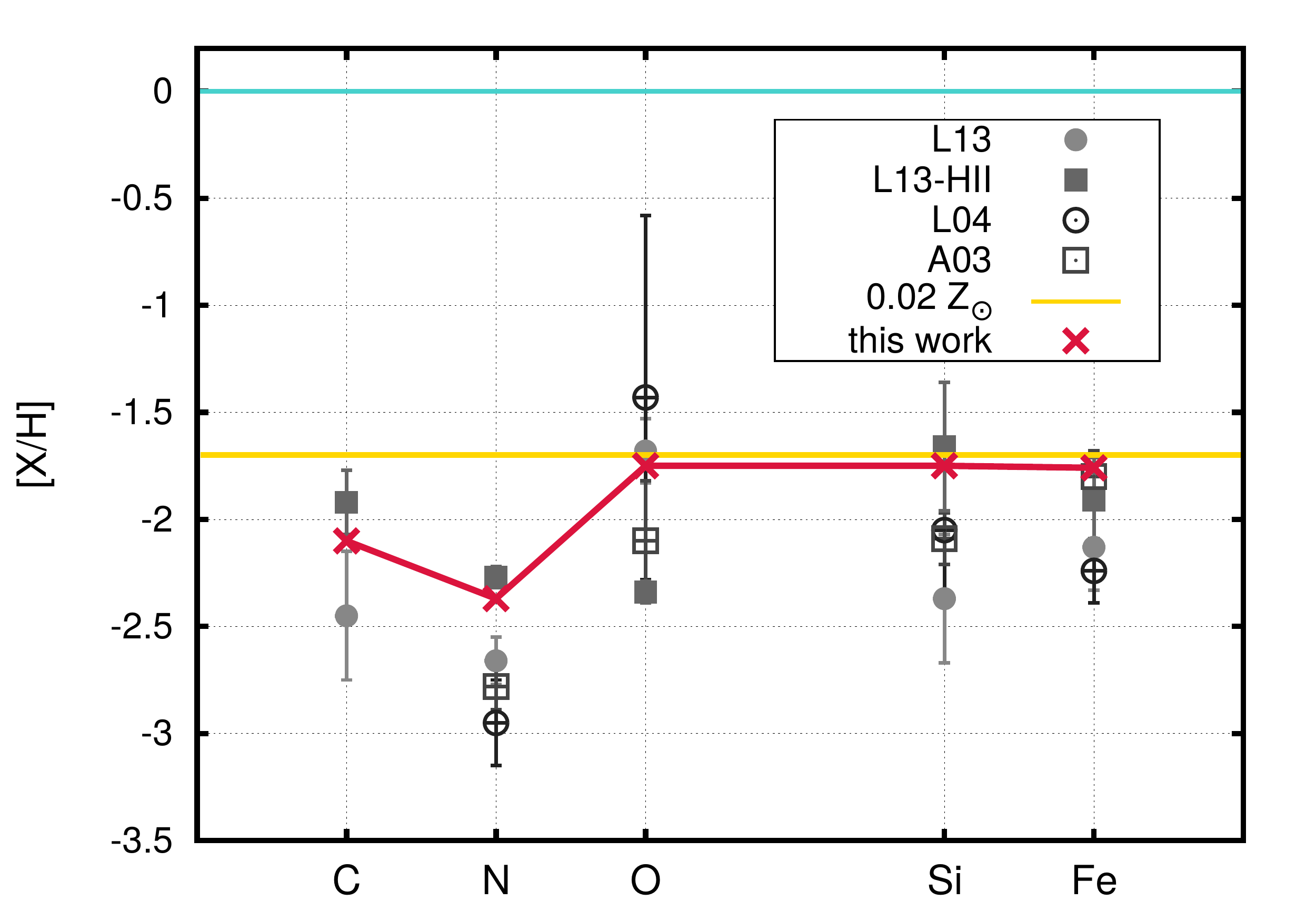}}
\caption{Recent measurements of abundances in \izw compared to our applied composition (i.e. SMC composition scaled down by ten; shown by a red line with crosses). Carbon, nitrogen, oxygen, silicon and iron abundances are given relative to solar \citep{Asplund:2009}: [X/H]=log(X/H)$-$log(X/H)$_\odot$. \textsl{L13}: first column of Table 7 in \citet{Lebouteiller:2013}. \textsl{L13-HII}: last column of the same table, composition of the HII regions. \textsl{L04} and \textsl{A03}: data of previous measurements, taken from \citet{LecavelierdesEtangs:2004} and \citet{Aloisi:2003}, respectively. \textsl{0.02 Z$_\odot$}: solar abundances of \citet{Asplund:2009} scaled down by a factor of 50.
}
\label{fig:szep}
\end{figure}

We compare recent observations of the metal abundance pattern of the Sun and the dwarf galaxy \izw in Fig.~\ref{fig:szep}. In particular carbon and nitrogen are under abundant compared to scaled solar abundances \citep[see also][]{Nicholls:2014}. 
We also plot the composition of the Small Magellanic Cloud (SMC) scaled down by ten. The metal abundance patterns of BCDGs in general are different from that of the Sun \citep{Izotov:1999,Vink:2001,Tramper:2011, Lebouteiller:2013,Nicholls:2014}, showing that the metal abundance pattern of the SMC, which is the nearest metal-poor irregular dwarf galaxy, is a better approximation for the composition of e.g. \izw than the solar abundance pattern. Hence, to obtain theoretical predictions for the massive stars in \izw, we take the abundance pattern of the SMC as in \citet{Brott:2011a}, scale it down by a factor of ten and calculate massive stellar evolutionary models with this composition. The metallicity of our models (i.e. the sum of all metals as mass fraction) is then Z=0.0002.

\begin{figure}[h!]
\resizebox{\hsize}{!}{\includegraphics[width=1.3\columnwidth]{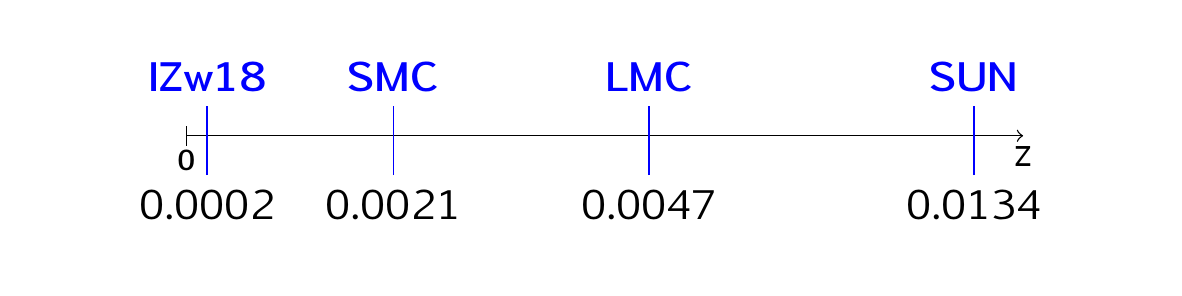}}
\caption{Metallicities on a linear scale. SUN: solar metallicity given by \citet{Asplund:2009}. LMC, SMC: Large and Small Magellanic Clouds \citep{Hunter:2007,Brott:2011a}. IZw18: Z$_{\mathrm{IZw18}}\simeq$~0.1$\times$Z$_{\mathrm{SMC}}$. The zero value corresponds to the metal-free Pop~III stars.}
\label{fig:linea}
\end{figure}

Fig.~\ref{fig:linea} shows the metallicities of the local group galaxies SMC and LMC \citep[as in][]{Brott:2011a}, and of the Sun \citep{Asplund:2009} as well as the metallicity of our \izw models. The zero value corresponds to the nearly metal-free first stars in the Universe called Population~III stars. The metallicity of \izw is very close to that of Pop~III stars on a linear scale; however, there are differences between our models and models of Pop~III stars \citep[see Sect.~\ref{sec:flux},][and Szécsi et al. in prep.]{Yoon:2012}.

For the initial value of helium, we assume that the mass fraction scales with the metallicity between the primordial helium mass fraction \citep{Peimbert:2007} and the solar value \citep{Grevesse:1996}. Thus, the initial helium abundance in our Z$_{\rm ini}$=0.0002 stellar models is Y$_{\rm ini}$=0.2477. 

Radiative opacities were interpolated from the OPAL tables \citep{Iglesias:1996} with solar-scaled metal abundances, with their iron abundance used as the interpolation parameter for metals. Our models thus correspond to a metallicity of \mbox{[Fe/H]=$-$1.7} and Z$\approx$Z$_{\odot}/50$. %using (Fe$_{IZw18}$/Fe$_{\odot}$)~$\times$~Z$_{\odot}$

\subsection{Physics of the stellar interior}\label{sec:mixing}

All the mixing processes considered here are modelled as diffusive processes. Convection is treated according to the mixing-length theory (MLT) \citep{Boehm:1958} with an MLT parameter of $\alpha_{\rm MLT}=1.5$ \citep{Langer:1991}. Semi-convection is considered with an efficiency parameter of $\alpha_{\rm SEM}=1$ \citep{Langer:1983,Langer:1991}, 
although it has minor effects on the models during the main-sequence evolution. As no calibration of the convective core overshooting parameter exists for stars of the considered metallicity, we rely on the work of \citet{Brott:2011a} who calibrated the overshooting against the rotational properties of B-type stars from the VLT-FLAMES survey \citep{Hunter:2008,Vink:2010} as $\alpha_{\rm over}=0.335 H_p$, where $H_p$ is the local pressure scale height. It has been suggested by \citet{Castro:2014} and confirmed by \citet{McEvoy:2015} that convective core overshooting of Galactic stars is probably mass-dependent and, at high mass ($\gtrsim$~15~M$_{\odot}$), stronger than previously thought. However, the metallicity dependence of this effect still needs to be investigated.

Rotationally-induced mixing of chemical elements is treated with an efficiency parameter $f_c=0.0228$ \citep{HegerI:2000,HegerII:2000}, calibrated by \citet{Brott:2011a}. Furthermore, transport of angular momentum by magnetic fields due to the Spruit--Taylor dynamo \citep{Spruit:2002,Heger:2005} is included, which is assumed here not to lead to additional transport of chemical elements \citep{Spruit:2006,Suijs:2008}.

\subsection{Mass-loss}\label{sec:massloss}

For the early evolutionary stages of our models, we use the mass-loss rate prescription of \citet{Vink:2000}, which includes a bi-stability jump at $\sim$25~kK. The dependence of mass-loss on the metallicity is additionally implemented according to \citet{Vink:2001} as $\dot{M}\sim Z^{0.86}$. Approaching the empirical Humphreys--Davidson limit (thought to be connected to the Eddington limit), O and B stars may experience an increase in mass-loss, which is taken into account by using the empirical mass-loss rate prescription of \citet{Nieuwenhuijzen:1990} 
(with the same $Z$ dependence as above) 
if its predicted mass-loss rate is higher than that of \citet{Vink:2000,Vink:2001} at any effective temperature smaller than $\sim$22~kK.

Since we find some of our models to evolve into cool supergiants (T$_{\rm eff}\lesssim 12$~kK) even during their main-sequence lifetime, we need to take the mass-loss of cool supergiant stars into consideration. In general, mass-loss of such stars is observed to be higher than that of O and B stars due to the low surface gravity at their large radii (>1000~$R_{\odot}$) and possibly due to dust formation in cool atmospheres \citep{Groenewegen:2009}. However, quantitative physical models of such winds still have deficiencies, hence we rely on the empirical parametrization of the mass-loss rate following the prescription of \citet{Nieuwenhuijzen:1990}. This prescription is a revised version of that of \citet{deJager:1988}, which has been shown by \citet{Mauron:2011} to be still applicable in the light of new observations of \textsl{bona-fide} red supergiants. The metallicity dependence of these winds is implemented as $\dot{M}\sim Z^{0.85}$ according to \citet{Vink:2001}. This formula is in accordance with the results of \citet{Mauron:2011} who find that the metallicity exponent should be between 0.5 and 1. 

Our calculations predict strong surface helium enrichment even during core hydrogen burning as a result of fast rotation (see Sects.~\ref{sec:wrHR} and \ref{sec:YcYs}). We use the prescription of \citet{Hamann:1995} for the winds of our models when the surface helium abundance is Y$_S \geq$~0.7 with reduction by a factor of 10 as suggested by \citet{Yoon:2006}.  This reduction gives a mass-loss rate comparable to the most commonly adopted one by \citet{Nugis:2000} \citep[see Fig.~1 of][]{Yoon:2015}. The \citet{Hamann:1995} prescription is applied together with a metallicity dependence of $\dot{M}\sim Z^{0.86 }$ \citep{Vink:2001}. For stars with surface helium abundances of 0.7~$\geq$~Y$_S \geq$~0.55, we interpolate linearly between the reduced \citet{Hamann:1995} mass-loss rate and the rate of \citet{Vink:2000,Vink:2001}.

A mass-loss enhancement is implemented for stars rotating near their critical rotation which includes their Eddington factor \citep{Langer:1997,Yoon:2005}. It remains unclear whether rapid rotation \textsl{per se} leads to an increase in mass-loss \citep{Mueller:2014}. However, as discussed in \citet{Mueller:2014}, it still appears reasonable to consider that the mass-loss rate does increase close to the Eddington limit \citep{Langer:1997,Grafener:2011}.

%\newpage
%---------------------------------------------------------------------------
% Classification

\section{The grid of stellar models}\label{sec:gridclass}

\begin{figure*}
\centering
\includegraphics[height=\twice\columnwidth,angle=270]{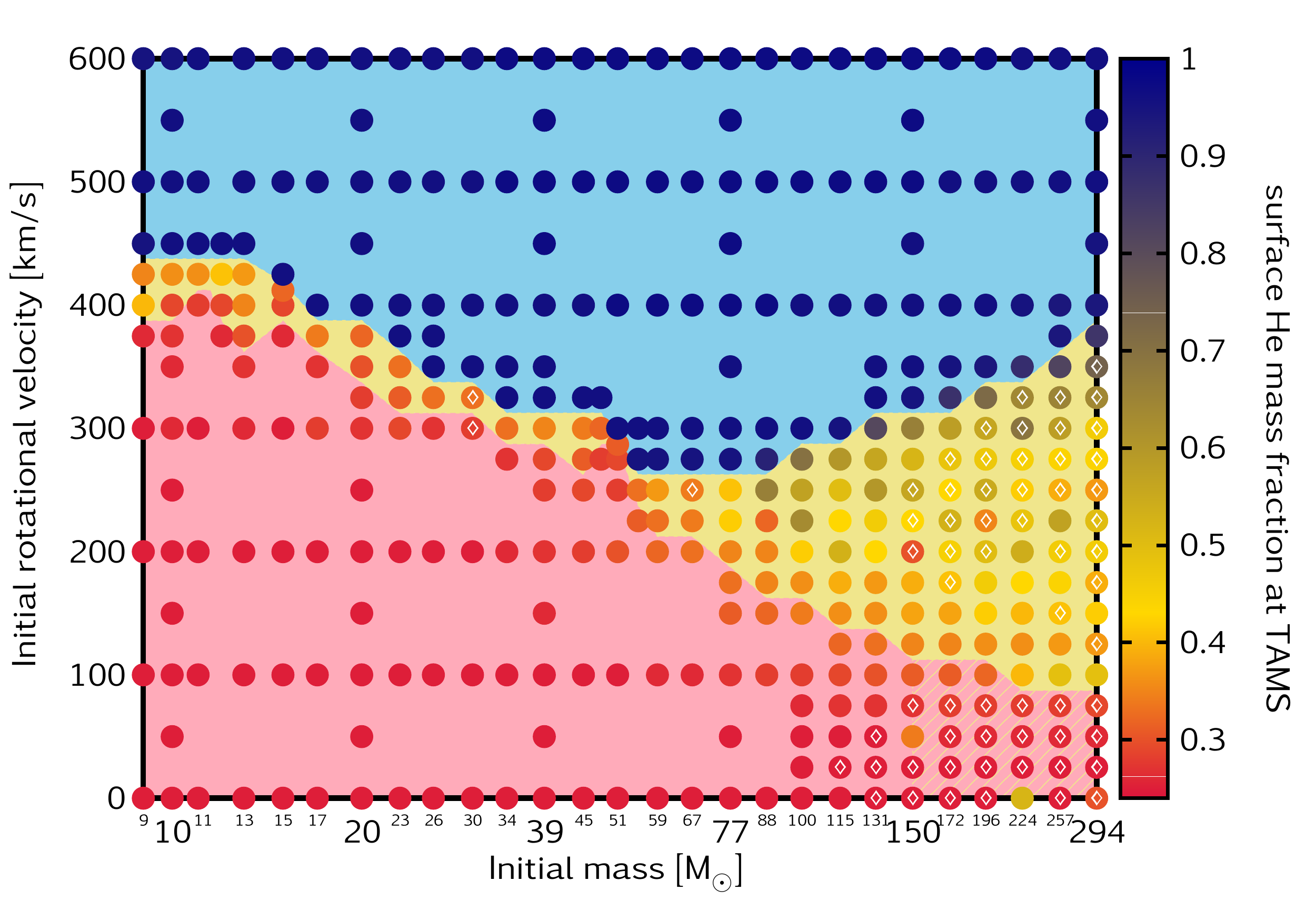}
\caption{Grid of 375 evolutionary sequences of single stars. Each evolutionary sequence of our grid is represented by one dot in this diagram. Sequences inside the blue shaded region follow chemically-homogeneous evolutionary paths evolving bluewards in the HR~diagram and having a surface helium abundance of Y$_S\simeq$~0.98 at the TAMS. Sequences inside the red region follow normal evolution, keeping Y$_{\mathrm{S}}$ close to the initial value of $\simeq$~0.24. Sequences inside the yellow region deviate from normal evolution: either they start their lives evolving chemically homogeneously and then switch to normal evolution, or they start normal evolution and mass-loss uncovers their helium-rich layers (cf. Sect.~\ref{sec:YcYs}). 
Diamonds mark the sequences that have not reached the TAMS (i.e. the calculation was stopped between 0.82<Y$_{\mathrm{C}}$<0.98), and the yellow-dashed pattern indicates that the separation line between the red and yellow regions is uncertain at the highest masses. 
}
\label{fig:mygrid}
\end{figure*}

Stellar model sequences were computed under the physical assumptions described in Sect.~\ref{sec:input}. Each sequence is represented by one dot in the diagram in Fig.~\ref{fig:mygrid}. 

	%density
The distribution of the sequences in the initial parameter space is chosen to support a study of synthetic populations. For such a study, an interpolation between the sequences would be needed, which is easier to do if the model grid is dense enough -- especially in the regions where the models are most varied. Therefore, we increased the number of computed models in the yellow region, which represents the transition between normal and chemically-homogeneous evolution, because these models show more variations. Additionally, we increased the number of computed models in the corner of the very massive slow rotators (which become core-hydrogen-burning cool supergiants) in order to study their evolution in more detail.

	%mass range
The initial masses of the models in our grid are chosen roughly equidistant on a logarithmic scale. The most massive stars found so far in the local universe are suggested to have an initial mass around 300~M$_{\odot}$ \citep{Crowther:2010,Schneider:2014}. Therefore, while stars more massive than this might be important in the presence of a top-heavy initial mass function \citep[e.g.][]{Ciardi:2003,Dabringhausen:2009} or in large starbursts \citep{Treu:2010,Sonnenfeld:2012,Chabrier:2014}, we use 294~M$_{\odot}$ here as an upper limit. 
%While some studies suggest a top heavy initial mass function \citep[IMF;][]{Salpeter:1955,Kroupa:2001} to be present in low-metallicity environments \citep[e.g.][]{Ciardi:2003,Dabringhausen:2009}, others are favouring a universal Salpeter IMF \citep{Treu:2010,Sonnenfeld:2012,Chabrier:2014}. 

	%colours and symbols
The colouring of the dots in Fig.~\ref{fig:mygrid} represents the surface helium mass fraction at the end of the main sequence (cf. Sect.~\ref{sec:YcYs}). The red, yellow and blue regions indicate the type of evolution a given model undergoes, as described in Sects.~\ref{sec:grid} and \ref{sec:YcYs}.

\subsection{Rotational velocities}\label{sec:zams}

\begin{figure}
\centering
\includegraphics[height=1\columnwidth,angle=270]{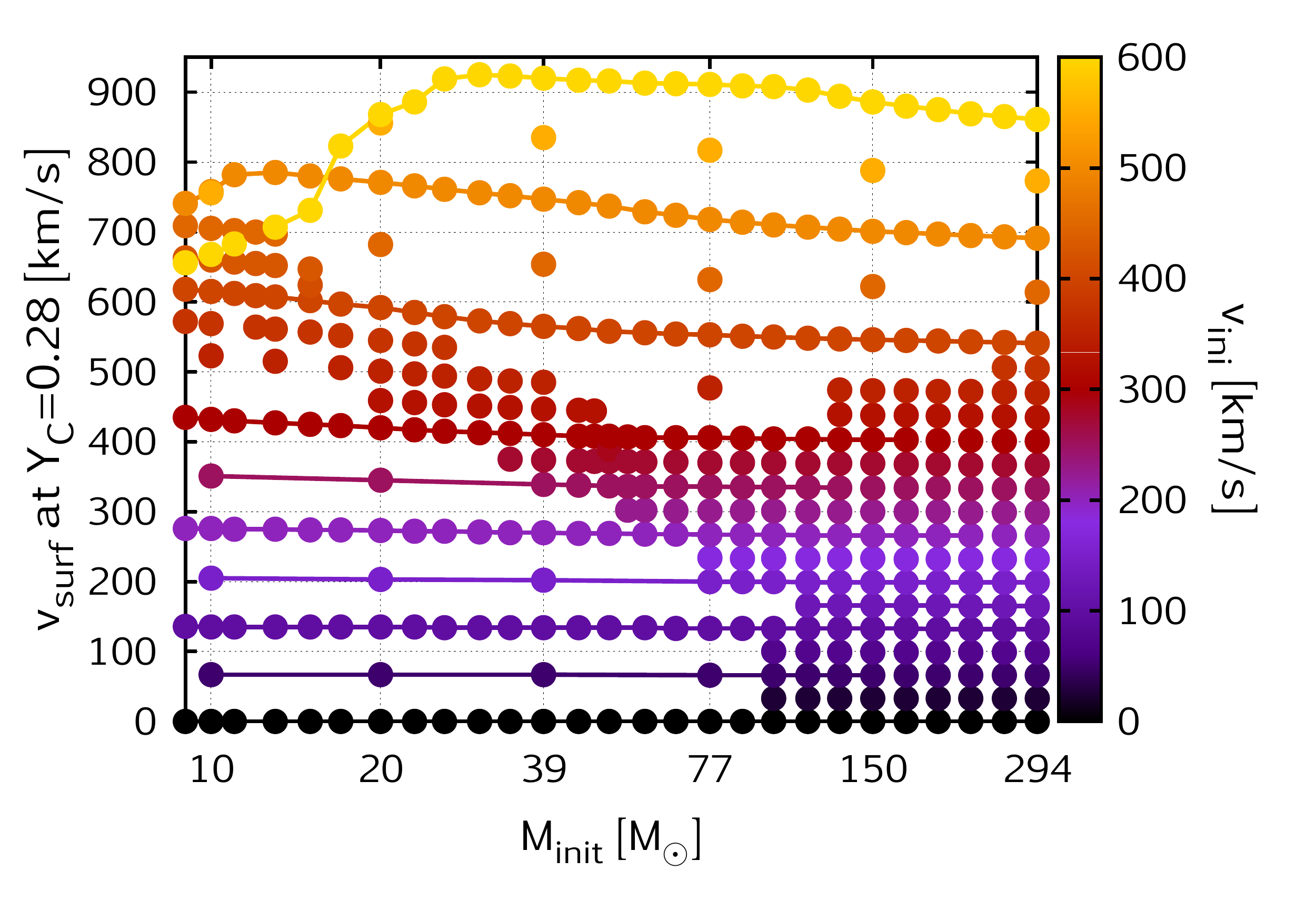}
\caption{
Surface rotational velocity at the ZAMS (cf. Sect.~\ref{sec:zams}). Every dot represents one evolutionary sequence, cf. Fig.~\ref{fig:mygrid}. The colours refer to the \textsl{initial} surface rotational velocity, v$_{\rm ini}$. Sequences with v$_{\rm ini}$=50, 100, 125, 150, 200, 250, 300, 400, 500 and 600~km~s$^{-1}$ are connected by lines. 
}
\label{fig:initzams}
\end{figure}

	%rotvel range
The Y-axis in Fig.~\ref{fig:mygrid} refers to the initial equatorial rotational velocity at the surface of our models. We chose to cover a wide range in initial rotational velocity from zero up to 600~km~s$^{-1}$. The models start out chemically homogeneous and in hydrostatic and thermal equilibrium initially. We emphasise that the initial rotational velocities refer to the values with which the calculations were started, and are generally significantly lower than the rotational velocity after hydrogen burning has reached CNO equilibrium, i.e. at central helium abundance Y$_{\mathrm{C}}\simeq 0.28$ (Fig.~\ref{fig:initzams}). We define this point in time as the zero-age main-sequence (ZAMS). The reason for the rotational velocity at the ZAMS being higher than initially is that at the beginning of the calculation, the star undergoes a short phase of structural changes while approaching CNO equilibrium. During this adjustment phase, the star contracts, spins up and thus continues its evolution with increased rotation. On average, our models rotate about 30\% faster than the nominal (i.e. Y-axis in Fig.~\ref{fig:mygrid}) initial rotational velocity indicates.

Stellar models in the left top corner of Fig.~\ref{fig:initzams} with M$\lesssim$17~M$_{\odot}$ and v$_{\rm ini}$=600~km~s$^{-1}$ rotate slower at the ZAMS than models with v$_{\rm ini}$=500~km~s$^{-1}$. This is because although the models with v$_{\rm ini}$=600~km~s$^{-1}$ also spin up initially, they nearly reach their Keplerian velocity during the early contraction phase. Stellar models close to the breakup rotation undergo enhanced mass-loss, so they lose mass and spin down at the beginning of the evolution. 
In this phase, the one-dimensional models only provide crude approximations of fast rotating stars
\citep[see e.g.][for a discussion of stars close to the breakup 
rotation]{Decressin:2007,Chiappini:2011,Krticka:2011,Espinosa:2013}. In particular, when the 
surface rotational velocity approaches the break-up velocity, angular momentum may be removed by losing mass into 
an equatorial, viscous decretion disc, as discussed by \citet{Krticka:2011}. The effects of the 
decretion disc on the evolution of our fast rotating massive stars still remains to be studied.

\subsection{Normal, homogeneous and transitionary evolution}\label{sec:evol}

The grid in Fig.~\ref{fig:mygrid} consists of 375 sequences, from which 
142 are classified as normal evolution (NE),
123 as (quasi) chemically-homogeneous evolution (CHE), and
110 as transitionary evolution (TE). 
The calculation of some sequences with NE and TE were stopped before reaching the TAMS due to numerical difficulties. However, all the sequences in the grid reached a core helium mass fraction of Y$_C\gtrsim 0.82$. In Fig.~\ref{fig:initzams}, the sequences that were not followed until the TAMS are marked. 

Models with NE develop a core-envelope structure: the core is chemically mixed through convection and fuses hydrogen into helium, while the envelope largely retains its initial composition. Their radii increase during the main-sequence lifetime because a chemical gradient develops and because the envelope inflates in the case of the highest-mass models (see the discussion in Sect.~\ref{sec:hbrs}). We also refer to Sect.~\ref{sec:HR} for the discussion of the HR~diagram, in which the models evolve towards lower effective temperatures (redwards).

Chemically-homogeneous evolution was first described by \citet{Maeder:1987} in the context of rotation. Several authors have since investigated this evolutionary behaviour \citep[see e.g][]{Yoon:2005,Yoon:2006,Cantiello:2007,Meynet:2007} and have reported observational support for it \citep{Walborn:2004,Eldridge:2012,Martins:2013}. Models with CHE develop only shallow chemical gradients between the core and the envelope and all the nuclear products are mixed throughout the star and reach the surface. We investigate their surface helium abundance and the optical depth of their winds in Sect.~\ref{sec:wrHR}. 

Transitionary evolution was introduced by \citet{Yoon:2012} for Pop~III sequences where the surface helium mass fraction Y$_{\mathrm{S}}$ becomes larger than 0.7 at the TAMS, but the post-main-sequence evolution proceeds redwards. However, in their grid of 51 stellar sequences, only three sequences were identified as TE. We decided to use this expression in a broader sense: to describe a behaviour when the model starts evolving homogeneously and, at some point of the main-sequence lifetime, turns to normal evolution due to angular momentum loss in the stellar wind (see also Sect.~\ref{sec:YcYs}). Note that this revised definition of TE considers only the main-sequence phase. 

\subsection{The structure of the grid}\label{sec:grid}

In this section, we analyse the grid of stellar sequences shown in Fig.~\ref{fig:mygrid}. A prominent feature for our grid is the shape of the transition region shown in yellow in Fig.~\ref{fig:mygrid}. This region is narrow in the lower-mass regime (9-55~M$_{\odot}$). For higher masses (55-294~M$_{\odot}$), however, it covers a larger range of initial rotational velocities. The higher the mass, the more sequences follow TE. 

At masses lower than $\sim$55~M$_{\odot}$ in Fig.~\ref{fig:mygrid}, the bifurcation between NE and CHE is sharp, and there is a very small transitionary region between them. For these masses, the initial rotational velocity at which a star evolves homogeneously decreases with the initial mass. This is consistent with the finding of \citet{Yoon:2012}, who showed for stars in the mass range of 13-60~M$_{\odot}$ that the ratio of the timescale of the Eddington--Sweet circulation $\tau_{\rm ES}$, which governs the mixing in our models, and the main-sequence lifetime $\tau_{\rm MS}$ is systematically smaller for a higher-mass star. The ratio $\tau_{\rm ES}$/$\tau_{\rm MS}$ becoming lower with mass is related to higher radiation pressure and lower density in higher-mass stars. Therefore, for a given initial rotational rate, CHE is favoured in higher-mass stars. Although \citet{Yoon:2012} applied this reasoning to metal-free massive stars, our low-metallicity stellar models nevertheless follow the same principles. 

In the regime above 55~M$_{\odot}$ in Fig.~\ref{fig:mygrid}, mass-loss effects are contributing significantly. Mass-loss influences the evolution at least in two ways. First, mass-loss removes angular momentum \citep{Langer:1998}. This can make an initially fast rotating star spin down and turn to normal, redwards evolution. Second, if enough mass is lost, deeper, helium-rich layers can be uncovered. This way the star appears more blue. Which effect of these two is more dominant, depends on the actual angular momentum and the size of the convective core, as explained below. 

Slow rotators follow normal redward evolution, and angular momentum loss has no significant effect on them. At masses $\gtrsim$80~M$_{\odot}$, slowly rotating (v$_{\rm ini}\lesssim$100~km~s$^{-1}$) models evolve into cool supergiants before core-hydrogen exhaustion due to envelope inflation (Sect.~\ref{sec:hbrs}). 
As we show in Sect.~\ref{sec:YcYs}, these supergiant models may expose helium-rich layers near the TAMS due to the strong mass-loss and the deep convective envelope. Therefore, some of them are marked by orange coloured dots in Fig.~\ref{fig:mygrid}. 

The normally-evolving models that are close to the yellow transition region also have orange colours. This implies that there is no clear separation between normally-evolving and transitionary-evolving models in the mass range 100-294~M$_{\odot}$. The transition here happens smoothly, and the separation line between the red and yellow regions that we draw in Fig.~\ref{fig:mygrid} in this mass range is somewhat arbitrary.

Additionally, as Fig.~\ref{fig:initzams} shows, for most of the sequences in the bottom right corner of the grid the calculation of the last model did not converge, so the simulations were stopped before reaching the TAMS. If these sequences were continued until Y$_{\mathrm{C}}$=0.98, they would probably expose deeper layers and would also appear more orange in Fig.~\ref{fig:mygrid}, and that would move the separation line between the red and yellow regions towards slower rotations, so we marked this uncertain part of the diagram with a dashed pattern. However, the fact that some of the models are unevolved does not explain all the diversity in the surface helium and the colours in the bottom right corner of the grid in Fig.~\ref{fig:mygrid}. The two consequences of mass-loss (the induced spin down due to angular-momentum loss and the uncovering of the deep-lying helium-rich layers) shape the surface properties of the models at the TAMS. Additionally, these models increase their radii, making the stars appear more red and, due to the effective core-envelope coupling (Sect.~\ref{sec:vsurf}), spin up. The consequence of these two competing mechanisms is that the models in the bottom right corner of the grid show diversity in the surface helium value at the TAMS (and also in the surface rotational velocity at the TAMS, as we discuss in Sect.~\ref{sec:red}).

At moderate initial rotation ($\sim$200-350~km~s$^{-1}$) angular momentum loss is important for very massive stellar models ($\gtrsim$88~M$_{\odot}$) and can turn the evolution from homogeneous to transitionary. The loss of angular momentum causes mixing to become inefficient. A star with inefficient mixing starts to possess a steep chemical gradient between the mixed core and a non-mixed envelope. This prevents CHE for the very massive stellar models in the upper part of the transitionary region. Their spindown behaviour shapes the boundary between the blue and yellow regions: models with TE in the yellow region would be models with CHE if there were no mass and angular momentum loss. The borderline velocity between the blue and the yellow region is increasing with mass above 55~M$_{\odot}$. 

If the rotation is fast enough, mass-loss cannot spin the star down enough to prevent the overall mixing. The fastest rotators therefore undergo CHE over their whole lifetime. They are enclosed in the blue region in Fig.~\ref{fig:mygrid}. 

Summarising, the slowest initial rotation (250~km~s$^{-1}$) showing chemically-homogeneous evolution occurs for stars of 55-88~M$_{\odot}$. Stars less and more massive than this need to rotate faster than 250~km~s$^{-1}$ initially to follow a homogeneous evolutionary path, because either the Eddington--Sweet timescale is too large (in the case of the lower-mass regime) or too much angular momentum is lost in the wind (in the case of the higher-mass regime). 

%\newpage
%---------------------------------------------------------------------------
% Evolutionary tracks

\section{Evolutionary tracks in the HR~diagram}\label{sec:HR}

\begin{figure*}[ht!]
\centering
\includegraphics[height=\zerosix\columnwidth,angle=0]{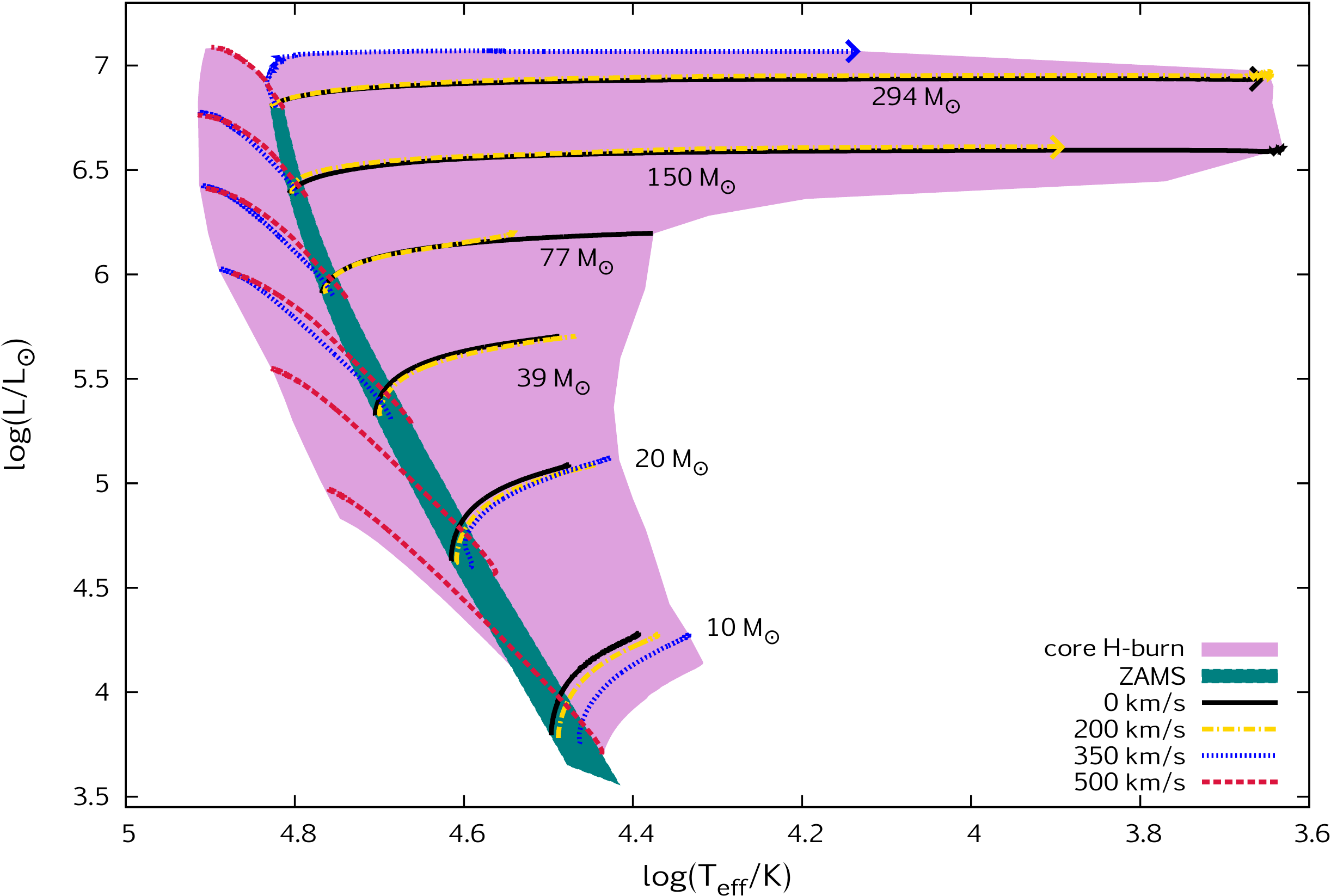}
\caption{Evolutionary tracks in the HR~diagram during core hydrogen burning for models with initial masses between 9-300~M$_{\odot}$ (see labels) and initial rotational velocities of 0, 200, 350 and 500~km~s$^{-1}$, with a composition of 1/10~Z$_{\mathrm{SMC}}$. The lighter (purple) shading identifies the region in which all models of our grid undergo core hydrogen burning. The darker (green) shading identifies the zero-age main-sequence. An arrow marks the end of the tracks for models that were stopped before the terminal age main-sequence was reached. Core-hydrogen-burning objects are expected to be found on both sides of the ZAMS, inside the purple coloured region. }
\label{fig:tracks}
\end{figure*}

In this section, we discuss the evolution of our low-metallicity massive stars in the HR diagram (see Fig.~\ref{fig:tracks}). The tracks that are plotted constitute a representative subset of our grid. Here we summarise their typical behaviour.

Slow rotators (v$_{\rm ini}$=0-200~km~s$^{-1}$ in Fig.~\ref{fig:tracks}, more precisely those in the red region of Fig.~\ref{fig:mygrid}) evolve from the ZAMS towards lower effective temperatures (i.e. \textsl{redwards}) and towards higher luminosities, which represents normal evolution. In contrast, fast rotating stars ($\gtrsim$500~km~s$^{-1}$ in Fig.~\ref{fig:tracks}, those in the blue region of Fig.~\ref{fig:mygrid}) turn towards higher temperatures (bluewards) from the beginning, following CHE \citep{Maeder:1987}. The bifurcation between redward NE and blueward CHE has been studied by e.g. \citet{Brott:2011a} who showed that the lower the metallicity, the more predominant the CHE becomes. 

Models shown in Fig.~\ref{fig:tracks} with intermediate initial rotational velocities ($\sim$200-350~km~s$^{-1}$) might evolve either normally or chemically-homogeneously, depending on their mass. In some cases, however, we can classify them neither NE nor CHE because the model shows properties of both evolutionary classes. For example, the 294~M$_{\odot}$ model with 350~km~s$^{-1}$ evolves first chemically-homogeneously then turns to normal evolution, which is defined as transitionary evolution (represented by the yellow region in Fig.~\ref{fig:mygrid}).

The type of evolution is not only a function of the rotational velocity but also of the initial mass. In Fig.~\ref{fig:tracks} one can observe the behaviour of the 350~km~s$^{-1}$ models: the lowest mass models (9-23~M$_{\odot}$) undergo NE, i.e. they evolve normally and redwards in the HR~diagram, while higher mass models (26-257~M$_{\odot}$) undergo CHE, i.e. they evolve chemically-homogeneously and bluewards. The 294~M$_{\odot}$ model with 350~km~s$^{-1}$ is a transitionary case. We investigate the dependence of the evolutionary types on initial mass and rotation in Sect.~\ref{sec:grid}.

The ZAMS positions of our models is shown by the green shaded region in Fig.~\ref{fig:tracks}. It is a broad region instead of a line due to the different rotation rates of the ZAMS models. Centrifugal acceleration reduces the effective gravity so while the radius of the rotating stellar model is higher, its temperature and luminosity are lower compared to a non-rotating stellar model of the same mass \citep[cf. Fig.3 in][]{Koehler:2015}.

Purple shading in Fig.~\ref{fig:tracks} represents the region which encloses all our models that burn hydrogen in their core. Due to the presence of the stars with CHE at this low metallicity, the purple main-sequence region encompasses the green ZAMS region. Our evolutionary calculations thus predict \textsl{hydrogen-burning massive stars to be found on both sides of the ZAMS}. 

Some of the tracks stopped at the upper red side of the purple region due to numerical instabilities (see also the white diamonds in Fig.~\ref{fig:initzams}). Therefore, the upper borderline of the main-sequence region is approximate and might change (however not significantly) if all models were continued until Y$_C=0.98$. 

There is a significant difference between the redwards evolving lower- and higher-mass stellar sequences. Lower-mass ($\lesssim$~80~M$_{\odot}$) models stay more or less close to the ZAMS, never reaching log(T$_{\mathrm{\rm eff}}$/K) values lower than $\sim$4.3. Higher-mass models, on the other hand, evolve all the way to the cool supergiant region (T$_{\rm eff}<12$~kK) before core-hydrogen exhaustion. These high-mass objects are, therefore, \textsl{core-hydrogen-burning cool supergiants} during the last 5-15\% of their main-sequence lifetimes. 

%---------------------------------------------------------------------------
% HBRS

\section{Core-hydrogen-burning cool supergiants}\label{sec:hbrs}

The models of $\gtrsim$80~M$_{\odot}$ in our grid with slow or intermediate rotation rates spend the last 5-15\% of their main-sequence evolution on the cool supergiant branch with T$_{\rm eff}<12$~kK.
We call this evolutionary phase the core-hydrogen-burning cool supergiant phase. 

\begin{figure}[h!]
\begin{center}
\resizebox{\hsize}{!}{\includegraphics[angle=270]{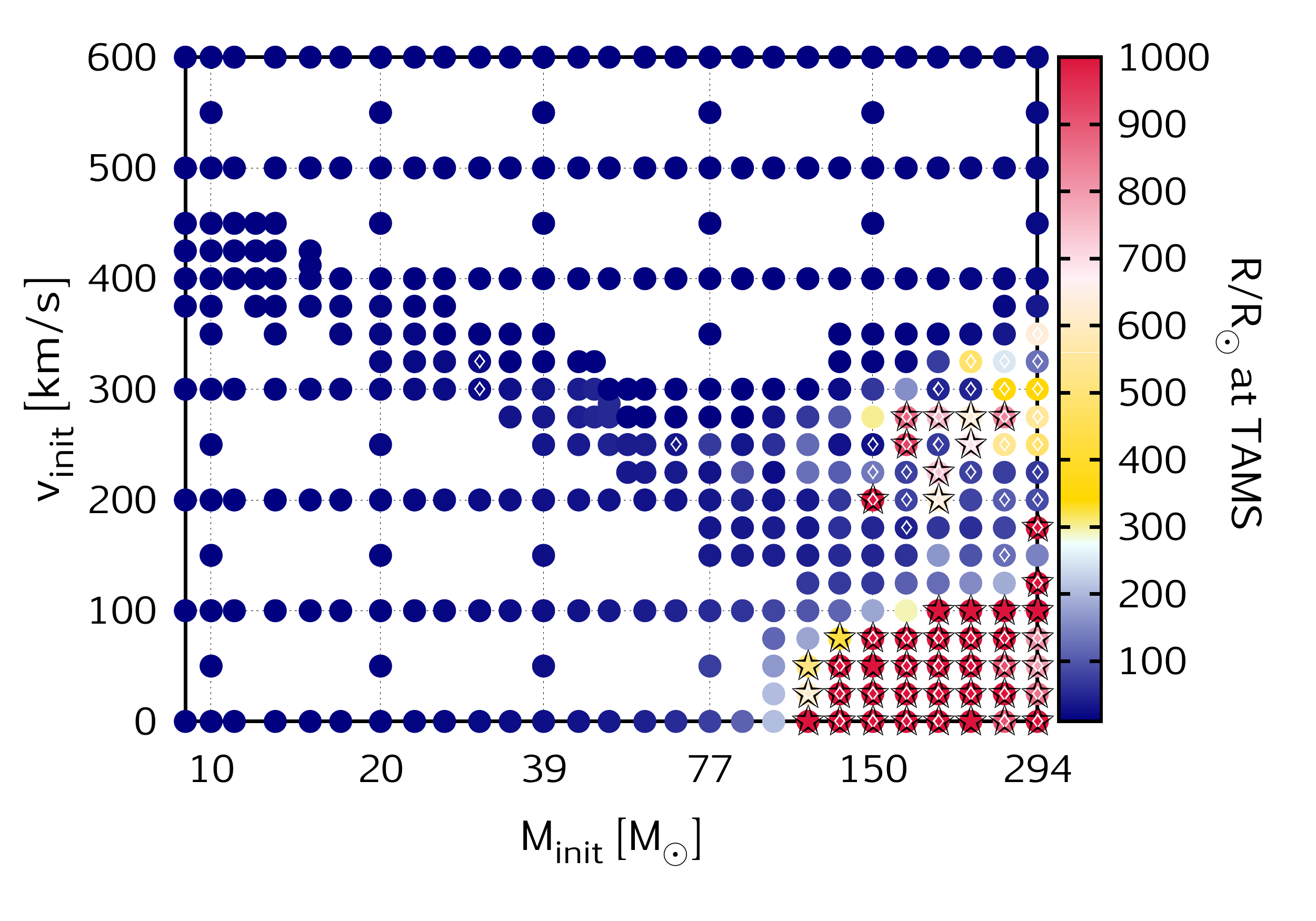}}
\end{center}
\caption{Radius at the end of the main-sequence evolution as a function of initial mass and rotational velocity. 
The core-hydrogen-burning cool supergiants (defined as T$_{\rm eff}^{\rm TAMS}<12$~kK) are found at high mass and slow or intermediate rotation. We mark them with a star symbol. White diamonds mark the sequences that have not reached the TAMS (i.e. the calculation was stopped between 0.82<Y$_{\rm C}$ <0.98).
}
\label{fig:radius}
\end{figure}

Fig.~\ref{fig:radius} shows the radius of our stellar models at the TAMS. 
The fast rotating, chemically-homogeneously-evolving models all remain compact and blue, while the massive (M$_{\rm ini}\gtrsim$80~M$_{\odot}$) models with normal and transitionary evolution expand during the main-sequence lifetime. They may reach T$_{\rm eff}$ values below $12$~kK and radii larger than 1000~R$_{\odot}$, and become core-hydrogen-burning cool supergiants near the TAMS.

The reason for the expansion of our massive unmixed models is their proximity to the Eddington limit. \citet{Koehler:2015} and \citet{Sanyal:2015} find that this occurs for stars above $\sim$50~M$_{\odot}$ in LMC models, whose mass-loss, however, removes the hydrogen-rich envelope such that stars above $\sim$100~M$_{\odot}$ do not become that cool. We note that even very massive zero-metallicity models have been shown to become red supergiants during core hydrogen burning \citep{Marigo:2003,Yoon:2012}.

\begin{figure}[h!]
\resizebox{\hsize}{!}{\includegraphics[angle=270,page=1]{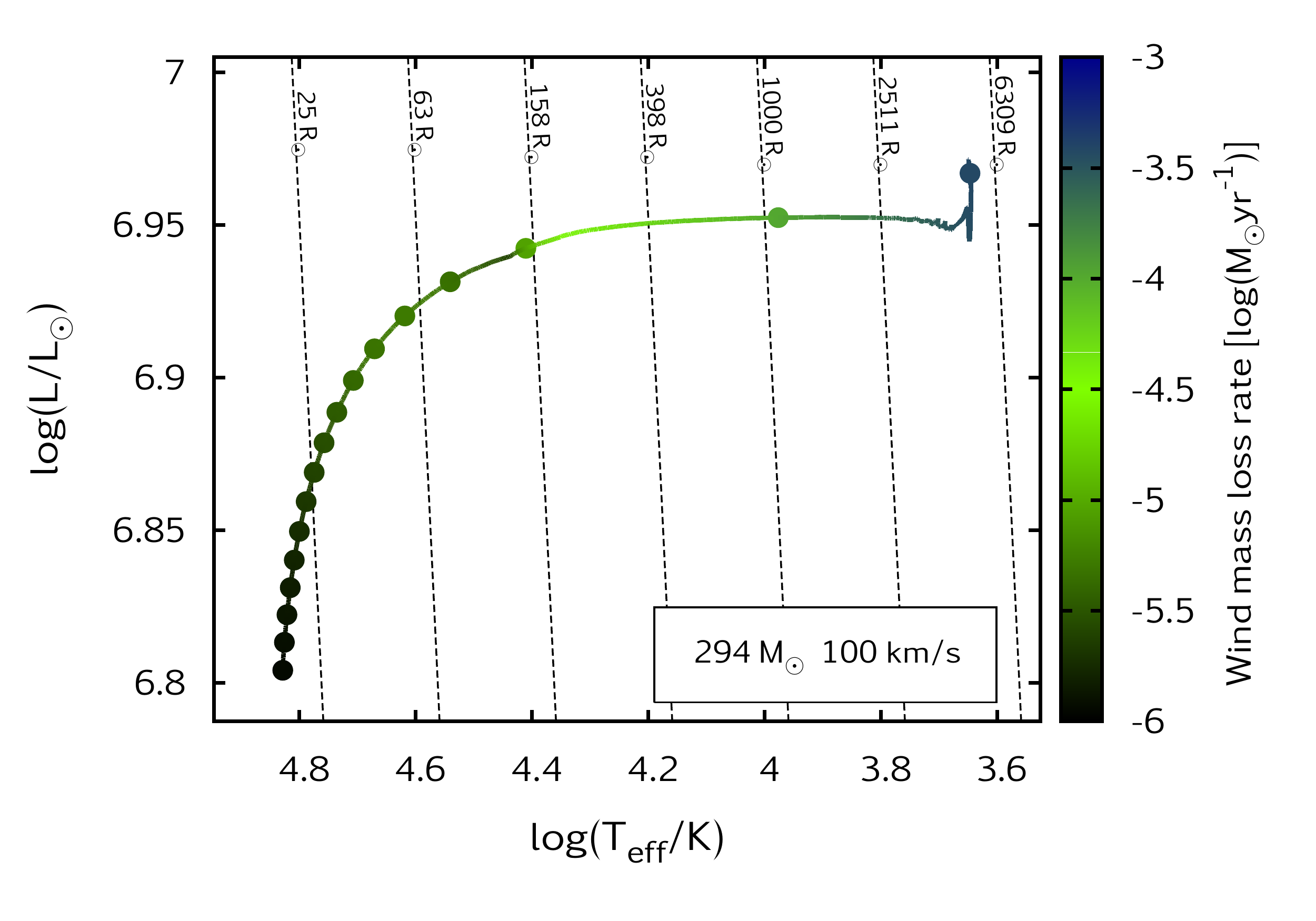}}
\caption{Evolutionary track of our model with M$_{\rm ini}$=294~M$_{\odot}$ and v$_{\rm ini}$=100~km~s$^{-1}$ during core hydrogen burning in the HR~diagram. Dots mark every 10$^5$~years of evolution. The stellar wind mass-loss rate is colour coded; black dashed lines of constant radii are labelled according to their radius value. The star becomes a core-hydrogen-burning cool supergiant during the last 15\% of its main-sequence evolution. 
}
\label{fig:kipMS}
\end{figure}

Fig.~\ref{fig:kipMS} shows the evolution of our slowly rotating stellar sequence with 294~M$_{\odot}$ in the HR~diagram. After the first 1.5~Myr, the radius inflates from 150~R$_{\odot}$ to 5100~R$_{\odot}$ within 0.2 Myr. Thus, the model spends $\sim$0.3~Myr (15\% of the total main-sequence lifetime) as a core-hydrogen-burning cool supergiant before hydrogen exhausts in the core. During this time, the mass-loss rate is very high (up to 4$\cdot$10$^{-4}$~M$_{\odot}/$yr). The star loses mass rapidly and ends up with 244~M$_{\odot}$ at the TAMS. However, it still retains a hydrogen-rich envelope of $\sim$60~M$_{\odot}$ at this time.

As seen in Fig.~\ref{fig:radius}, several sequences evolve similarly to the 294~M$_{\odot}$ sequence discussed above, reaching surface temperatures below 12~kK. There are two distinct regions containing core-hydrogen-burning supergiants, one at high mass and slow rotation, and the other at high mass and around 275~km~s$^{-1}$ initial rotation. The slow rotators evolve normally during the first part of their main-sequence lifetimes, while those at intermediate rotation rates evolve homogeneously initially, and turn to normal evolution due to angular momentum loss (transitionary evolution). 

The stability of the extended envelopes of the core-hydrogen-burning cool supergiants is uncertain. \citet{Moriya:2015} suggest that their likely pulsational instability may lead to enhanced mass-loss. This may significantly shorten this evolutionary stage.

Nevertheless, should they exist, they may be extremely bright stars. As their bolometric correction is essentially zero, the cool supergiants predicted by our model grid with log(L/L$_{\odot}$)=6.3...7 would have visual magnitudes in \izw, adopting a distance of 18~Mpc \citep{Aloisi:2007}, in the range of 20.3~mag...18.6~mag. Brightness variations with periods of the order of months to years due to pulsations may reveal them as stars rather than star clusters in photometric multi-epoch observations.

There may also be other ways to look for core-hydrogen-burning cool supergiants in nature.
According to our simulations, core-hydrogen-burning supergiants lose a significant amount of mass during the red supergiant phase. In the case of the 294~M$_{\odot}$ star analysed above, for example, as much as $\sim$40~M$_{\odot}$ of material is lost in the red supergiant wind. As the material lost in the wind has undergone CNO processing, the material that returns to the circumstellar gas pollutes the environment with hydrogen-burning products. The low wind velocity may allow this gas to be retained in the vicinity of the star-forming region which produced the cool supergiants, and thus pollute the gas from which further stars in the same region may form. E.g., our cool supergiants may have an impact on the understanding of abundance anomalies in globular clusters \citep[][and Szécsi et al. in prep.]{Caretta:2010,Bastian:2013}.

%--------------------------------------
% TWUIN

\section{Transparent Wind Ultraviolet Intense stars}\label{sec:wrHR}

Stars of all masses that evolve homogeneously mixed during their main-sequence lifetime occupy the left purple region in Fig.~\ref{fig:tracks}, i.e. blueward from the ZAMS. 

\begin{figure}
\centering
\includegraphics[height=\columnwidth,angle=270]{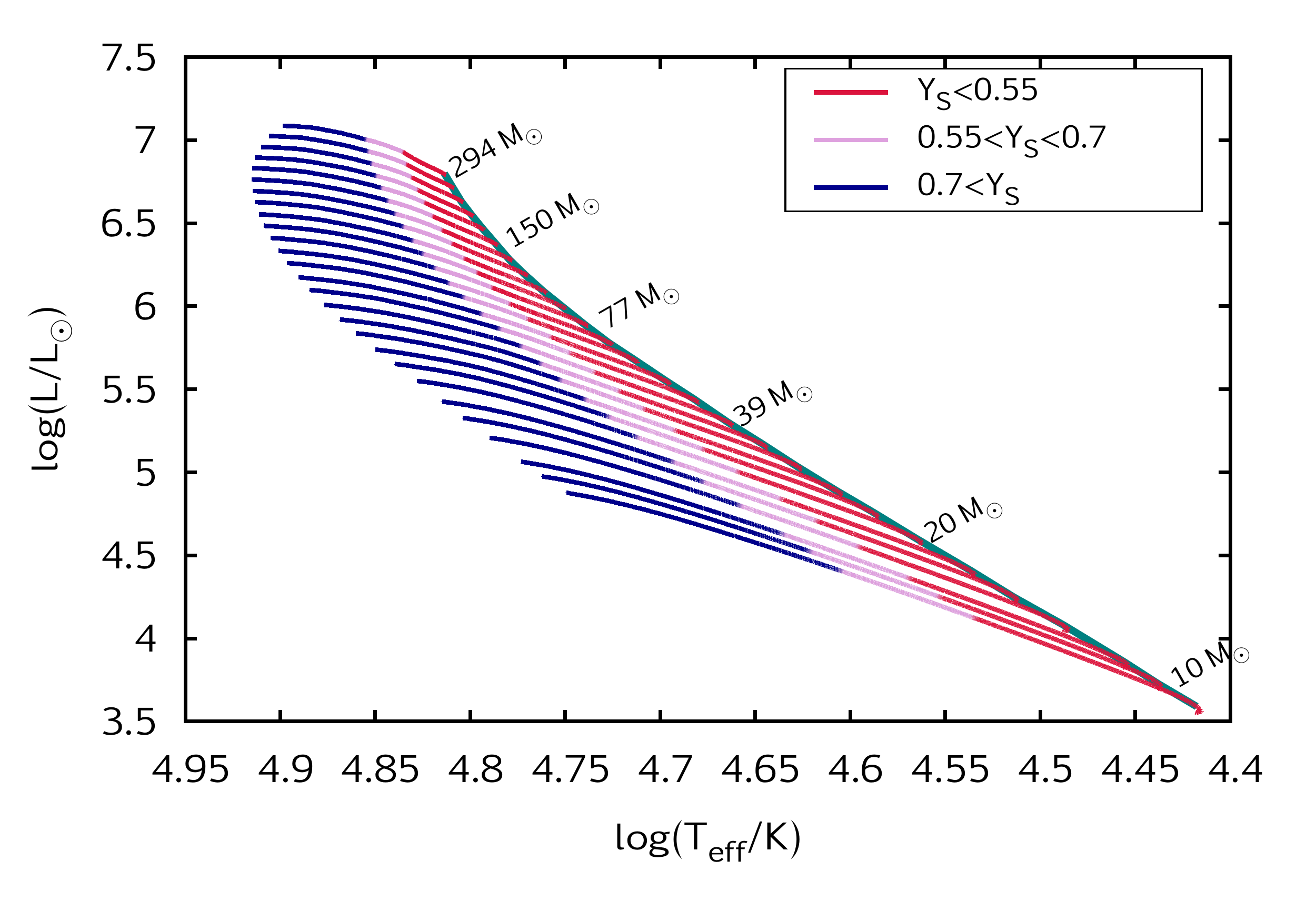}
\includegraphics[height=\columnwidth,angle=270]{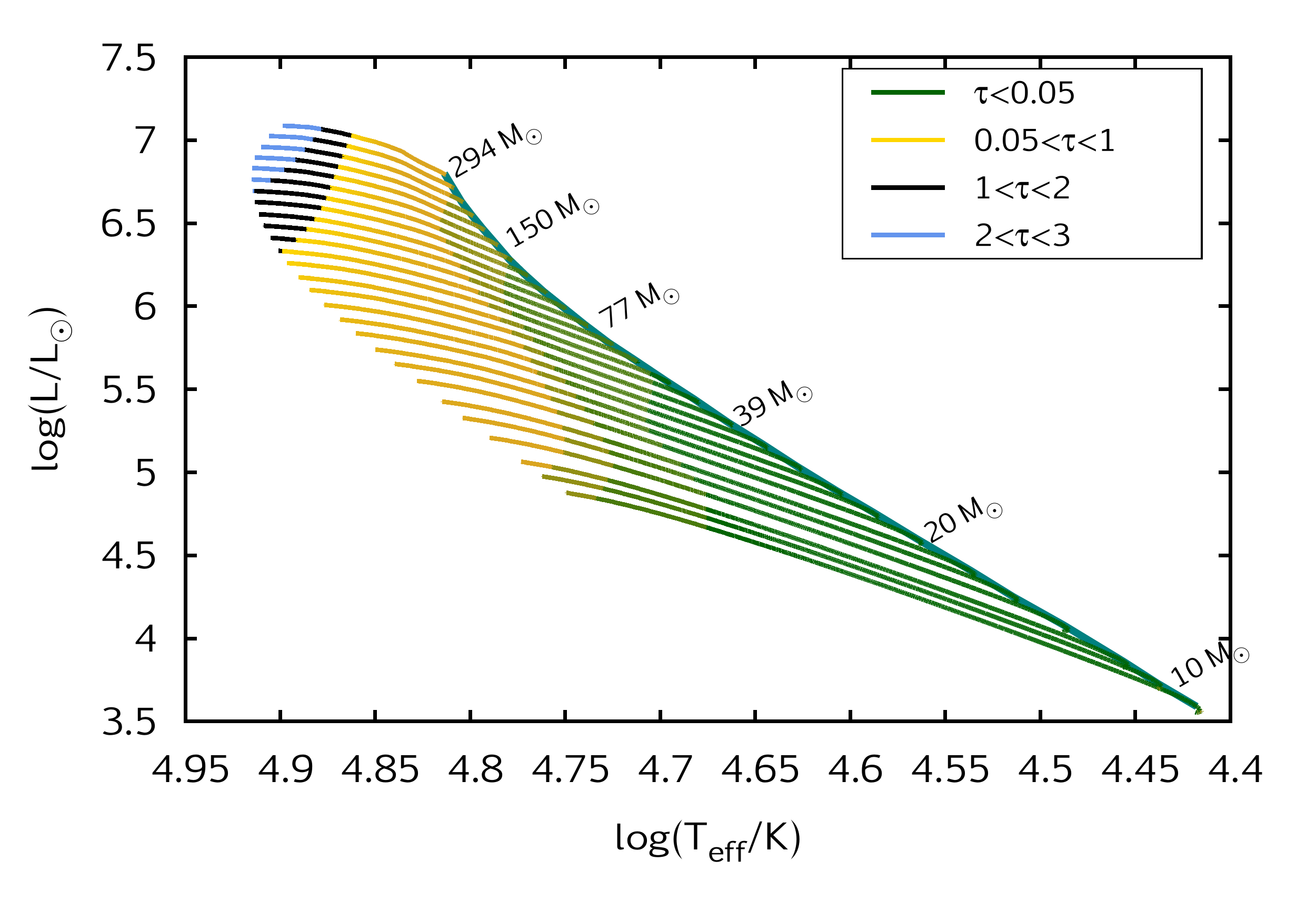}
\caption{
\textsl{Top:} HR diagram of models with v$_{\rm ini}$=500~km~s$^{-1}$ (chemically-homogeneous evolution) and masses between 9-294~M$_{\odot}$. The thick green line marks the ZAMS. The colouring marks the surface helium mass fraction as indicated by the legend. For Y$_{\mathrm{S}}$<0.55, OB-type mass-loss is applied; for Y$_{\mathrm{S}}$ between 0.55 and 0.7, an interpolation between OB- and WR-type mass-loss is applied; and for Y$_{\mathrm{S}}$>0.7, WR-type mass-loss is applied (cf. Sect.~\ref{sec:massloss}). 
\textsl{Bottom:} HR diagram of the same collection of models as above. The colouring marks the wind optical depth $\tau$ according to Eq.~(\ref{eq:ttaa}). 
}
\label{fig:tau}
\end{figure}

These models have OB-type mass-loss initially. WR-type mass-loss is adopted for Y$_{\mathrm{S}}$>0.7, see the top panel in Fig~\ref{fig:tau}. Therefore, from the evolutionary point of view, these models might be considered as core-hydrogen-burning WR stars. However, from the observational point of view, WR stars are characterized by the presence of strong emission lines, which indicate optically-thick winds. We estimate the optical depth of their winds following Eq.~(14) of \citet{Langer:1989a} as:
\begin{equation}
\tau(R)=\frac{\kappa\dot{M}}{4\pi R(v_{\infty}-v_{0})} \ln\frac{v_{\infty}}{v_{0}},\label{eq:ttaa}
\end{equation}
where $R$ designates the radius of the stellar model without taking the wind into account. This equation is derived from a $\beta$-velocity law with $\beta$=1. In that, we use the electron scattering opacity $\kappa=\sigma (1+X)$, $\sigma$ being the Thomson scattering cross-section, an expansion velocity of $v_0$=20~km~s$^{-1}$ at the surface of the stellar model, and a terminal wind velocity of $v_{\infty}=\sqrt{\frac{GM}{R}}$. 

Fig~\ref{fig:tau} (bottom panel) shows the optical depth of the stellar winds as calculated from Eq.~(\ref{eq:ttaa}) for our homogeneously-evolving stellar models. The behaviour of the wind optical depth seen in this figure is mostly related to the mass-loss rate (cf. Sect.~\ref{sec:masshist}), which is increasing with mass. 
While these numbers are only approximate, they show that the winds of the lower-mass (M$_{\rm ini}\lesssim$80~M$_{\odot}$) models with CHE, even when applying WR-type mass-loss, remain optically thin ($\tau<1$) throughout their main-sequence lifetime. Even the higher-mass models (M$_{\rm ini}\gtrsim$80~M$_{\odot}$) keep optically-thin winds for most of core hydrogen burning, and the wind optical depth does not exceed $\tau\simeq 3$ even up to core hydrogen exhaustion for the most luminous stars. 

Our fully mixed stars are extremely hot (up to T$_{\rm eff}\simeq80$~kK) and bright (up to 10$^7$~L$_{\odot}$) objects which have an optically-thin wind. Additionally, they emit intense mid- and far-UV radiation (see also Sect.~\ref{sec:flux}), so we call them Transparent Wind Ultraviolet INtense stars or TWUIN stars. 

We emphasize that TWUIN stars are only expected at very low-metallicity. 
Their mass-loss, which depends on the metallicity, is not strong enough to spin them down to prevent homogeneous evolution (Szécsi et al. in prep.). They remain compact, i.e. the radii remain small, typically around 10-20~R$_{\odot}$. Additionally, they develop no core-envelope structure, so most of the hydrogen in the envelope is mixed into the burning regions and converted into helium. 
TWUIN stars therefore finish their main-sequence evolution as massive fast-rotating helium stars which make them strong candidates for long-duration gamma-ray bursts \citep{Yoon:2005,Woosley:2006}. 
Their rotational rate at the TAMS is discussed in Sect.~\ref{sec:chem}.

          %-----------------------------------------------------------------
          % He-fraction in the surface and the core
          
\section{The helium abundance at the surface and in the core}\label{sec:YcYs}

In the surface helium vs. central helium mass fraction (\mbox{Y$_{\mathrm{S}}$-Y$_{\mathrm{C}}$}) diagram, every stellar evolutionary sequence can be represented by one line, and the core helium mass fraction merely serves as a clock. During the core-hydrogen-burning stage, the slope of the line tells us about the efficiency of mixing helium from the core through the radiative envelope to the surface by rotation-induced turbulence. Thus, the steeper the slope, the more helium reaches the surface. Tracks of normally-evolving stellar models form a horizontal line while tracks of homogeneously-evolving models lie close to the diagonal in the \mbox{Y$_{\mathrm{S}}$-Y$_{\mathrm{C}}$}~diagram. Furthermore, tracks of models with transitionary evolution lie between the horizontal and the diagonal lines. Consequently, it is easy to distinguish these three evolutionary behaviours in the \mbox{Y$_{\mathrm{S}}$-Y$_{\mathrm{C}}$}~diagram.

\begin{figure}
\centering
\includegraphics[width=\columnwidth,angle=0]{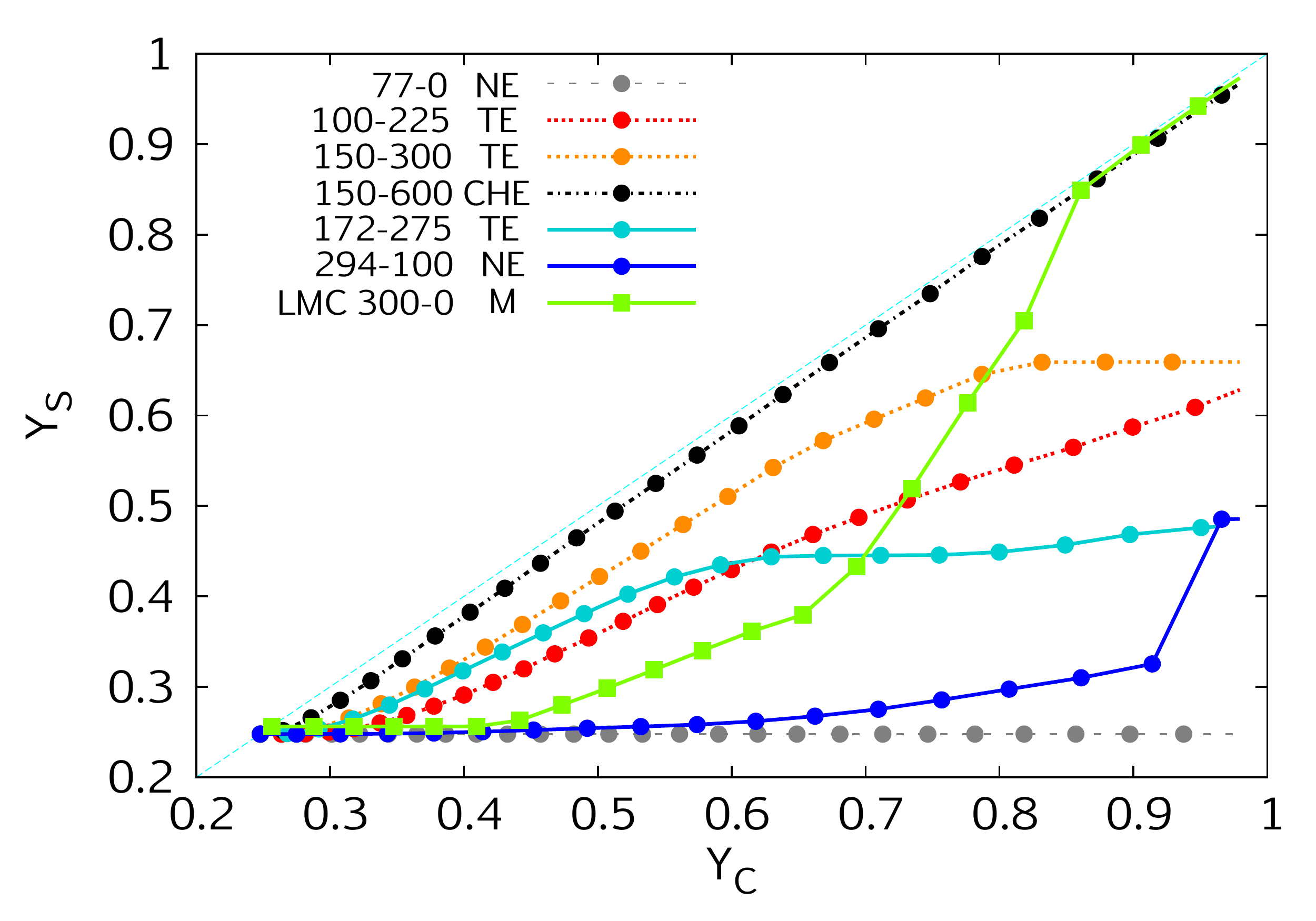}
\caption{Mass fraction of helium at the stellar surface (Y$_{\mathrm{S}}$) as a function of that in the core (Y$_{\mathrm{C}}$) for sequences of different initial masses and rotational velocities as indicated by the legend, in units of M$_{\odot}$-km~s$^{-1}$. Dots mark every 10$^5$~years of the evolution. The diagonal line (Y$_{\mathrm{C}}$=Y$_{\mathrm{S}}$) is marked by a lightblue (dashed) line. 
Stars that evolve chemically-homogeneously (CHE) lie close to the diagonal, while those that undergo normal evolution (NE) trace a horizontal line; stars with transitionary evolution (TE) lie between (see also Fig.~\ref{fig:mygrid}).
One non-rotating sequence (of type~M) with M$_{\rm ini}$=300~M$_{\odot}$ from the LMC grid of \citet{Koehler:2015} is shown for comparison. 
}
\label{fig:YcYsALL}
\end{figure}

Fig.~\ref{fig:YcYsALL} presents some of our stellar sequences in the \mbox{Y$_{\mathrm{S}}$-Y$_{\mathrm{C}}$}~diagram. The non-rotating sequence of 77~M$_{\odot}$ evolves close to the X-axis of the \mbox{Y$_{\mathrm{S}}$-Y$_{\mathrm{C}}$}~diagram, which indicates that there is no mixing between the core and the surface. 

Sequences of intermediate rotational velocities (such as the models of 100~M$_{\odot}$-225~km~s$^{-1}$, 150~M$_{\odot}$-300~km~s$^{-1}$ and 172~M$_{\odot}$-275~km~s$^{-1}$) start their life homogeneously and with a slight rise in the \mbox{Y$_{\mathrm{S}}$-Y$_{\mathrm{C}}$}~diagram, but after a while they lose enough angular momentum so they turn to normal evolution and show a horizontal slope in the diagram. Therefore, we consider these sequences having transitionary evolution. 

Fast rotating sequences of 600~km~s$^{-1}$ undergo CHE, turning bluewards in the HR~diagram and following the diagonal line in the \mbox{Y$_{\mathrm{S}}$-Y$_{\mathrm{C}}$}~diagram. In these models, the ashes of nuclear burning are mixed between the core and the surface, enhancing the surface with burning products (e.g. helium) and supplying unprocessed material to the hydrogen-burning region. 

\citet{Koehler:2015}, who analysed stellar models with LMC composition, introduced Type~M evolution, which stands for an evolutionary behaviour during which mass-loss is so efficient that the homogeneous layers of the stellar interior are uncovered. Fig.~\ref{fig:YcYsALL} shows one LMC sequence which is of Type~M.

None of our sequences undergo evolution classified as Type~M. This is simply because the mass-loss at our low-metallicity is less effective than at LMC metallicity. Although some sequences (e.g. the one with  294~M$_{\odot}$ and 100~km~s$^{-1}$ in Fig.~\ref{fig:YcYsALL}) show effects of mass-loss near the TAMS, this effect is not strong enough to make the model homogeneous (i.e. Y$_{\mathrm{C}}\simeq$~Y$_{\mathrm{S}}$).

The evolution leading to a core-hydrogen-burning cool supergiant star (Sect.~\ref{sec:hbrs}) is represented by the track of 294~M$_{\odot}$ with 100~km~s$^{-1}$ initial rotation in Fig.~\ref{fig:YcYsALL}. During the last $\sim$10$^5$~years of the simulated evolution, the surface helium abundance increases rapidly for two reasons. The first reason is that the supergiant mass-loss takes over, leading to a significant increase in the surface helium abundance. The second reason is that a deep convective envelope develops in the outer layers of the star, which dredges out helium from the core. This model spends the last phase of its main-sequence evolution (between Y$_C\gtrsim 0.92$ and the TAMS) as a core-hydrogen-burning red supergiant with T$_{\rm eff}\approx 4500$~K.

Another example of a core-hydrogen-burning supergiant is given by the track of 172~M$_{\odot}$ with 275~km~s$^{-1}$ initial rotational velocity. This model is categorised as transitionary evolution, since Y$_{\rm S}$ increases with Y$_{\rm C}$ initially (as in the case of the homogeneously-evolving models). Between $0.6\lesssim{\rm Y}_{\mathrm{C}}\lesssim 0.8$, however, Y$_{\rm S}$ stays constant (typical for normal evolution). At Y$_{\mathrm{C}}\sim0.8$, a slight increase in the surface helium abundance happens again as the sequence proceeds towards lower effective temperatures and the mass-loss becomes more effective. Amongst our core-hydrogen-burning cool supergiant models, the highest surface helium mass fraction we find is 0.52. 
%###############################################
% mass-loss

\section{Mass-loss history}\label{sec:masshist}

\begin{figure}[h!]
\begin{center}
\resizebox{\hsize}{!}{\includegraphics[angle=270]{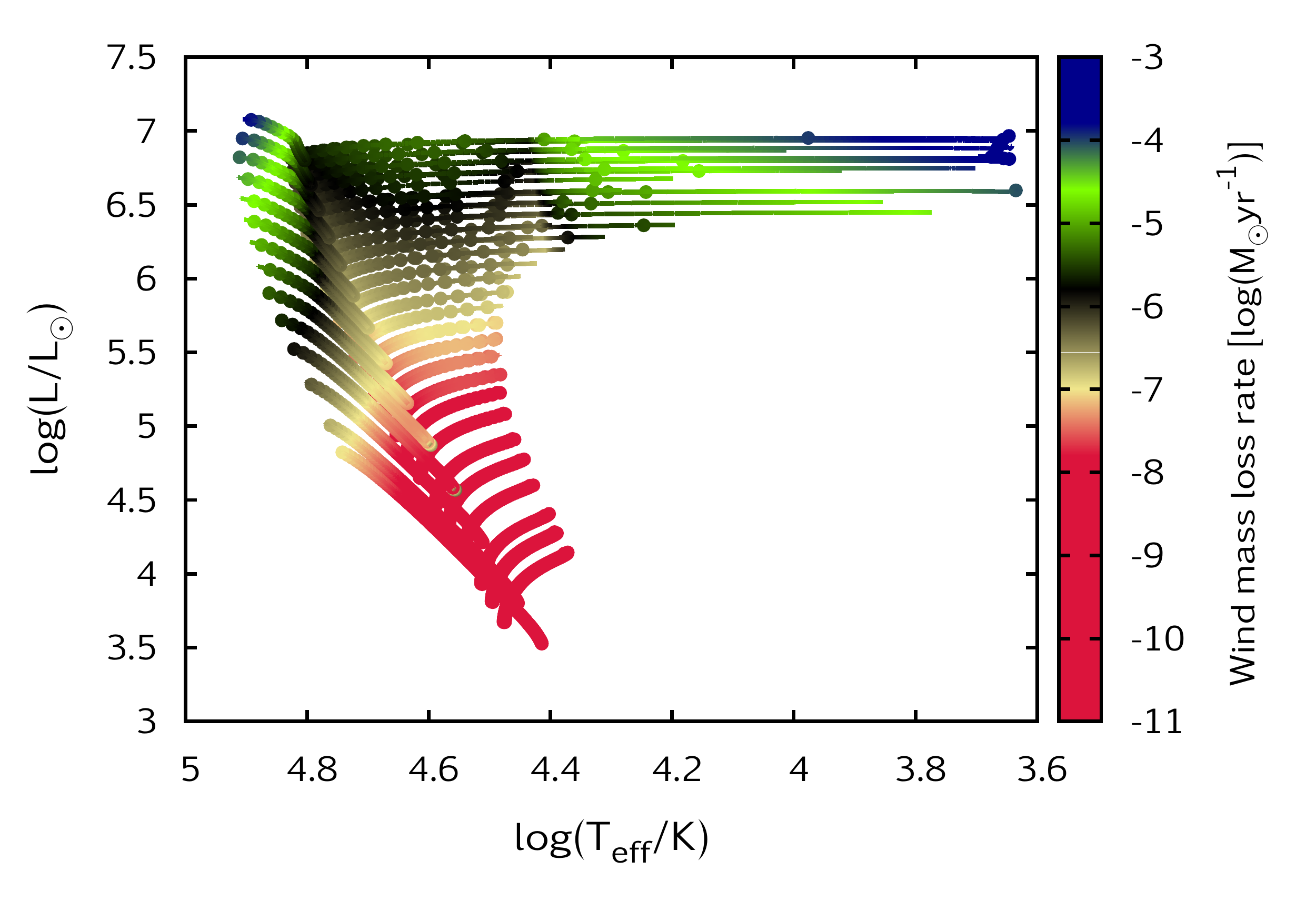}}
\end{center}
\caption{
HR diagram showing sequences with v$_{\rm ini}$=0 and 500~km~s$^{-1}$ for all masses of our grid. Mass-loss rates are colour coded. Dots mark every 10$^5$~years of evolution.
}
\label{fig:Mdot}
\end{figure}

While the mass-loss rates adopted for our models (Sect.~\ref{sec:massloss}) depend strongly on the initial metallicity \citep[][and Szécsi et al., in prep.]{Vink:2001}, and our models lose less mass than their counterparts at, for example, LMC composition \citep{Koehler:2015}, in the most extreme cases of the most massive TWUIN stars and the core-hydrogen-burning supergiants, our stellar models reach mass-loss rates as high as 4$\times$10$^{-4}$~M$_{\odot}$~yr$^{-1}$. This is demonstrated by Fig.~\ref{fig:Mdot}, which shows the mass-loss rate of some of our models in the HR diagram.

\begin{figure}[h!]
\begin{center}
\resizebox{\hsize}{!}{\includegraphics[angle=270]{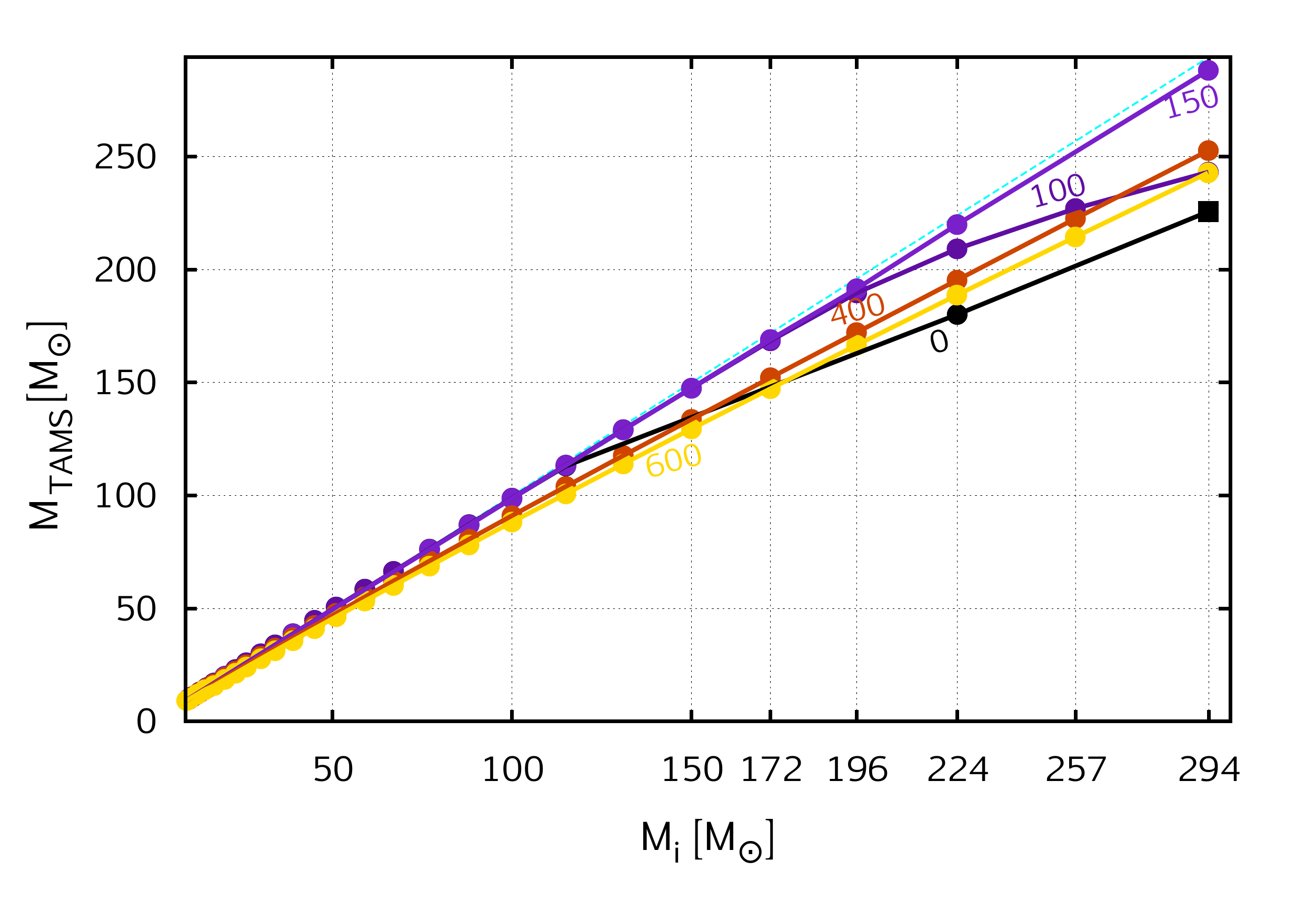}}
\end{center}
\caption{
Initial mass vs. final mass (at the TAMS). Every dot marks one evolutionary sequence. Sequences with the same initial rotational velocity v$_{\rm ini}$ are connected, labels indicate the value of the corresponding v$_{\rm ini}$ in km~s$^{-1}$. Only models that evolved until the TAMS are shown, except for the non-rotating one with M$_{\rm ini}$=294~M$_{\odot}$ (marked by a rectangle), for which the final mass is extrapolated.
}
\label{fig:mass}
\end{figure}

Fig.~\ref{fig:mass} shows the M$_{\rm ini}$ vs. M$_{\rm TAMS}$ relation for our stellar sequences. Overall, none of the tracks deviate much from the diagonal line, meaning that the mass-loss is quite weak for our models. However, there are some differences in how much mass the models with different evolutionary paths retain during their main-sequence lifetimes. 

In the lower-mass regime ($\lesssim$100~M$_{\odot}$), the fast rotating, chemically-homogeneously-evolving sequences of 400 and 600~km~s$^{-1}$ end up having less mass than the slow rotating, normally-evolving sequences. This is due to the WR-type mass-loss rate that applies for the sequences with CHE during the second part of their main-sequence evolution. 

For the high-mass ($\gtrsim$100~M$_{\odot}$) sequences, however, another behaviour is present: the very massive slow rotators (represented by the models of 0-100~km~s$^{-1}$ in Fig.~\ref{fig:mass}) become core-hydrogen-burning cool supergiants. The efficiency of mass-loss increases when a star approaches the cool supergiant phase because the mass-loss prescription applied here has a radius dependence of $\dot{M}\sim R^{0.81}$. Thus, the mass-loss in this phase may be even stronger than the WR-type mass-loss, which means that stellar models that evolve to the cool supergiant phase during the main sequence may end up less massive than models with CHE of the same mass. Note that the model marked with a rectangle in Fig.~\ref{fig:mass} is stopped at Y$_{\mathrm{C}}=0.87$. For this model, we predicted the final mass based on the mass-loss rate in the last computed model and on the remaining hydrogen-burning lifetime.

Intermediate rotation rates (150-350~km~s$^{-1}$) are represented by the models at 150~km~s$^{-1}$ in Fig.~\ref{fig:mass}. These models eventually evolve normally but stay bluer due to enhanced surface helium abundance by rotational mixing (cf. models with TE in Fig.~\ref{fig:YcYsALL}). They therefore undergo neither WR-type mass-loss nor cool supergiant mass-loss and only lose small amounts of mass due to the OB-type mass-loss that applies to them during their whole main-sequence lifetime.

%******************************************
% Evolution of rotation
\section{Rotation}\label{sec:rotation}

At higher metallicity (e.g. Solar or LMC), rotating massive stars would be spun down during the main-sequence evolution because of mass and angular momentum loss via winds \citep{Langer:1998,Koehler:2015}. At the metallicity of \izw, in contrast, mass-loss is less efficient and the stars can retain a more or less constant amount of angular momentum. If there are efficient mechanisms transporting angular momentum in the interior between the core and the envelope, the surface rotational velocity might increase during the main-sequence evolution even when the star evolves towards lower surface temperature and larger radius \citep[][]{Ekstroem:2008,deMink:2013}. In this section we show how this core-envelope coupling plays a role in shaping the rotational history of our stars.

\subsection{Evolution of the surface rotational velocity}\label{sec:vsurf}

\begin{figure}
\centering
\includegraphics[height=0.95\columnwidth,angle=270]{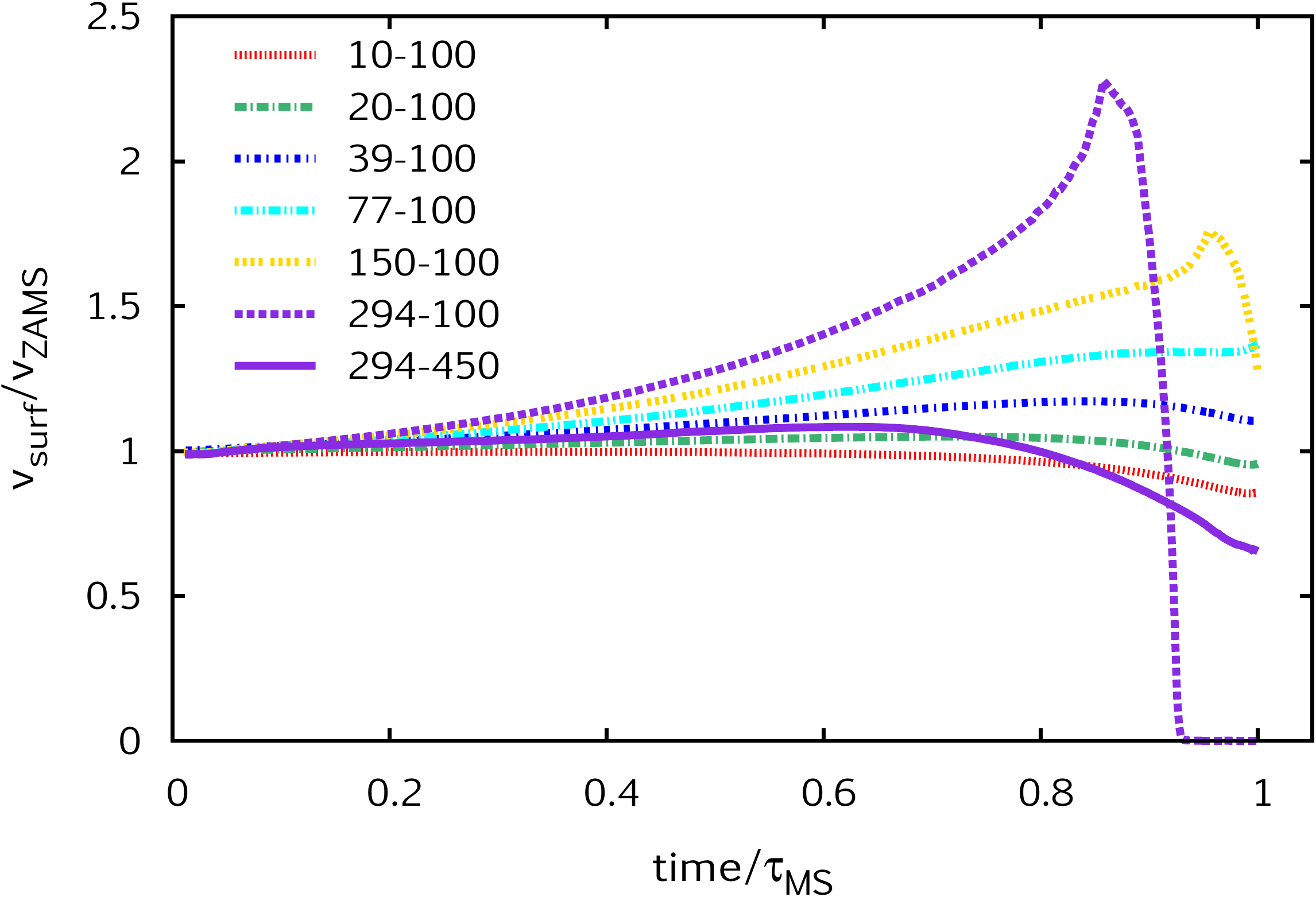}
\caption{Surface rotational velocity as function of time for models with different initial masses and rotational velocities as indicated by the legend (units are [M$_{\odot}$]-[km~s$^{-1}$]). The time is normalised to the main-sequence lifetime of the stars and the rotational velocity is normalised to the ZAMS value (see Sect.~\ref{sec:zams}). The track of the 294~M$_{\odot}$--450~km~s$^{-1}$ model which evolves chemically homogeneously is plotted with a dotted line for comparison.
}
\label{fig:vsurf}
\end{figure}

The evolution of the surface rotational velocity for some of our models is presented in Fig.~\ref{fig:vsurf}. The surface rotation of the 10~M$_{\odot}$-100~km~s$^{-1}$ model gradually decreases. Higher-mass models at 100~km~s$^{-1}$ from our grid, however, all increase their surface rotational velocity during the first $\sim$80\% of their main-sequence lifetime. The most massive models then reach a maximum and start a rapid decrease and spin down -- in case of the 294~M$_{\odot}$-100~km~s$^{-1}$ model all the way to zero. This sequence evolves into a core-hydrogen-burning cool supergiant.

To understand this behaviour we need to consider the following mechanisms. In our stellar models, angular momentum can be transported from the core to the envelope due to meridional circulations and shear turbulence, as well as by magnetic torques (Sect.~\ref{sec:mixing}). The angular momentum transport aims to make the whole star rotate with constant angular velocity. During the main-sequence phase of a normally-evolving model, the stellar core contracts, the envelope expands and the star evolves redwards in the HR diagram. Although the radius increases, angular momentum can be effectively transported from the contracting core outwards, at least during the first $\sim$80\% of the main-sequence lifetime. As a result, the surface rotational velocity of the star must increase during this evolutionary phase.

The star therefore spins up. According to Fig.~\ref{fig:vsurf}, the maximum velocity depends on the initial mass, being greater for higher-mass objects. 
The reason of this mass dependence is that the higher the mass the more massive the stellar core. The angular momentum which is transported from this more massive core to the envelope is therefore higher. The rotation rate can increase to more than twice the initial value in the case of the 294~M$_{\odot}$ model. In contrast, the 10~M$_{\odot}$ model does not show any increase of surface rotation because its core is relatively small.

\begin{figure}
\centering
\includegraphics[height=1\columnwidth,angle=270]{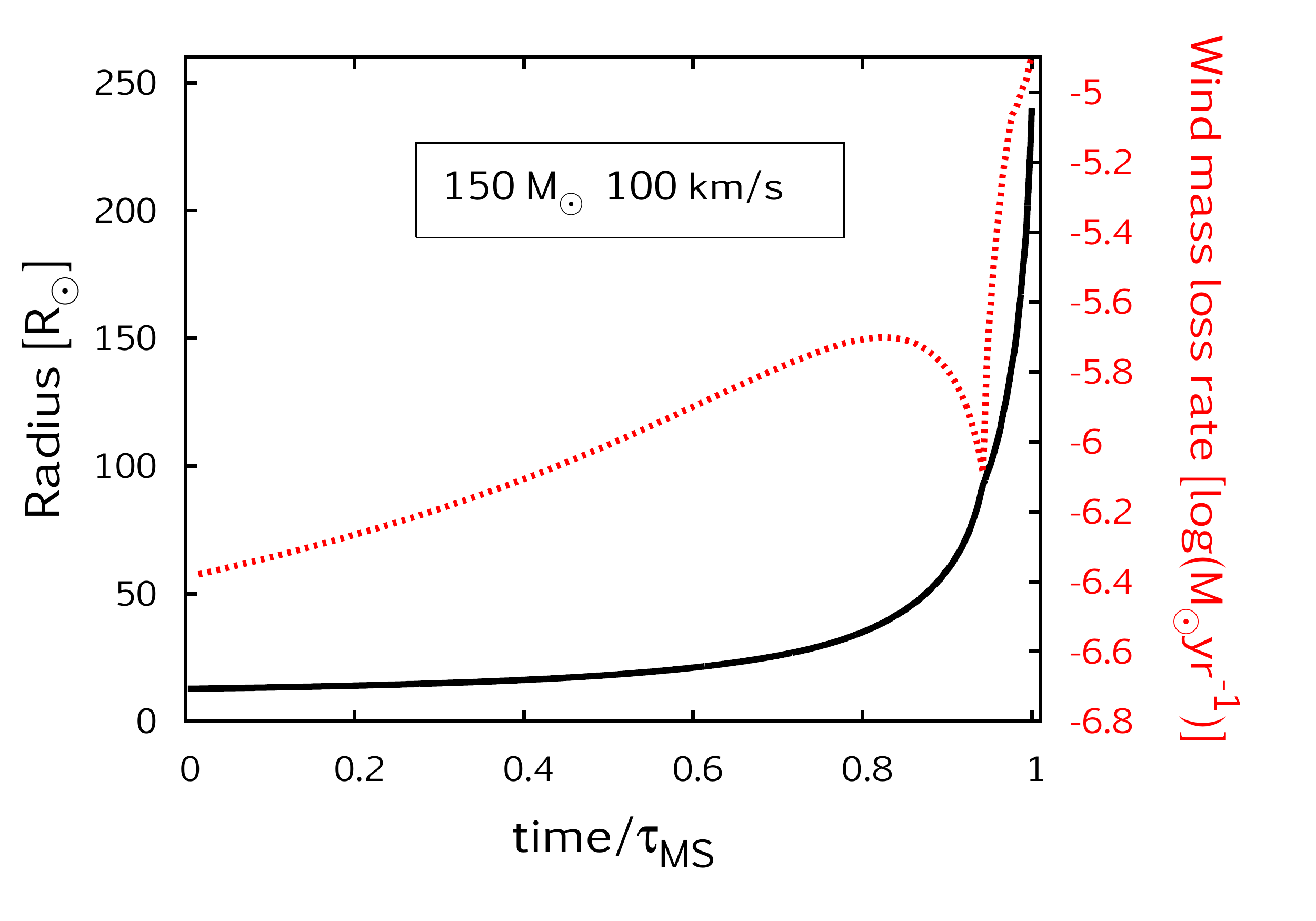}
\includegraphics[height=1\columnwidth,angle=270]{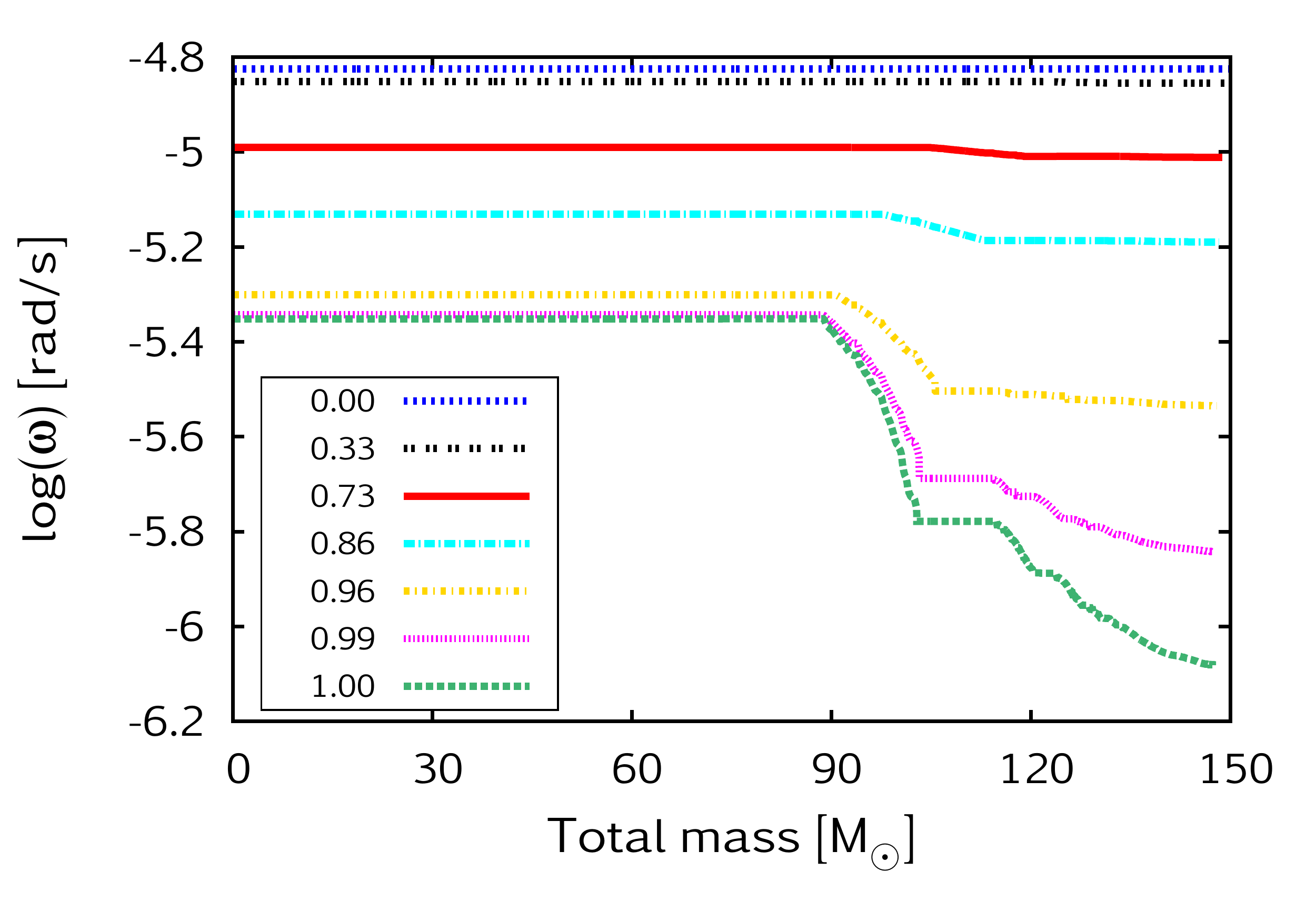}
\caption{
\textsl{Top:} Radius and mass-loss rate as a function of the main-sequence lifetime for the stellar sequence with M$_{\rm ini}$=150~M$_{\odot}$ and v$_{\rm ini}$=100~km~s$^{-1}$ (yellow track in Fig.~\ref{fig:vsurf}).
\textsl{Bottom:} Angular velocity ($\omega$) distribution inside the same sequence at the fractions of main-sequence lifetime, $t/\tau_{\rm MS}$, indicated by the legend, 
where $\tau_{\rm MS}=t^{Y_{\mathrm{C}}=0.98}- t^{Y_{\mathrm{C}}=0.28}$.
}
\label{fig:angvel}
\end{figure}

For the highest mass models of 150~M$_{\odot}$--100~km~s$^{-1}$ and 294~M$_{\odot}$--100~km~s$^{-1}$, a sudden drop happens at $\sim$85\% and $\sim$95\% of the main-sequence lifetime, respectively. 
This is further illustrated by Fig.~\ref{fig:angvel} which shows the evolution of the radius and the mass-loss rate, as well as the angular velocity distribution inside the 150~M$_{\odot}$ model. The angular velocity is approximately constant until t$\approx$0.80\,$\tau_{\rm MS}$. 
Then the radial expansion becomes so pronounced that the angular momentum transport through the core-envelope coupling cannot keep the star rigidly rotating and the surface layers slow down. Additionally, the mass-loss increases at $\sim 0.94$~$\tau_{\rm MS}$ when the star encounters the bi-stability jump at T$_{\rm eff}\sim 25$~kK \citep{Vink:2000}. As a result, a significant amount of mass and angular momentum is lost contributing to the fast decrease of the surface rotational velocity. 

At the TAMS, stars that have not become cool supergiants during the main-sequence still rotate rapidly (see also Sect.~\ref{sec:vTAMS}). The core-hydrogen-burning supergiants, represented by the 294~M$_{\odot}$--100~km~s$^{-1}$ model in Fig.~\ref{fig:vsurf}, have negligible surface rotation at the TAMS in our calculations.

Our results support the finding of \citet{Meynet:2002} and \citet{Ekstroem:2008} that normally-evolving stars may increase their surface rotation during the main-sequence evolution if the mass-loss is low. 
This implies that the rotational velocity distribution of hydrogen-burning massive stars to be observed in low-metallicity environments is expected to be significantly different than that in higher-metallicity environments.

For fast rotators, the angular velocity distribution inside the star is always close to constant during the main-sequence lifetime. 
They are represented by the 294~M$_{\odot}$--450~km~s$^{-1}$ model in Fig.~\ref{fig:vsurf}. 
This model evolves with only a slight radius increase during the first half of the main-sequence.
When the WR-type mass-loss turns on at $\sim 0.65$~$\tau_{\rm MS}$, the mass-loss increases and spins the star down. However, WR-type mass-loss at this metallicity is not strong enough to remove all the angular momentum. Therefore these stars still rotate rapidly at the TAMS (see Sect.~\ref{sec:chem}) and, if there is little angular momentum loss afterwards, also during their post-main-sequence phases (see Sect.~\ref{sec:grb}).

\subsection{Surface rotational velocity at the TAMS}\label{sec:vTAMS}

\begin{figure*}[ht!]
\centering
\includegraphics[width=\oneseven\columnwidth,angle=0]{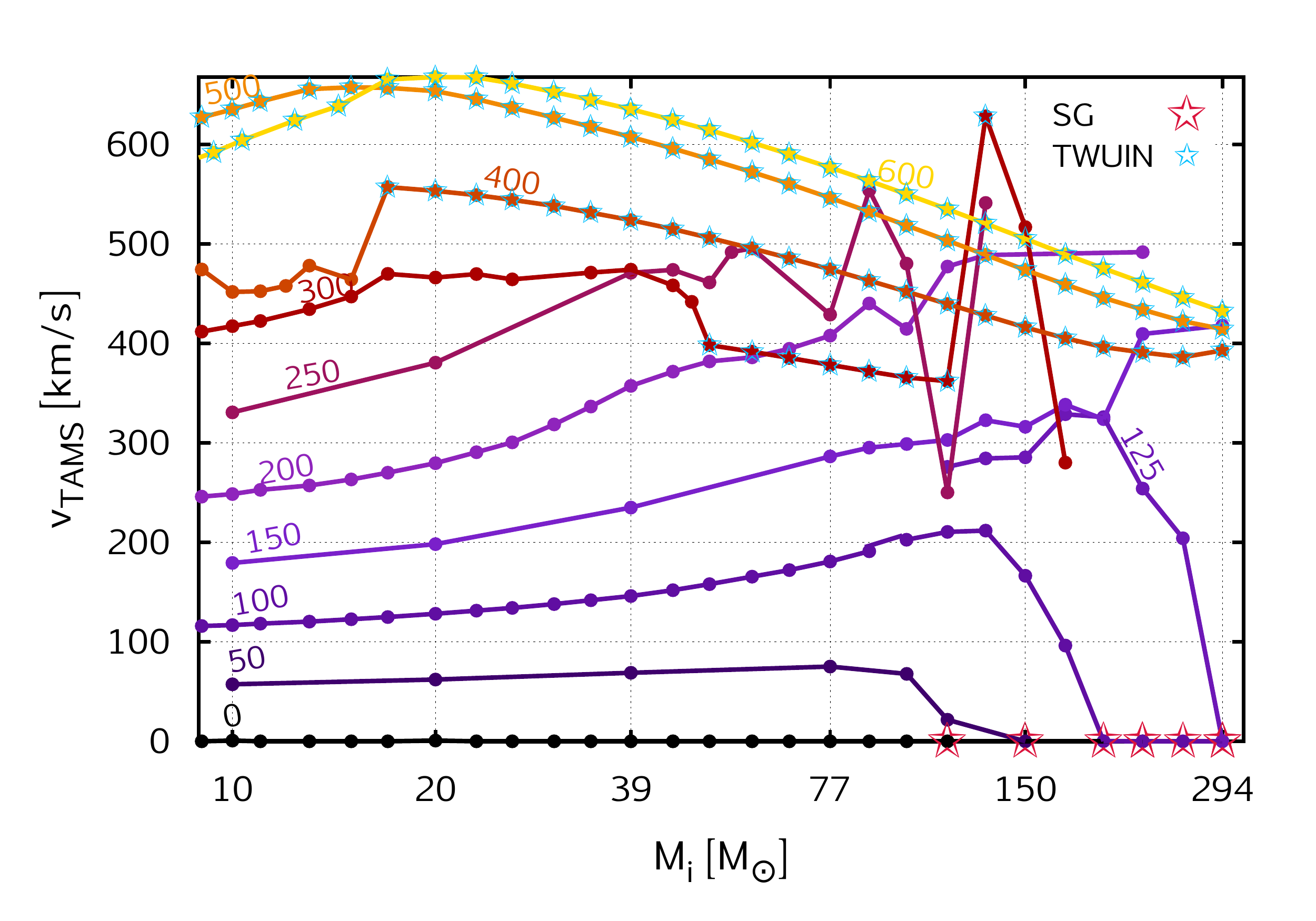}
\caption{Rotational velocity at the terminal age main-sequence (TAMS) as a function of initial mass. Every dot represents one evolutionary sequence. The colours refer to the initial rotational velocity; dots of model sequences with the same initial rotational velocity are connected and labelled (units in km~s$^{-1}$). Only those sequences that have reached Y$_{\mathrm{C}}$=0.98 have been plotted.
(See also Fig.~\ref{fig:initzams} which shows the rotational velocity at the ZAMS: here the same models are connected as in Fig.~\ref{fig:initzams}.) 
Core-hydrogen-burning cool supergiants (SG) are marked with large red stars, chemically-homogeneous TWUIN stars (Y$^{\rm TAMS}_{\mathrm{S}}$>0.7) are marked with small blue stars. 
}
\label{fig:vTAMS}
\end{figure*}

Here we discuss the rotational rates of our models at the end of the main sequence. The rotational velocity of a model at the TAMS depends on its rotational behaviour during the main-sequence evolution, which in turn depends on the actual evolutionary path (normal, transitionary or homogeneous evolution, as seen in Sect.~\ref{sec:vsurf}). Also, we refer to Sect.~\ref{sec:zams} where we discuss the rotational velocities of our models at the ZAMS. 

\subsubsection{Redward evolving stars}\label{sec:red}

Fig.~\ref{fig:vTAMS} presents the surface rotational velocity of our models at the TAMS. Sequences that are marked by red star symbol or not marked with any symbol undergo normal or transitionary evolution. In both cases, the models evolve redwards in the HR~diagram. 

Sequences in the left bottom corner of the figure with initial masses $\lesssim$~26~M$_{\odot}$ and initial rotational velocities $\lesssim$~150~km~s$^{-1}$ reduce their rotational velocity gradually, the same way as the sequence 10~M$_{\odot}$--100~km~s$^{-1}$ in Fig.~\ref{fig:vsurf}. The reason that their rotational velocities are still above 100~km~s$^{-1}$ at the TAMS is the spin up during the adjustment phase at the beginning of their evolution, as discussed in Sect.~\ref{sec:zams}.

The rotational velocity at the TAMS of the higher-mass models (between 26-131~M$_{\odot}$ in the case of the 100~km~s$^{-1}$ line) with slow initial rotation ($\lesssim$~150~km~s$^{-1}$) is an increasing function of the initial mass. As we have explained in the discussion of Fig.~\ref{fig:vsurf}, this is because the higher the mass, the more angular momentum can be released by the contracting core and transported to the envelope. 
Stars more massive than 80~M$_{\odot}$ have, in the last $\sim$20\% of the main-sequence lifetime, undergone envelope inflation. They have encountered the bi-stability jump and spun down in the same way as the models of 150~M$_{\odot}$--100~km~s$^{-1}$ and 294~M$_{\odot}$--100~km~s$^{-1}$ shown in Fig.~\ref{fig:vsurf}. Those which rotate very slowly at the TAMS have evolved into core-hydrogen-burning cool supergiants. Accordingly, sequences with T$_{\mathrm{\rm eff}}^{\rm TAMS}<$~12~kK are marked with large red star symbols.

Lower-mass ($\lesssim$~50~M$_{\odot}$) sequences with initial velocities of 200-300~km~s$^{-1}$ evolve normally. Their behaviour is similar to those of lower mass at 100~km~s$^{-1}$: the surface rotational velocity at the TAMS is an increasing function of the initial mass.
Higher-mass ($\gtrsim$~50~M$_{\odot}$) sequences with initial velocities of 200-300~km~s$^{-1}$, on the other hand, have variable values of surface rotational velocity at the TAMS. They undergo transitionary evolution. The following effects contribute significantly in shaping their evolution. 
(1) mass-loss uncovers helium-rich layers, and the star appears bluer due to the lower opacities at the surface. Furthermore, the mass-loss, which depends on the surface composition and the effective temperature, removes angular momentum.
(2) The radius increases during the main-sequence evolution, and the star appears redder. At the same time, the core contracts, and the star increases the surface rotational velocity due to the core-envelope coupling. 
The net effect of these competing mechanisms can be that the model at the TAMS is either fast rotating and blue, or slow rotating and red, or somewhere in between. 

\subsubsection{Homogeneously-evolving stars}\label{sec:chem}

The fastest rotators (400-600~km~s$^{-1}$) are chemically-homogeneously-evolving TWUIN stars (marked in Fig.~\ref{fig:vTAMS} with small blue stars). They undergo WR-type mass-loss at the TAMS (i.e. Y$_{\mathrm{S}}^{\rm TAMS}\geq$~0.7), mostly have optically thin winds (as seen in Sect.~\ref{sec:wrHR}) and emit intense UV radiation (as seen in Sect.~\ref{sec:flux}).

Low-mass ($\lesssim$~20~M$_{\odot}$) stars with CHE reach breakup rotational rates early during the main-sequence evolution because the breakup velocity is less for lower masses. When the models spin up to close to breakup, they manage to spin down again by losing mass through rotationally-enhanced stellar winds. As a consequence, they rotate slower at the TAMS than they would if they had not reached breakup rotation or if there was no rotationally-enhanced mass-loss included in the calculations. This is why the surface rotational velocity at the TAMS of the fast rotators increases at low mass and does not follow the decreasing trend of the more massive stars which undergo CHE. Although rotating massive stars at breakup have gained some interest in the past \citep{Decressin:2007}, theoretical suggestions by \citet{Mueller:2014} disfavour the concept of rotationally-enhanced stellar winds. Since the physical assumptions in these stellar models are currently under debate, we do not analyse this issue further at this point. %--> put this sentence to Sect.2.4 ??

High mass (>20~M$_{\odot}$) sequences which undergo CHE, on the other hand, are hardly influenced by the rotational mass-loss enhancement. Their surface rotational velocity at the TAMS is decreasing as a function of the initial mass. This decreasing trend is a consequence of the stellar wind being more efficient at higher mass (but not efficient enough to turn them back to TE). Their typical behaviour is presented in Fig.~\ref{fig:vsurf} by the sequence 294~M$_{\odot}$--450~km~s$^{-1}$: their rotational velocity does not change much during the first two-thirds of their main-sequence lifetime, and then they slightly spin down due to the WR-type mass-loss. However, they still rotate at least as fast as 350~km~s$^{-1}$ at the TAMS. This fast rotation, if not reduced during the post-main-sequence evolutionary phases, might lead to the formation of a long-duration gamma-ray burst in the collapsar scenario (Sect.~\ref{sec:grb}).

\subsection{Surface nitrogen abundance and internal mixing}\label{sec:nitrogen}

\begin{figure}
\centering
\includegraphics[width=0.9\columnwidth,angle=0,page=1]{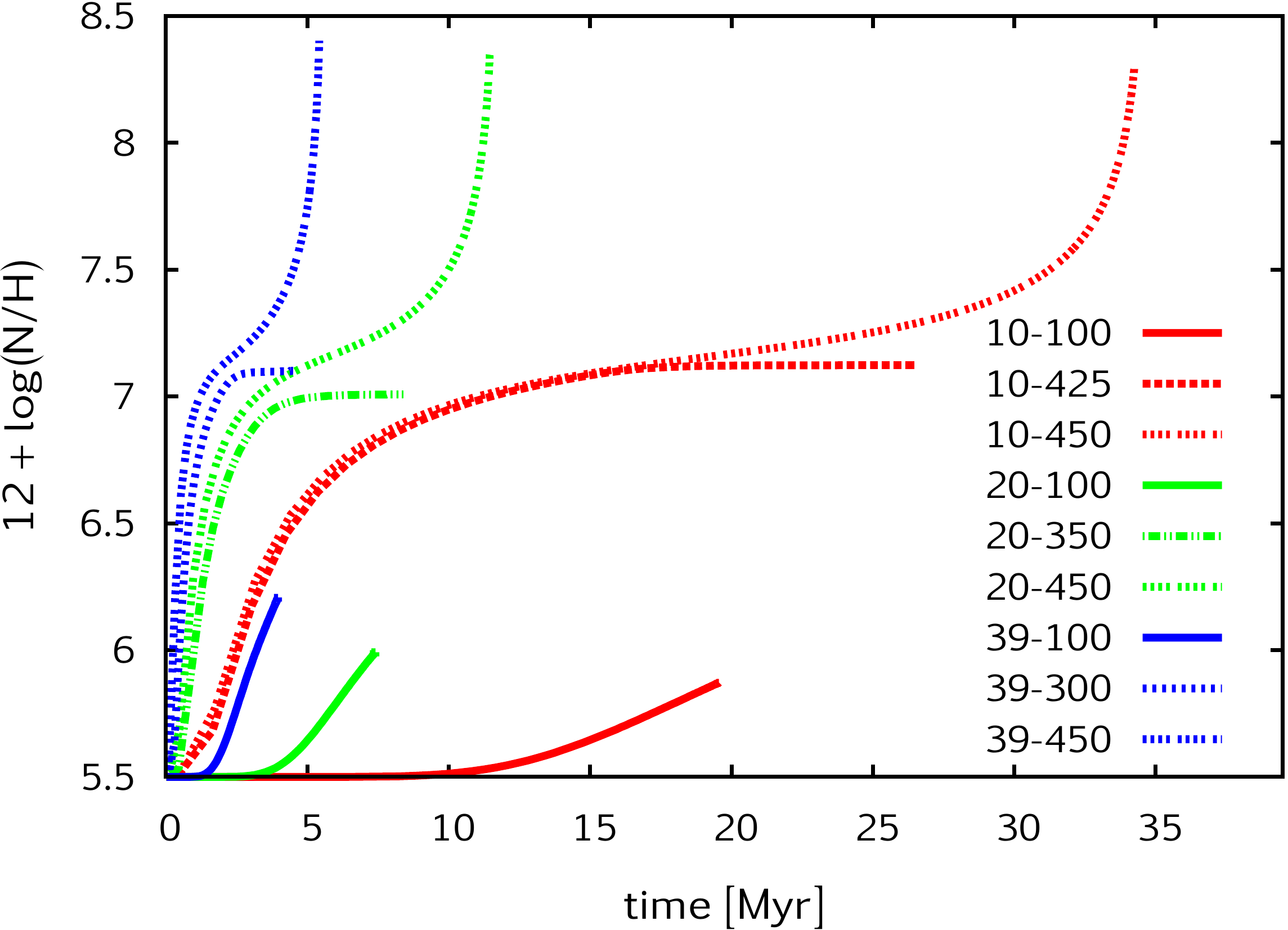}
\includegraphics[width=0.9\columnwidth,angle=0,page=1]{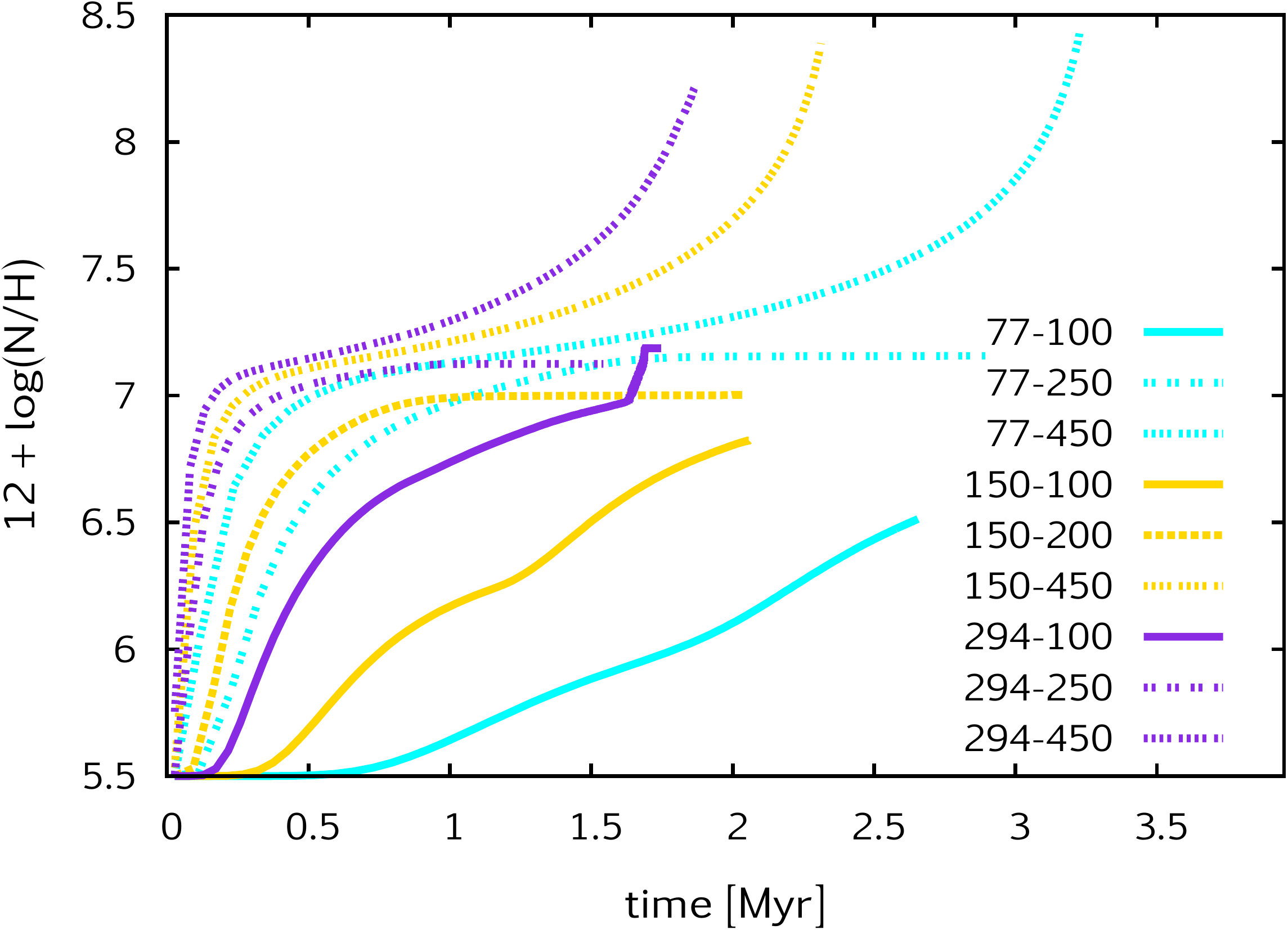}
\caption{Surface nitrogen abundance as a function of time for models with different initial masses and rotational velocity as indicated by the legend (units are [M$_{\odot}$]-[km~s$^{-1}$]). Models are chosen to represent all mass ranges and the three classes of evolution.}
\label{fig:vN}
\end{figure}

The surface nitrogen abundance of a star traces the internal mixing efficiency \citep[e.g.][]{Koehler:2015}. Fig.~\ref{fig:vN} shows the evolution of the surface nitrogen mass fraction for some of our stellar models relative to their surface hydrogen mass fraction. Three tracks are plotted for every mass representing the three classes of evolution (NE, TE, CHE). Due to hydrogen-burning, the N/H ratio cannot decrease during the main-sequence lifetime plotted here. The CNO equilibrium abundance of nitrogen for normal hydrogen and helium abundances corresponds roughly to 12+log(N/H)=7. Higher values in Fig.~\ref{fig:vN} imply a reduced hydrogen abundance.

Stars which undergo NE (represented by the tracks of 100~km~s$^{-1}$ in Fig.~\ref{fig:vN}) are slow rotators. They mix some amount of nitrogen but the surface nitrogen abundance remains far from the CNO equilibrium abundance, indicating that rotational mixing is not very efficient. However, the higher the mass the larger the convective core and the more nitrogen appears at the surface. 

The 294~M$_{\odot}$--100~km~s$^{-1}$ sequence becomes a core-hydrogen-burning red supergiant near the end of the main-sequence evolution. When the model approaches the red supergiant branch, the mass-loss becomes higher and deeper layers are uncovered. This causes a rapid increase of the N/H ratio at the stellar surface near the TAMS. 

The intermediate rotators (represented by the tracks of 200-425~km~s$^{-1}$ in Fig.~\ref{fig:vN}) undergo TE (cf. Sect.~\ref{sec:YcYs}). These models are mixed during the first part of their evolution, but then a chemical gradient develops between the core and the envelope which prevents further mixing. % \red Discuss an example here 
The fastest rotators (represented by the tracks of 450~km~s$^{-1}$ in Fig.~\ref{fig:vN}) are chemically-homogeneously-evolving stars. They are, per definition, mixed throughout: every chemical change in the core is apparent at the surface as well. 

\citet{Brott:2011a} computed the N/H ratio for stellar models with Galactic, LMC and SMC composition. The initial abundance of nitrogen in our models is much lower than that in the Brott models. However, our rotating models with NE and TE reach surface N/H ratios at the TAMS which can be higher than that in the adopted (initial) LMC and SMC compositions. Moreover, our models with CHE, even the less massive ones, have surface N/H abundance ratios as high as 8.4. This value is higher than any value predicted by the Galactic, LMC and SMC models without CHE. 
The reason of this high N/H value in our models is the homogeneous mixing which transports all the hydrogen supply into the burning regions where it is destroyed. 
Consequently, observing surface N/H abundance ratios as high as 8.4 for a massive single star in a low-metallicity environment might imply that the star evolved chemically-homogeneously.

%---------------------------------------------------------------------------
% Ionizing flux

\section{Photoionizing fluxes}\label{sec:flux}

Massive stars ionise their surroundings through their intense UV radiation \citep{Schaerer:1999b,Peters:2010}. 
To estimate the amount of ionizing radiation released by low-metallicity main-sequence stars,  we discuss the ionizing fluxes of our stellar models based on the black body approximation (see also Table~\ref{tab:ion}). 

In this section, we first present the ionizing fluxes and photon numbers calculated in the Lyman continuum (i.e. $\lambda < 912~\AA$), in the HeI continuum (i.e. $\lambda < 504~\AA$) 
and in the He\,II continuum (i.e. $\lambda < 228~\AA$). We then analyse the time evolution of the emission and the validity of the black body approximation. Finally, we discuss two aspects of our stellar models in terms of observational constraints: the total He\,II flux measured in \izw and the connection of our models to gamma-ray bursts and superluminous supernovae.

\subsection{Time-integrated ionizing fluxes}

\begin{figure}[h!]
\resizebox{\hsize}{!}{\includegraphics[angle=270]{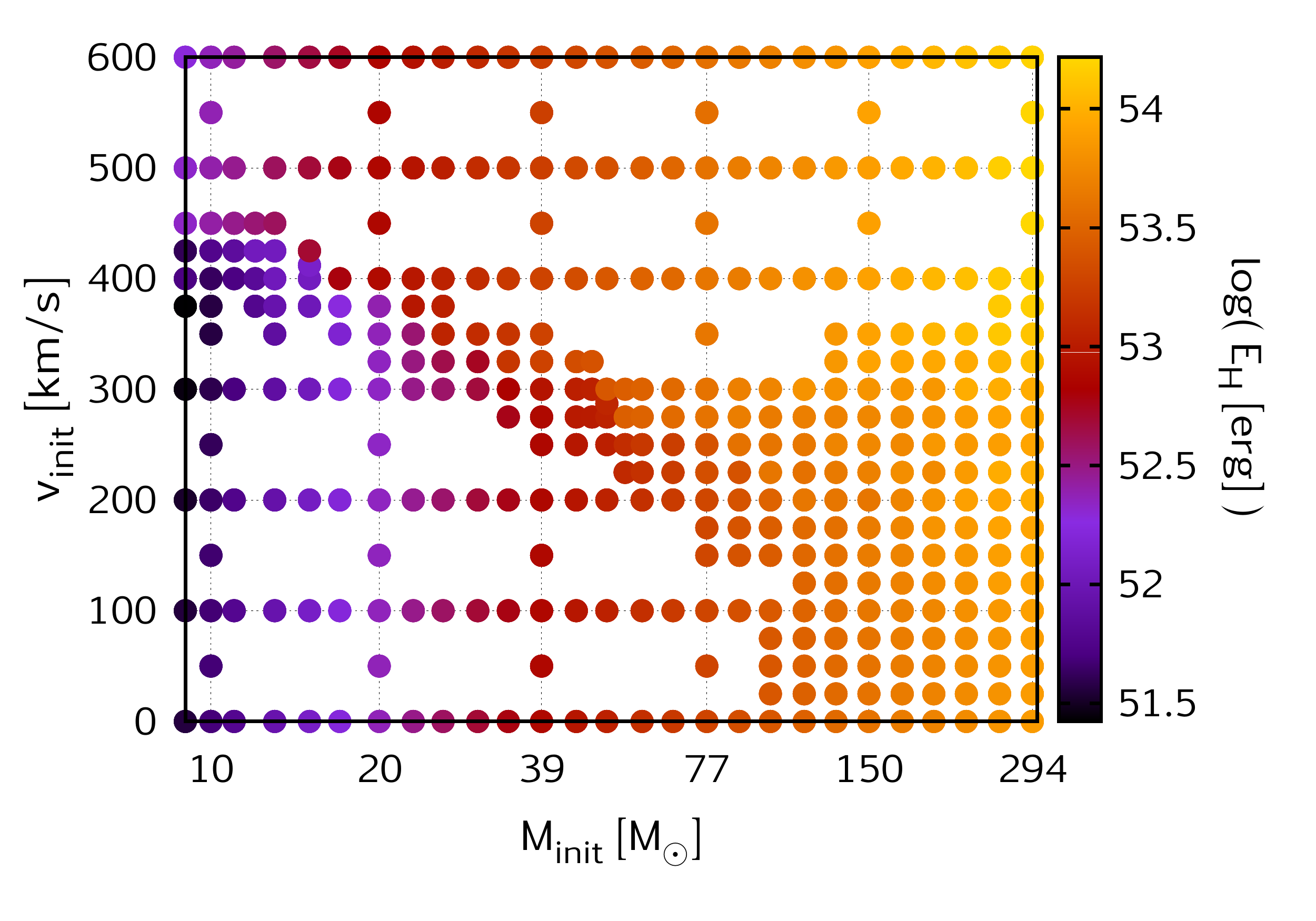}}
\resizebox{\hsize}{!}{\includegraphics[angle=270]{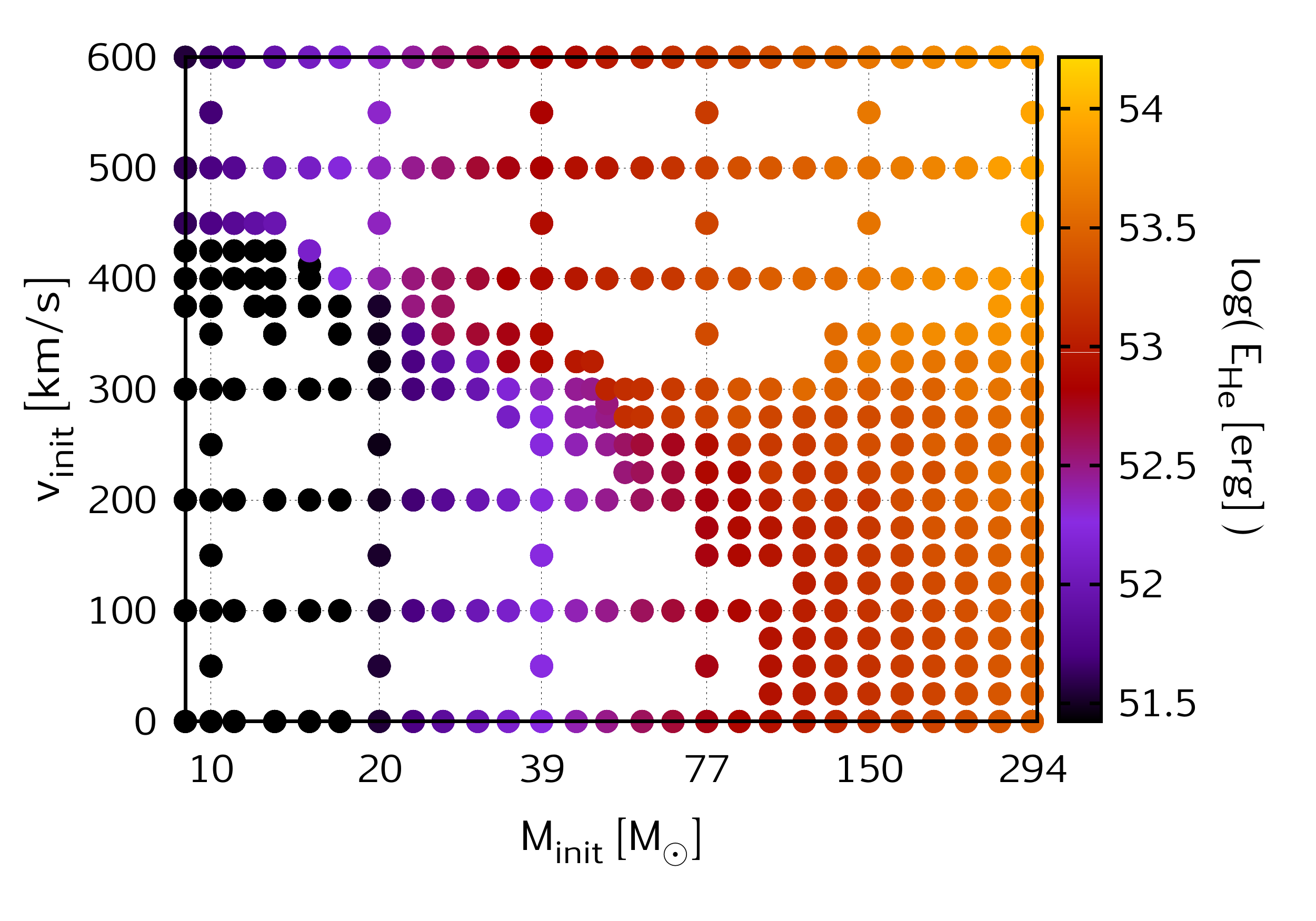}}
\resizebox{\hsize}{!}{\includegraphics[angle=270]{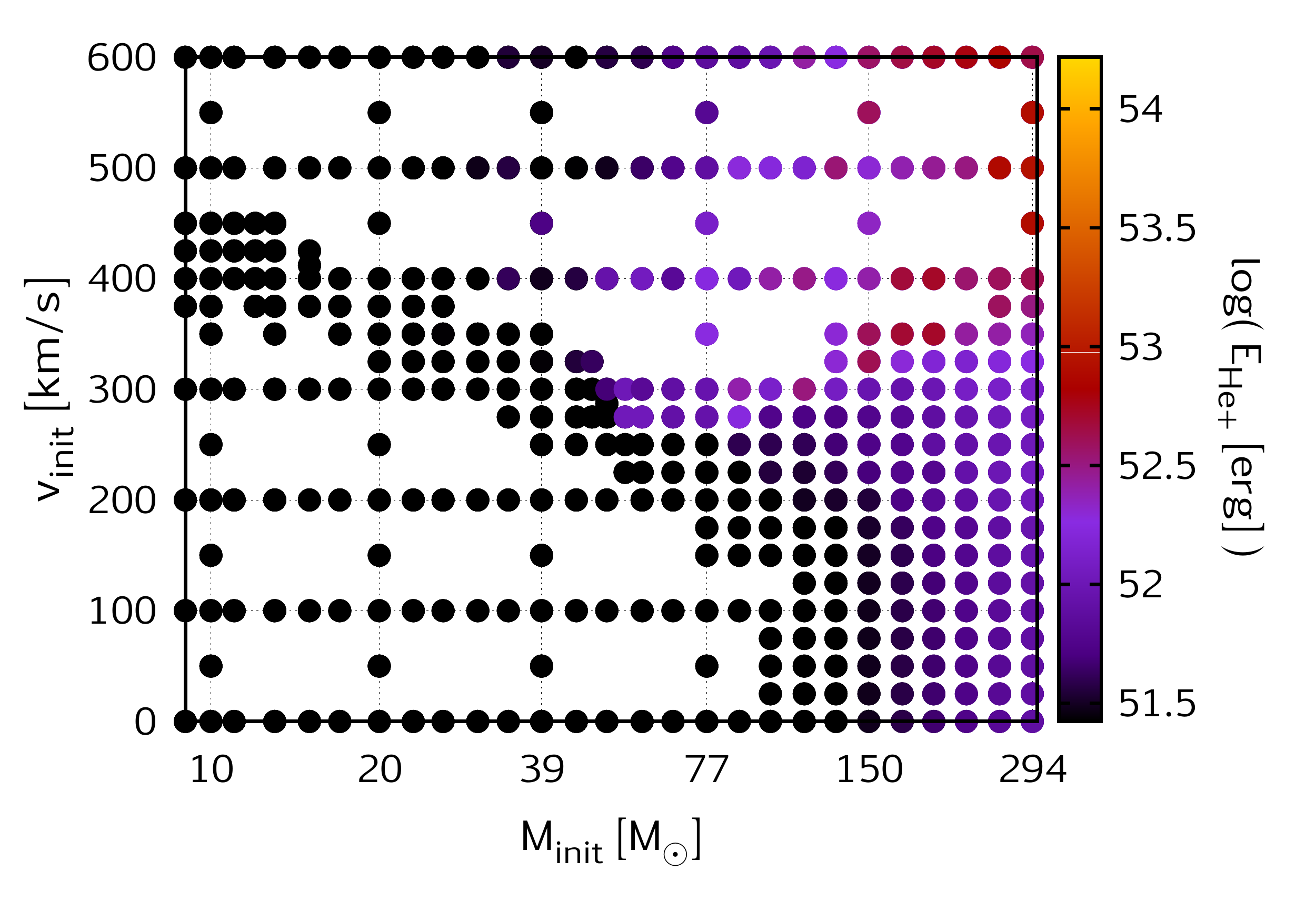}}
\caption{\textsl{Top:} Ionizing energy emitted by our stellar sequences in the Lyman continuum during their main-sequence lifetime. Each evolutionary sequence of our grid is represented by one dot in this diagram and the total amount of energy emitted in the Lyman continuum (in ergs) is colour coded in a logarithmic scale.
\textsl{Middle and bottom:} The same as the top figure but for the HeI and He\,II continua, respectively.
}
\label{fig:Lyman}
\end{figure}

	%Lym grid
The top panel in Fig.~\ref{fig:Lyman} shows the time-integrated energy, i.e. the total energy that is emitted by our models in the Lyman continuum during their core-hydrogen-burning lifetimes. The total emitted flux is an increasing function of the initial mass for the following reason. Although the main-sequence lifetime becomes shorter for a higher-mass model, both the luminosity and the surface temperature increase with the mass so much that the most massive model is able to radiate $\sim$10$^3$~times more ionizing energy during its main-sequence lifetime than the lowest mass one. 

According to Fig.~\ref{fig:Lyman}, the emitted flux also depends on the initial rotation rate. On one hand, rotation increases the lifetime of a model (more fuel is mixed into the core due to rotational mixing), therefore increasing the time-integrated energy. 
Amongst the less massive, normally-evolving models, on the other hand, the lowest amount of ionizing energy is produced in the sequence with 9~M$_{\odot}$-375~km~s$^{-1}$ (see Table~\ref{tab:sumflux}), while the non-rotating model with the same mass produces somewhat more energy. This is because the rotating model has a larger radius than the non-rotating one due to the centrifugal acceleration, hence its luminosity and effective temperature are lower. 

The fast rotators evolve chemically-homogeneously towards higher luminosities and higher surface temperatures. They generally produce $\sim$5-10 times more ionizing energy than their normally-evolving counterparts of the same mass during the main-sequence lifetime.

	%He grid
The time-integrated fluxes in the HeI and the He\,II continua are influenced by the mass and the rotation a similar way as those in the Lyman continuum. Consequently, the minimum and maximum time-integrated flux values correspond, respectively, to the lowest- and highest-mass models, at intermediate rotation rates. Table~\ref{tab:sumflux} gives the minimum and maximum values of the energy and photon numbers emitted by our models in all three continua.

\begin{table*}[ht!]
\caption{Minimum and maximum values of time-integrated energy (E) in ergs, 
time-integrated number (Q) of the ionizing photons and the time average value of the photon numbers per seconds 
for hydrogen (H), neutral helium (He) and singly ionised helium (He$^+$) emitted during the main-sequence phase by our stellar evolutionary models (cf. Table~\ref{tab:ion}). Black body radiation is assumed. For comparison, the ionizing energy and photon numbers of the Pop~III models of \citet{Yoon:2012} are shown: these values are systematically higher than those of our corresponding models due to the higher effective temperatures of the metal-free models (see also Fig.~\ref{fig:flux}).}
\label{tab:sumflux}
\centering
\begin{tabular}{l||c|ccc||c|ccc}
%\hline
& \textsl{this work} & E$_{\rm tot}$ [erg] & Q$_{\rm tot}$ & Q$_{\rm avr}$ [s$^{-1}$] & \textsl{Pop~III models} & E$_{\rm tot}^{PopIII}$ [erg] & Q$_{\rm tot}^{PopIII}$ & Q$_{\rm avr}^{PopIII}$ [s$^{-1}$] \\ \hline
H & 9~M$_{\odot}$-375~km/s & 2.65e51 & 1.00e62 & 1.21e+47  & 10~M$_{\odot}$-0~v$_{\rm k}$ & 1.53e52 & 4.95e62 & 6.41e+47 \\
 & 100~M$_{\odot}$-500~km/s & 4.87e53 & 1.32e64 & 1.52e+50  & 100~M$_{\odot}$-0.4~v$_{\rm k}$ & 7.31e53 & 1.66e64 & 2.48e+50 \\
 & 294~M$_{\odot}$-500~km/s & 1.49e54 & 3.89e64 & 6.75e+50  & 300~M$_{\odot}$-0.4~v$_{\rm k}$ & 1.96e54 & 4.55e64 & 7.19e+50 \\ \hline
He & 9~M$_{\odot}$-375~km/s & 9.22e49 & 2.11e60 & 2.55e+45  & 10~M$_{\odot}$-0~v$_{\rm k}$ & 3.36e51 & 7.10e61 & 9.20e+46 \\
 & 100~M$_{\odot}$-500~km/s & 2.28e53 & 4.33e63 & 4.99e+49  & 100~M$_{\odot}$-0.4~v$_{\rm k}$ & 4.78e53 & 8.18e63 & 7.32e+49 \\
 & 294~M$_{\odot}$-500~km/s & 7.58e53 & 1.42e64 & 2.46e+50  & 300~M$_{\odot}$-0.4~v$_{\rm k}$ & 1.24e54 & 2.15e64 & 2.92e+50 \\ \hline
He$^{+}$ & 9~M$_{\odot}$-375~km/s & 1.81e45 & 1.98e55 & 2.39e+40  & 10~M$_{\odot}$-0~v$_{\rm k}$ & 1.20e49 & 1.27e59 & 1.65e+44 \\
 & 100~M$_{\odot}$-500~km/s & 9.63e51 & 9.72e61 & 1.12e+48  & 100~M$_{\odot}$-0.4~v$_{\rm k}$ & 6.24e52 & 6.03e62 & 5.39e+48 \\
 & 294~M$_{\odot}$-500~km/s & 3.93e52 & 3.95e62 & 6.85e+48  & 300~M$_{\odot}$-0.4~v$_{\rm k}$ & 1.41e53 & 1.37e63 & 1.86e+49 \\ \hline
\end{tabular}
%\tablefoot{}
\end{table*}

\subsection{Time evolution of the emission}

\begin{figure}[h!]
\resizebox{\hsize}{!}{\includegraphics[angle=270]{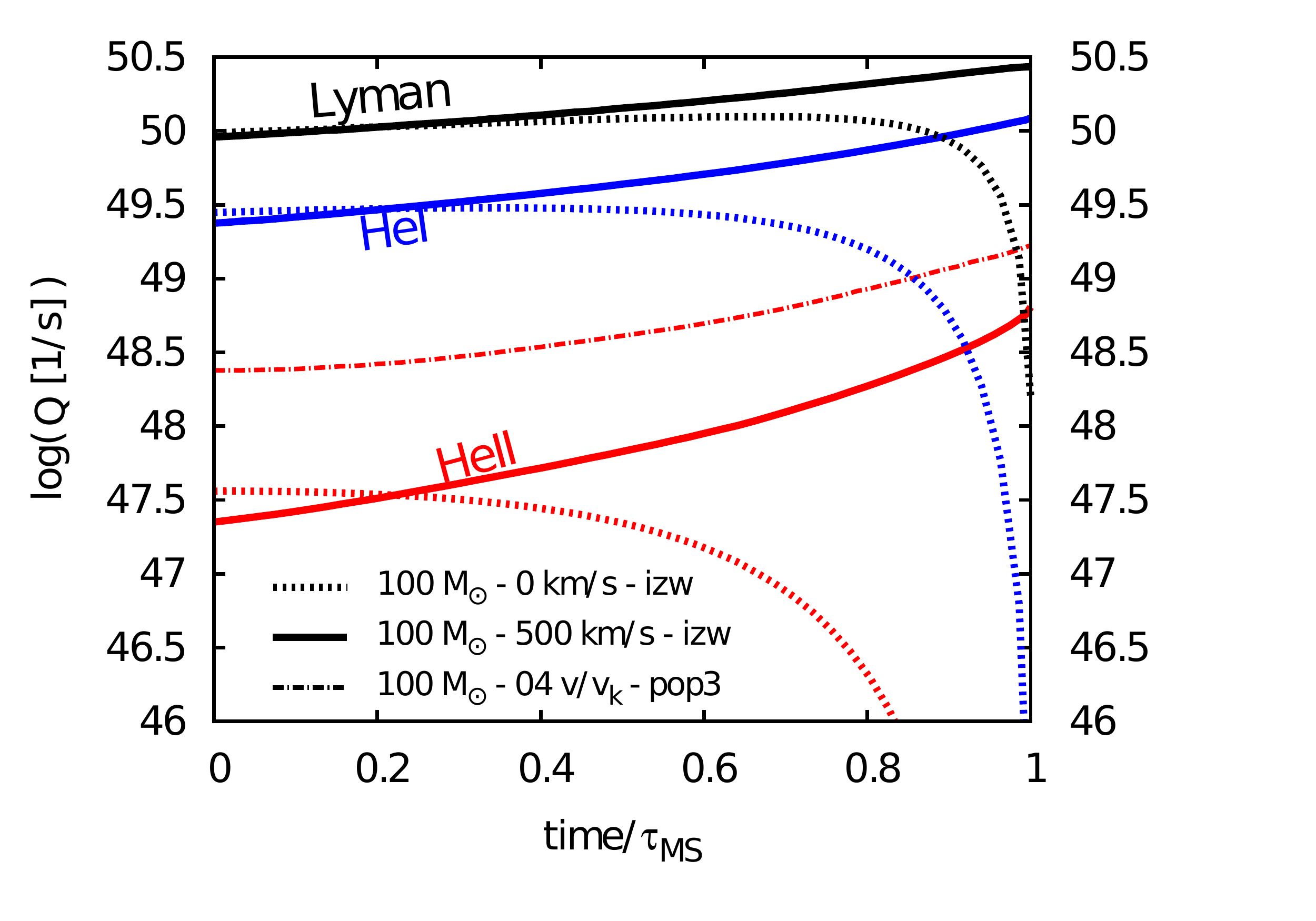}}
\caption{
Time evolution of the number of the ionizing photons for models with an initial mass of 100~M$_{\odot}$. 
Photons emitted in the Lyman, HeI and He\,II continua by a normally-evolving model (without rotation) and a chemically-homogeneously-evolving model (with an initial rotational velocity of 500~km~s$^{-1}$) from our grid are plotted. 
For comparison, the number of photons emitted in the He\,II continuum by a chemically-homogeneously-evolving Pop~III model from \citet{Yoon:2012} with similar initial mass and initial rotational velocity (in units of the critical rotation, which refers to v$_{\rm ini}\sim$520~km~s$^{-1}$) is 
plotted by the dotted-dashed line. Note that the He\,II flux of the Pop~III model may, however, be overestimated by a factor of three (see text). 
}
\label{fig:flux}
\end{figure}

	%time evol
Fig.~\ref{fig:flux} shows the time evolution of the emission from both normal and chemically-homogeneous models with M$_{\rm ini}=$100~M$_{\odot}$. According to the plot, the emission from the model with NE decreases while that with CHE increases during their main-sequence evolution. This is expected since the model with CHE evolves towards higher luminosities and higher effective temperatures. 

The average photon flux in the He\,II continuum is 2.00$\times$10$^{47}$~s$^{-1}$ for 
the model with NE and 1.12$\times$10$^{48}$~s$^{-1}$ for the model with CHE during their main-sequence phase. 
Considering somewhat lower masses, we find that the time-average He\,II photon flux from the chemically-homogeneous models is higher than that of the normally-evolving models by factors of
9, 12, and 15 at 77$\mso$, 51$\mso$ and 39$\mso$, respectively. 
The order of magnitude of these ratios implies that the contribution of the models with 
CHE to the total emitted He\,II ionizing flux of a low-metallicity galaxy may be significant. 

Moreover, Fig.~\ref{fig:flux} demonstrates that towards the end of the main-sequence evolution, 
the ionizing fluxes of the chemically-homogeneous models can be an order of magnitude larger 
than those of normally-evolving models. In fact, comparing the peak ionizing He\,II fluxes
from the chemically-homogeneous models to those of the normally-evolving models, we find ratios of
20, 27, 50, and 92 for stars of 100$\mso$, 77$\mso$, 51$\mso$ and 39$\mso$, respectively.
We therefore expect that ionizing fluxes predicted by starburst models will drastically 
change when the TWUIN stars are taken into account.

	%popIII evol
For comparison, a corresponding metal-free model with CHE is plotted in Fig.~\ref{fig:flux}. The ratio of the 
time-integrated fluxes of the two models with CHE for the main-sequence lifetime is 
E($\gamma_{He^{+}}$)$^{IZw18}$/E($\gamma_{He^+}$)$^{PopIII}$=0.15 in the He\,II continuum.  Table~\ref{tab:sumflux} compares the time-integrated ionizing energy and photon numbers 
in all three bands between our sequences and the Pop~III sequences. The differences between the metal-poor and metal-free models derive from the latter evolving at systematically higher effective temperatures. Note, however, that the He\,II fluxes of the metal-free models may be overestimated by a factor of three (see below).

\subsection{Validity of the approximations}\label{sec:valid}

At the considered metallicity, the mass-loss of massive stars is generally sufficiently weak to make the wind transparent in the continuum. As mentioned in Sect.~\ref{sec:wrHR}, only our most massive chemically-homogeneously-evolving models are expected to develop winds with a continuum optical depth of order unity (cf., Fig.\,\ref{fig:tau}).

\citet{Kudritzki:2002} investigated the dependence of the ionizing photon fluxes of low-metallicity massive main-sequence stars on the mass-loss rate. He found the ionizing fluxes in general to be reduced at the highest considered metallicities due to the correspondingly stronger stellar winds. However, below a threshold metallicity, Kudritzki found that the fluxes are not affected by the winds anymore as they become too weak. While in the quoted work, the threshold metallicity is close to the one used in our models, the mass-loss rates adopted by Kudritzki are significantly larger than what is assumed in our work.

\begin{figure}[h!]
\resizebox{\hsize}{!}{\includegraphics[angle=270,page=3]{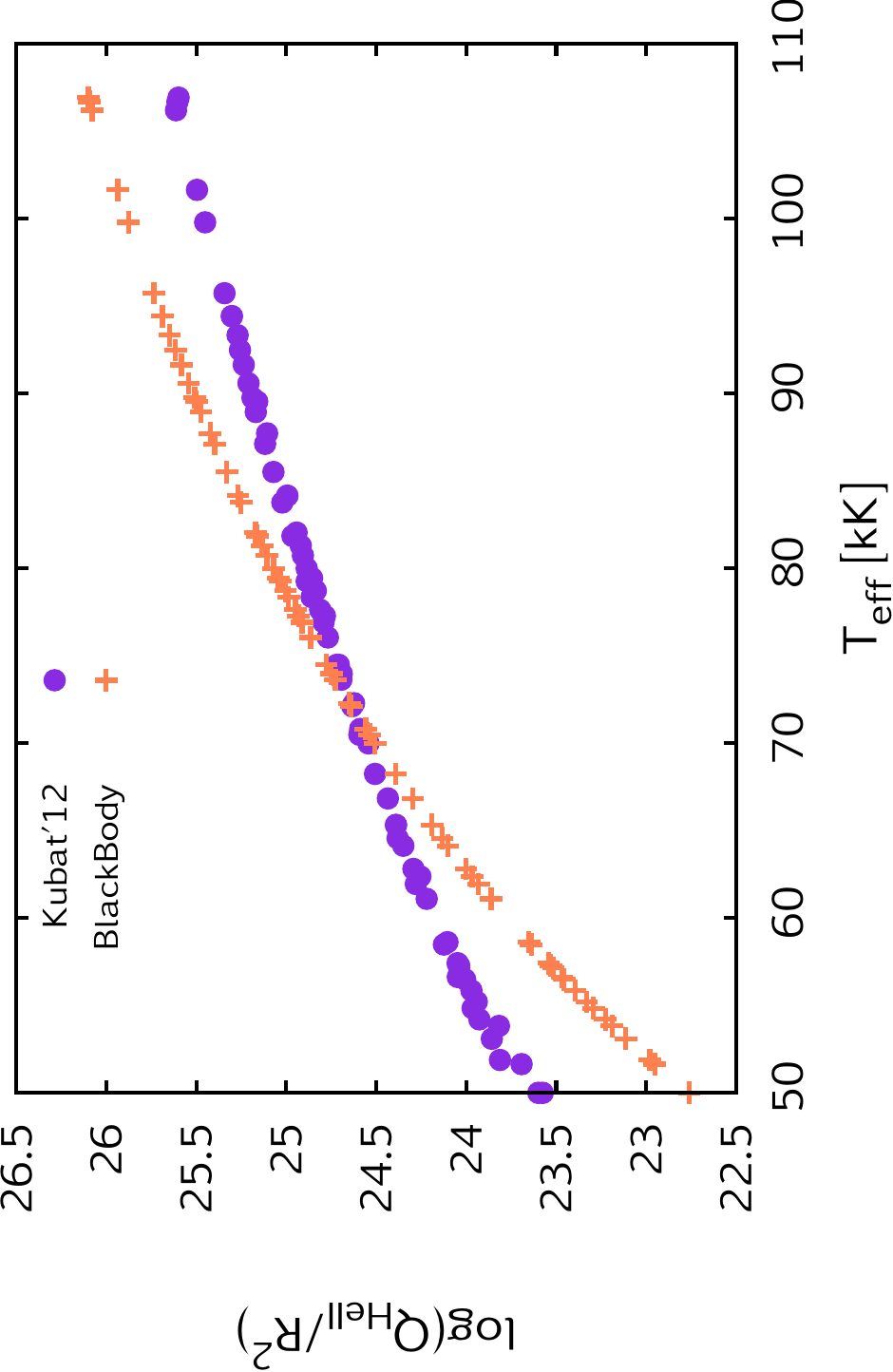}}
\caption{
Ionizing flux (normalised to one square centimetre of stellar surface) in the He\,II continuum provided by \citet{Kubat:2012} based on non-LTE spectra of metal-free massive stars and ionizing flux values using a black body approximation for the same collection of metal-free massive stars.
}
\label{fig:kubat}
\end{figure}

For He\,II ionizing photons, and only for those, Kudritzki found slightly more complex behaviour. At 50\,kK and below, he found that the He\,II flux~per~cm$^2$ can increase for stronger winds. However, at the highest effective temperature, 60\,kK, this effect was not seen anymore. In fact, the He\,II flux~per~cm$^2$ for the most luminous main-sequence stars at 60\,kK is predicted by Kudritzki to be approximately $10^{23.5}\,$s$^{-1}$ in the case of both our metallicity and lower values. This value is very close to the black body prediction, as shown in Fig.~\ref{fig:kubat}. 

        %Kubat
\citet{Kubat:2012} calculated the ionizing flux of metal-free massive hot stars based on simulated stellar atmospheres, for a large range of effective temperatures. In Fig.~\ref{fig:kubat}, we show the He\,II fluxes from \citet{Kubat:2012} for a large number of stellar models. As the models cover a wide mass range, Fig.~\ref{fig:kubat} demonstrates that gravity effects, which cause the scatter in the plot, are very small. 

When comparing Kubat's results to the black body prediction, we note that while our metallicity is different from zero, \citet{Kudritzki:2002} finds that the metallicity dependence of the ionizing fluxes at the considered low-metallicity is generally quite weak, as discussed above. We find that in the temperature range of our most massive and hot stellar models, between 60~kK<T$_{\rm eff}$<85~kK (i.e. where they provide most of the ionizing radiation), Fig.~\ref{fig:kubat} shows that the ionizing flux in the He\,II continuum calculated from the black body approximation matches that calculated from the stellar atmospheres to within $\lesssim$50\% (0.3 dex). Although this comparison is limited by the number of stellar spectra provided by \citet{Kubat:2012}, as well as the difference between the composition of their metal-free models and our metal-poor models, it implies that the ionizing energy coming from our stellar evolutionary models using the black body approximation is indeed a good estimate for their He\,II ionizing fluxes.

We note that the Pop~III models are significantly hotter than our low-metallicity models during core-hydrogen-burning. While this is the reason of their He\,II fluxes based on black body spectra being almost one order of magnitude larger than those of our models, Fig.~\ref{fig:kubat} indicates that they may over-predict the true He\,II fluxes by a factor of three, whereas at the effective temperatures of our models this appears not to be the case. Thus, the step from the metallicity considered by our models to zero may increase the He\,II fluxes not by a factor of ten, as one might grasp from Fig.~\ref{fig:flux}, but just by a factor of three.
%“We note that the Pop III models from Yoon et al. (2012) are significantly hotter than our low-metallicity models during core hydrogen burning. This results in significantly greater estimates of He II fluxes (e.g. Fig. 17) as they also used black body models. At the lower temperatures of our models, this discrepancy is less severe. Thus, the step from our low-metallicity calculations to the zero-metallicity models may only increase the He II fluxes by around a factor of three (rather than the factor of ten which one might conclude from inspection of Fig. 17)” 

In summary, the neglect of wind effects and the black body approximation both introduce uncertainties into the predicted ionizing fluxes. However, in the mass and metallicity regime which we consider here, the uncertainty of both effects appears to be within a factor of two. Conceivably, other uncertainties may be larger. Indeed, the mass-loss rates we consider, in particular those for WR stars, may be more uncertain. E.g. \citet{Vink:2011}, found the mass-loss rate to jump to a steeper relation once the winds become optically thick. This effect, which is observationally confirmed for very luminous Of/WN and WNh stars in the LMC \citep{Bestenlehner:2014}, is not implemented into our stellar models. On the other hand, \citet{Graefener:2008} and \citet{Muijres:2012} predict the winds of the hottest helium-rich stars to become weaker or even to break down for increasing temperature. However, the investigated wind models are largely restricted to effective temperatures below 50\,kK, whereas our TWUIN stars reach values of 80\,kK, and higher. 

Clearly, the ionizing fluxes which we provide are only approximate. Our work demonstrates the need for model atmosphere calculations for very hot stars (50...100~kK) at low (but finite) metallicity. At the same time, self-consistent mass-loss rate predictions are required, to place firmer constraints on the predicted ionizing fluxes.

\subsection{He\,II ionizing flux of star-forming dwarf galaxies}\label{sec:Qtot}

As found in the comprehensive study by \citet{Shirazi:2012}, a large fraction of star-forming dwarf galaxies display strong He\,II emission, which is difficult to understand based on previously published evolutionary models of low-metallicity massive stars (cf. also Sect.~\ref{sec:geneva}). While WR stars are thought to have the potential to produce He\,II ionizing photons, most of the He\,II emitting dwarf galaxies below a certain metallicity do not show WR features in their spectra \citep{Crowther:2006}. We suggest that TWUIN stars (Sect.~\ref{sec:wrHR}) could potentially resolve this discrepancy. 

\citet{Kehrig:2015} reported a He\,II ionizing photon flux, Q(He\,II)$_{\rm obs} \simeq 1.3\times 10^{50}$~photons~s$^{-1}$ measured by integral field spectroscopy for \izw. They also suggested that WR stars are not responsible for most of this emission, and speculated about the presence of very massive, metal-free, chemically-homogeneously-evolving stars in this galaxy. 
Indeed, about 10-15 massive chemically-homogeneously-evolving Pop~III stars with fluxes of 10$^{49}$~photons~s$^{-1}$ in the He\,II continuum could emit the amount of ionizing photons observed \citep{Yoon:2012}. However, the gas in \izw is very metal-poor but not primordial, so the presence of actual Pop~III stars in \izw may be debatable.

As we have shown above, our simulations of massive stars with the composition of \izw predict chemically-homogeneous evolution even for moderately fast rotating stars. Based on the empirical distribution of rotational velocities for O\,stars in the SMC by \citet{Mokiem:2006}, up to 20\% of the very massive stars could undergo CHE. Possibly, at the ten-times smaller metallicity of \izw, massive stars rotate even faster.

\begin{figure}[h!]
\resizebox{1.05\hsize}{!}{\includegraphics[angle=270]{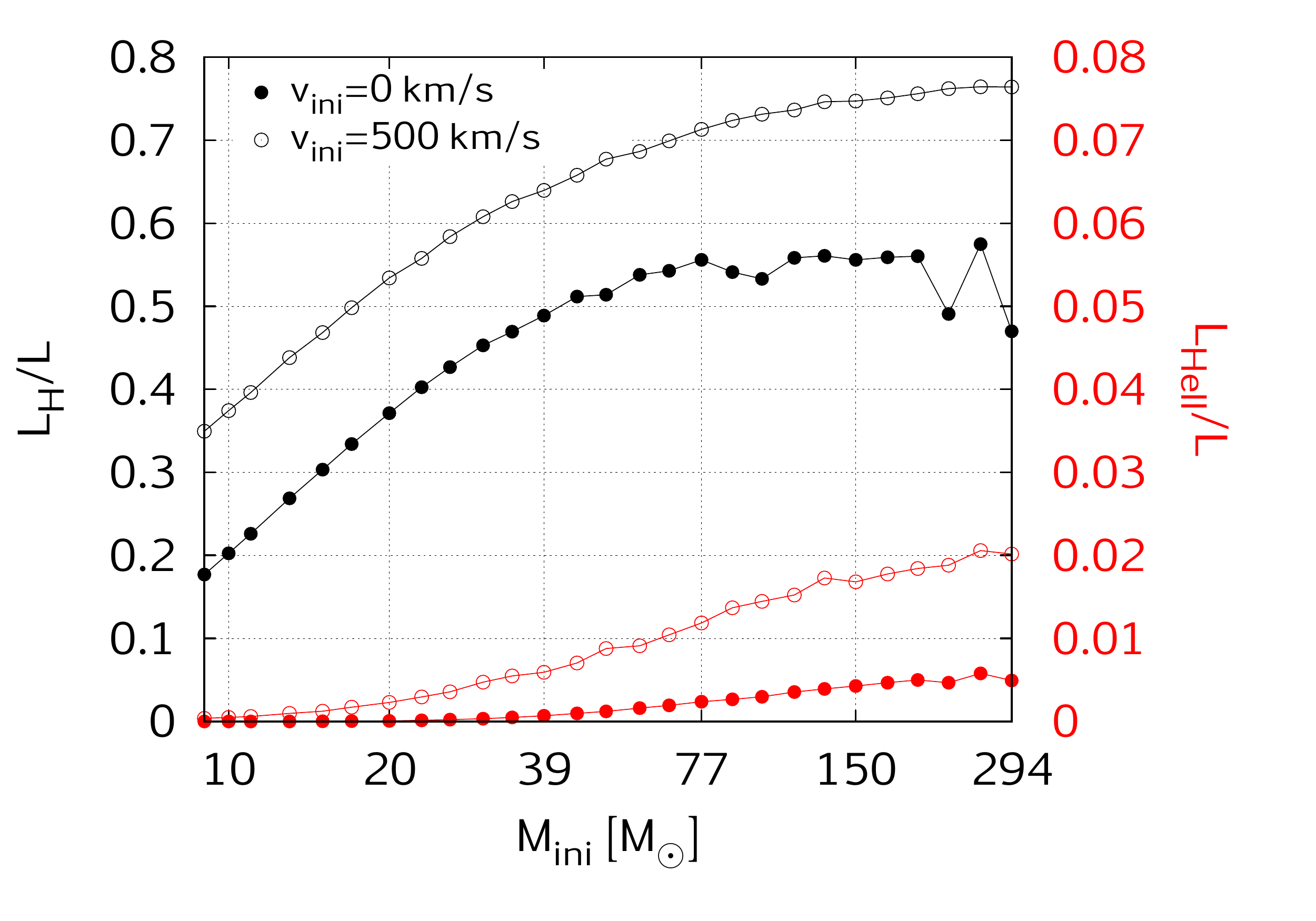}}
\caption{
Time-averaged luminosity in hydrogen and helium\,II ionizing photons, relative to the time-averaged total stellar luminosity of our models without rotation and with $v_{\rm ini}=500\,$km/s, as function of the initial stellar mass.
}
\label{fig:l-ratios}
\end{figure}

Figure\,\ref{fig:l-ratios} shows that the fraction of the stellar luminosity which is emitted as
He\,II ionising photons above $\sim 20\mso$ is weakly increasing with mass. This, together with the number of stars of given mass decreasing as $M^{-\alpha}$, with $\alpha \simeq 2.35$, and the mass-luminosity relation $L\sim M^{\beta}$ having an exponent of $\beta\simeq 2.5\dots 1.5$ for stars in the mass range $20\mso ... 200\mso$ \citep[cf. Fig.~17 of][]{Koehler:2015}, implies that all mass bins in the considered mass range provide similar contributions to the total He\,II flux of a stellar generation. Integration over a Salpeter initial mass function \citep[IMF;][]{Salpeter:1955,Kroupa:2001} from $0.5\mso$ to $500\mso$ and assuming a constant star-formation rate of $0.1\mso$ \citep{Lebouteiller:2013} giving $300\,000\mso$ of stars within 3\,Myr results in a time-averaged He\,II flux of $1.6\times 10^{50}\,$s$^{-1}$ when 20\% of the stars are assumed to undergo chemically-homogeneous evolution. While this simple estimate can not replace proper population synthesis calculations, it indicates that TWUIN stars of finite metallicity may indeed explain the He\,II flux found for \izw, especially given the fact that the maximum He\,II fluxes are about five times higher than the time-averaged values. 

As discussed in Sect.~\ref{sec:wrHR}, TWUIN stars have \mbox{optically-thin ($\tau\lesssim3$)} winds. 
Therefore, they do not contribute to the broad emission signatures that characterise galaxy spectra with WR stars, 
but they still emit sufficient radiation to explain the observed He\,II ionizing photon flux in \izw. 
This may imply that chemically-homogeneous evolution, which leads to TWUIN stars in our calculations, 
is a phenomenon that is indeed happening in nature. 

As a consequence of their high temperature and the lack of optically-thick winds,
TWUIN stars are expected to radiate at ultraviolet wavelengths. This means that their
optical brightness is quite faint, with bolometric corrections estimated from
assuming a black body spectrum of the order of $5\dots 6^{\rm mag}$ for effective temperatures
in the range 70\dots 90\,kK. On the other hand, they may contribute significantly to the
observed optical spectra (rest-frame UV) of high-redshift galaxies.

\subsection{The connection to GRBs, superluminous supernovae and high-z galaxies}
\label{sec:grb}

Our rapidly-rotating models become TWUIN stars due to quasi-chemically-homogeneous evolution, which was identified as a promising road toward long-duration gamma-ray bursts (GRBs) by Yoon \& Langer (2005) and Woosley \& Heger (2006). Indeed, our results are consistent with the study of \citet{Yoon:2006}, who found a very similar threshold rotational velocity for chemically-homogeneous evolution for stars below $60\mso$ as the present work. While we shall present the post-main-sequence evolution of our models in a forthcoming paper, from our models we can expect a similar ratio of GRBs to supernovae (SNe) of the order of 1\%\dots 3\% as \citet{Yoon:2006}. This is consistent with the GRB/SN-ratio in the local Universe being significantly smaller \citep{Podsiadlowski:2004} due to the observed preference for GRBs to occur in low-metallicity dwarf galaxies \citep{Langer:2006,Niino:2011}. As a consequence, we can consider large He\,II-emission in low-metallicity star-forming dwarf galaxies (Sect.~\ref{sec:Qtot}) as a signpost for upcoming GRBs in the same objects.

Similar to GRBs, the recently discovered hydrogen-poor superluminous supernovae \citep[SLSNe;][]{Quimby:2013} also
occur preferentially in low-metallicity dwarf galaxies \citep{Leloudas:2015}. 
While pair-instability explosions \citep{Kozyreva:2014} and massive circumstellar interactions \citep{Moriya:2013,Mackey:2014} 
have been proposed to explain some of these events, the magnetar model \citep{Thompson:2004,Woosley:2010} appears currently favoured \citep{Inserra:2013}. Within the magnetar model, the enormous luminosities as observed
in SLSNe are produced by heating due to the spin-down of a millisecond magnetar. 
Consequently, again similar to GRBs \citep{Thompson:2004}, the progenitor stars
need to produce extremely-rapidly-rotating iron cores. Within this scenario, TWUIN stars could also
be considered as progenitors of SLSNe. While a quantitative connection requires the investigation
of their post-main-sequence evolution, a qualitative connection of SLSNe with low-metallicity
dwarf galaxies appears likely in this context. 

Recently, \citet{Sobral:2015} observed CR7, the most luminous Lyman-$\alpha$ emitter found at $z>6$. They explained the high Lyman-$\alpha$ and He~II emission with a combination of two populations of stars: a `normal', red stellar population which dominates the mass, and a Pop~III population which dominates the nebular emission. While comparing our theoretical predictions to the observational properties of CR7 falls outside of the scope of present work, we emphasize that our stellar models inherently predict two populations of stars: the normally, redwards-evolving ones with slow rotation, and the chemically-homogeneous, bluewards-evolving ones with fast rotation. This latter type, the TWUIN stars, emits intense ionizing radiation but show no WR features, similar to the supposed Pop~III stars. Consequently, two chemically-distinct populations may not be required in CR7, because massive stellar evolution at low metallicity inherently produces the two types of stars observed.

%---------------------------------------------------------------------------
% Discussion

\section{Comparison to previous results}\label{sec:geneva}

We discuss the similarities and differences between our stellar models and two grids of models at similar metallicities, one published by \citet{Meynet:2002} (from now on, MM02) and \citet{Ekstroem:2008}, the other by \citet{Georgy:2013} (from now on, G+13). Both grids have subsolar initial compositions. The grid from G+13 (with Z=0.002) consists of stellar sequences with initial masses between 9-120~M$_{\odot}$ and initial rotational velocities of 0 and 0.4 v/v$_k$ (v$_k$ being the critical velocity at the ZAMS). The grid from MM02 (with Z=0.00001) consists of sequences with initial masses between 9-60~M$_{\odot}$ and initial rotational velocities of 0 and 300~km~s$^{-1}$. 

\subsection{HR diagram}

\begin{figure}[h!]
\resizebox{\hsize}{!}{\includegraphics[angle=270]{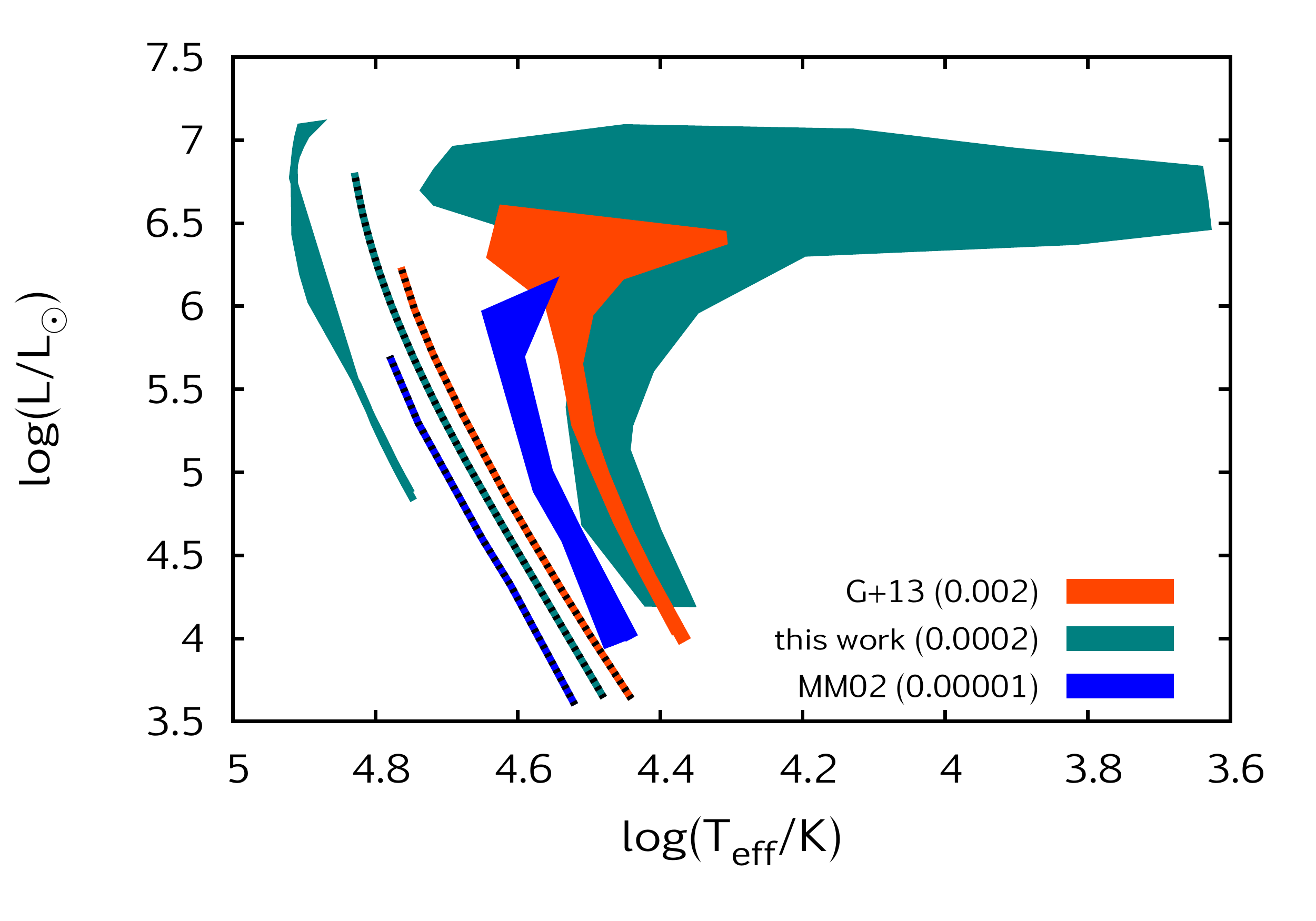}}
\caption{
HR diagram showing where the stellar sequences of three different grids begin the main-sequence evolution (ZAMS, marked with black dashed lines) and where they finish the main-sequence evolution (TAMS, shaded regions). \textsl{G+13}: \citet{Georgy:2013}; \textsl{MM02}: \citet{Meynet:2002}; the initial metallicity of the grids is also indicated by the legend. 
}
\label{fig:HRDD}
\end{figure}

Figure~\ref{fig:HRDD} shows the HR diagram of the three low-metallicity grids. The ZAMS regions of the grids move towards higher effective temperatures when the metallicity is lower. The ZAMS region of our grid extends to higher luminosities because it contains masses up to 294~M$_{\odot}$. 

The TAMS regions populated by the MM02 and G+13 grids are on the red side of the corresponding ZAMS regions, meaning that all the sequences evolve redwards. In the case of our grid, however, there are two separate TAMS areas corresponding to the normally-evolving and the chemically-homogeneously-evolving sequences. The prediction of chemically-homogeneously-evolving sequences at the TAMS is the first important difference between the previous results and our work.

Another important difference between the three grids derives from the value of the overshooting parameter utilised. The grid with Z=0.00001 was computed without taking overshooting into consideration \citep{Meynet:2002}, while the grid with Z=0.002 included an overshooting parameter $\alpha_{\rm over}=0.1 H_p$ \citep{Georgy:2013}. Convective core overshooting gives larger cores, and has been shown to extend the main sequence to lower effective temperatures \citep{Langer:1995}. This is why some of the sequences of our grid (with $\alpha_{\rm over}=0.335 H_p$) finish their main-sequence evolution at lower T$_{\rm eff}$ than the corresponding sequences with Z=0.002. 

The broadening of the TAMS of the normally-evolving sequences of our grid at the very high masses is related to the envelope inflation (Sect.~\ref{sec:hbrs}, also see Fig.~\ref{fig:tracks}). Although the grid with Z=0.002 also shows a broadening around the highest masses (indeed, the non-rotating sequence of the Z=0.002 grid with initial mass of 120~M$_{\odot}$ finishes the main-sequence evolution at logT$_{\rm eff}=4.3$, while the corresponding rotating sequence finishes at logT$_{\rm eff}=4.6$), this effect was linked to the efficiency of the stellar wind \citep{Meynet:2002}. If the winds are strong and the mass lost during the main-sequence evolution is significant (but not strong enough to remove the hydrogen envelope), the mass fraction of the core increases with respect to the total mass. This increases the ratio of the core mass vs. total mass (similarly to the effect of overshooting) and the stellar models appear more red. 

\subsection{Mass-loss history and rotation}

\begin{figure}[h!]
\resizebox{\hsize}{!}{\includegraphics[angle=270]{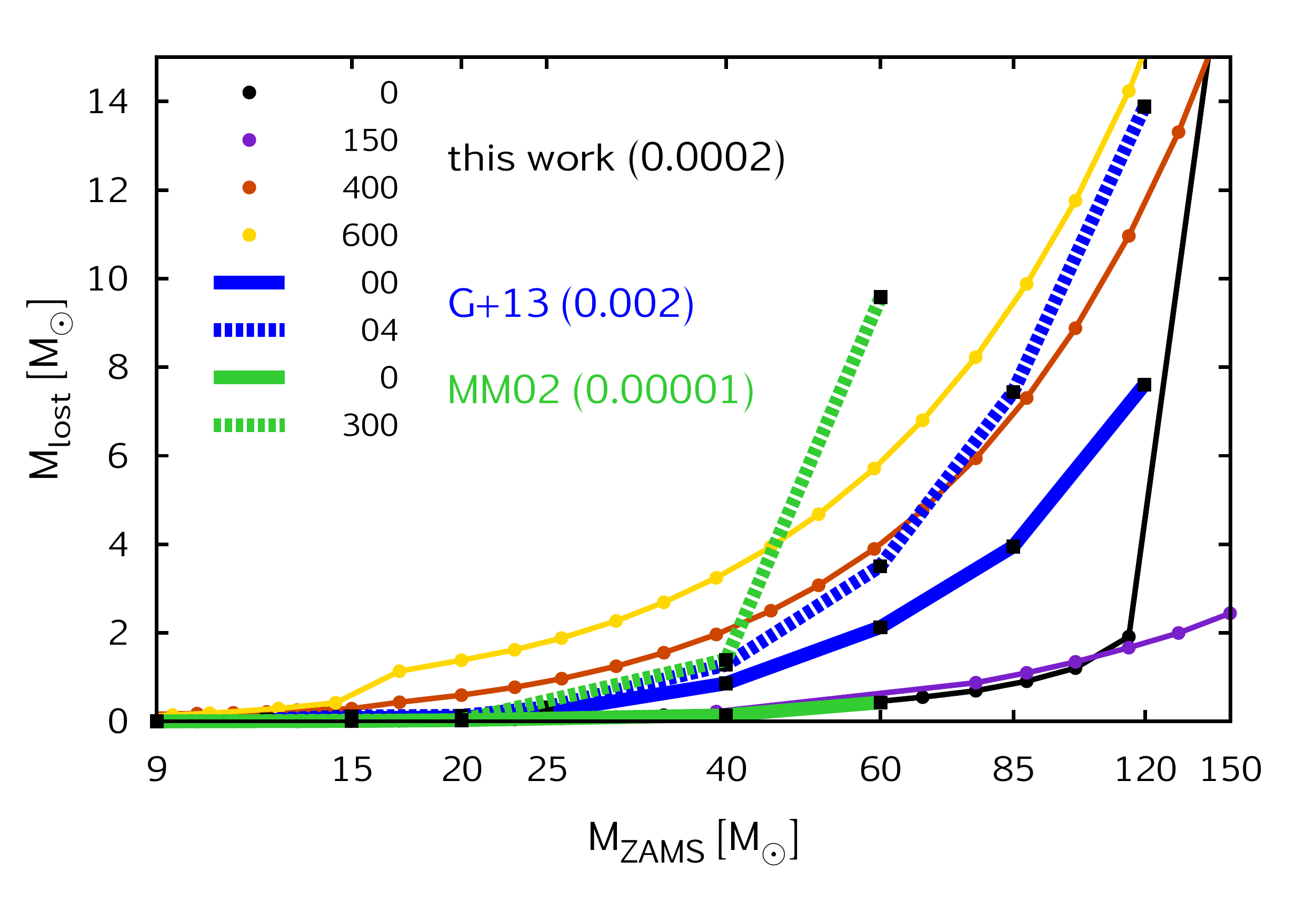}}
\caption{
Mass lost during the evolution of the stellar sequences in the three grids with different initial compositions. Sequences with four initial rotational rates of our grid, as indicated by the legend with units in km~s$^{-1}$, are shown by connected dots. Sequences of the Z=0.002 grid \citep{Georgy:2013} and the Z=0.00001 grid \citep{Meynet:2002} are shown by black rectangles connected with lines as indicated by the legend with units in v$_k$ and km~s$^{-1}$, respectively. Note that the M$_{lost}$ values of the Z=0.00001 grids correspond to the end of the helium-burning phase and are, therefore, an overestimate of the mass lost during the main sequence. The M$_{lost}$ values of the Z=0.002 and \izw grids correspond to the end of the main sequence.
}
\label{fig:massgeneva}
\end{figure}

Fig.~\ref{fig:massgeneva} compares the mass that is lost during the evolution of the stellar sequences in the three different grids as a function of initial mass and rotation. For the analysis of the mass-loss history of our models, we refer to Sect.~\ref{sec:masshist}. While the mass-loss rate prescriptions used by MM02 and G+13 are not exactly the same as ours, they nevertheless result in mass-loss rates comparable to those of the prescriptions used here. This is apparent from Fig.~\ref{fig:massgeneva}. The mass that is lost during the evolution of stars in the mass range of 9-120~M$_{\odot}$ is between 0 and 14~M$_{\odot}$ for models from all three grids, depending strongly on the initial mass. 

The non-rotating Z=0.002 sequences lose more mass during the main-sequence lifetime as our non-rotating sequences due to the metallicity dependence of the mass-loss rates. The non-rotating Z=0.00001 sequences end up having similar mass at the end of the helium-burning phase as our non-rotating sequences at the TAMS.  

Our fast rotating sequences in Fig.~\ref{fig:massgeneva} evolve chemically-homogeneously and undergo WR-type mass-loss during the last few Myr of the main-sequence evolution. Therefore, although their initial metallicity is lower, they might end up less massive at the TAMS than the rotating Z=0.002 sequences, that do not evolve chemically-homogeneously \citep{Georgy:2013}. The rotating Z=0.00001 sequences of 40 and 60~M$_{\odot}$ lose more mass than the rotating Z=0.002 sequences of the same masses; and the rotating Z=0.00001 sequence of 60~M$_{\odot}$ loses even more mass than our chemically-homogeneously-evolving sequences of the same mass. However, this is because the M$_{lost}$ values of the Z=0.00001 grid shown in Fig.~\ref{fig:massgeneva} correspond to the end of helium burning and are, therefore, overestimating the mass lost during their main-sequence evolution.

In Sect.~\ref{sec:rotation}, we analysed the evolution of the rotational velocity in our models and noted that their behaviour is consistent with the findings of \citet{Meynet:2002} and \citet{Ekstroem:2008} for their Z=0.00001 models. These authors concluded that the massive ($\gtrsim$30~M$_{\odot}$) stellar sequences increase their surface rotation due to the strong core-envelope coupling and the low mass-loss rates. Also, if the initial mass function at low metallicity extends up to high-mass stars, as often supposed, rotation is likely to be a major effect in the course of the evolution of massive stars, since many of them are likely to reach high velocities. 

Our calculations supports these conclusions since our normally-evolving massive ($\gtrsim$20~M$_{\odot}$) sequences also increase their surface rotational velocity during the main sequence, as shown in Sect.~\ref{sec:rotation}. A quantitative comparison of our models with the Z=0.00001 models is less meaningful because of a limited overlap of the initial parameter space investigated.

%---------------------------------------------------------------------------
% Summary

\section{Conclusions}\label{sec:Summary}

% the grid and the classification
We presented a grid of stellar evolutionary models in the mass range of 9-300~M$_{\odot}$ 
with initial rotational velocities between 0-600~km~s$^{-1}$, which is dense enough
to be well suited for population-synthesis studies.
The initial mass fraction 
of metals in our models is chosen to be 10\% of that found in the SMC, 
which is probably appropriate for metal-poor blue compact dwarf galaxies such as \izw. 
We found that our models evolve qualitatively differently compared to models of
solar metallicity in several respects. 
We summarize the most important new results below.

\begin{enumerate}

% HDR
\item \textbf{Massive main-sequence stars populate both sides of the ZAMS.}
Apart from the normal (i.e. redwards) evolution of the slow rotating models, fast rotation induces chemically-homogeneous evolution in our low-metallicity massive stars. As these objects evolve bluewards from the zero-age main-sequence (cf. Sect.~\ref{sec:HR}), we predict core-hydrogen-burning objects to be found on both sides of the ZAMS. 
This finding might be relevant to explain observations of high-$z$ galaxies such as CR7 \citep{Sobral:2015} which apparently contains two different types of objects: a normal, red stellar population and a hot and luminous stellar population that dominates the ionizing radiation of the galaxy. As we have shown, low-metallicity massive stellar evolution inherently produces both type of objects. 

% CHB RSG
\item \textbf{Core-hydrogen-burning cool supergiants}. 
We find the majority of our massive (>80\,M$_{\odot}$) models evolve into cool
supergiants while still burning hydrogen in the core, and spend up to 10\%
of their life time as such (cf. Sect.~\ref{sec:hbrs}). This evolutionary outcome is a consequence of 
the low mass-loss rate and the envelope inflation close to the Eddington limit \citep[][]{Sanyal:2015}. Although short lived and rare, they may provide an important 
contribution to the chemical evolution of their environments due to the large amount of 
nuclearly-processed material that is lost in their slow stellar wind.

% WR
\item \textbf{Transparent Wind Ultraviolet Intense (TWUIN) stars.} 
Our fast rotating stars, which may comprise 10\dots 20\% of all massive stars, evolve chemically 
homogeneously and bluewards in the HR~diagram during core hydrogen burning (cf. Sect.~\ref{sec:wrHR}). Due to their 
extremely high effective temperatures, and the expectation that their winds remain 
optically thin, we show that these TWUIN stars may have very high ionizing fluxes (cf. Sect.~\ref{sec:flux}). 
E.g., their maximum He\,II ionizing photon flux is about 20 to 100 times larger than
that of their non-rotating counterparts (cf. Sect.~\ref{sec:Qtot}). We find that the measured He\,II flux of \izw
\citep{Kehrig:2015} as well as the weakness or absence of Wolf-Rayet features 
in \izw and other low-metallicity dwarf galaxies \citep{Shirazi:2012} is compatible
with a population of TWUIN stars in these objects. 

% rotation
\item \textbf{Increasing surface rotational velocity.}
Consistent with previous models of low-metallicity massive stars (cf. Sect.~\ref{sec:geneva}), our calculations show that the normally evolving models increase their surface rotational velocity during the main-sequence lifetime due to strong core-envelope coupling and low mass-loss rates (cf. Sect.~\ref{sec:rotation}). 
Therefore, the rotational velocity distribution of core-hydrogen-burning massive stars to be observed in low-metallicity environments might be different to that in higher-metallicity environments.

% GRB
\item \textbf{Connection to lGRBs and superluminous SNe.}
It has been argued previously that chemically-homogeneous evolution is a promising
path towards long-duration gamma-ray bursts (cf., Sect.\,\ref{sec:grb}). Their preference to
occur in low-metallicity dwarf galaxies, together with the spectroscopic features of dwarf galaxies 
mentioned above, provides increasing evidence for chemically-homogeneous evolution indeed
occurring at low metallicity. 
As superluminous supernovae may also require low metallicity and rapid rotation (see Sect.\,\ref{sec:grb}), chemically-homogeneous evolution may also be a factor in these dramatic final explosions and in the evolutionary path that leads to them.

\end{enumerate}

% outlook
Observations of massive stars in nearby compact dwarf 
galaxies, of massive-star populations in high-redshift galaxies, or even of
stellar explosions in the far Universe can
provide promising avenues to expand and improve our knowledge 
of massive star evolution at low metallicity. Here we provided 
a complementary view from theoretical models
considering the main-sequence evolution, while the post-main-sequence phase will be presented in a forthcoming work.
It will thus require further efforts
on both sides, observational and theoretical, before we are able to understand low-metallicity massive stars
as well as those in our Milky Way, and thus obtain 
a deeper understanding of metal-poor environments in the Universe. 

%---------------------------------------------------------------------------
\begin{acknowledgements}
The authors thank Jorick Vink and the VLT FLAMES Tarantula consortium for their comments.
S.-C. Y was supported by the Basic Science Research (2013R1A1A2061842) program through the National Research Foundation of Korea (NRF). We thank the referee, Dr Sally Heap, for her constructive report on our draft.
\end{acknowledgements}

\bibliographystyle{aa} % style aa.bst
\bibliography{Yourfile} % your references Yourfile.bib

\begin{thebibliography}{162}
\expandafter\ifx\csname natexlab\endcsname\relax\def\natexlab#1{#1}\fi

\bibitem[{Abel {et~al.}(2002)Abel, Bryan, \& Norman}]{Abel:2002}
Abel, T., Bryan, G., \& Norman, M. 2002, Science, 295, 93

\bibitem[{{Aloisi} {et~al.}(2003){Aloisi}, {Savaglio}, {Heckman}, {Hoopes},
  {Leitherer}, \& {Sembach}}]{Aloisi:2003}
{Aloisi}, A., {Savaglio}, S., {Heckman}, T.~M., {et~al.} 2003, \apj, 595, 760

\bibitem[{Aloisi {et~al.}(1999)Aloisi, Tosi, \& Greggio}]{Aloisi:1999}
Aloisi, A., Tosi, M., \& Greggio, L. 1999, ApJ, 118, 302

\bibitem[{Aloisi {et~al.}(2007)}]{Aloisi:2007}
Aloisi, A. {et~al.} 2007, ApJ, 667, L151

\bibitem[{{Annibali} {et~al.}(2013){Annibali}, {Cignoni}, {Tosi}, {van der
  Marel}, {Aloisi}, {Clementini}, {Contreras Ramos}, {Fiorentino}, {Marconi},
  \& {Musella}}]{Annibali:2013}
{Annibali}, F., {Cignoni}, M., {Tosi}, M., {et~al.} 2013, \aj, 146, 144

\bibitem[{{Arcavi} {et~al.}(2010){Arcavi}, {Gal-Yam}, {Kasliwal}, {Quimby},
  {Ofek}, {Kulkarni}, {Nugent}, {Cenko}, {Bloom}, {Sullivan}, {Howell},
  {Poznanski}, {Filippenko}, {Law}, {Hook}, {J{\"o}nsson}, {Blake}, {Cooke},
  {Dekany}, {Rahmer}, {Hale}, {Smith}, {Zolkower}, {Velur}, {Walters},
  {Henning}, {Bui}, {McKenna}, \& {Jacobsen}}]{Arcavi:2010}
{Arcavi}, I., {Gal-Yam}, A., {Kasliwal}, M.~M., {et~al.} 2010, \apj, 721, 777

\bibitem[{Asplund {et~al.}(2009)}]{Asplund:2009}
Asplund, M. {et~al.} 2009, ARA\&A, 47, 481

\bibitem[{Bastian {et~al.}(2013)}]{Bastian:2013}
Bastian, N. {et~al.} 2013, MNRAS, 436, 2398

\bibitem[{{Beers} \& {Christlieb}(2005)}]{Beers:2005}
{Beers}, T.~C. \& {Christlieb}, N. 2005, \araa, 43, 531

\bibitem[{{Bestenlehner} {et~al.}(2014){Bestenlehner}, {Gr{\"a}fener}, {Vink},
  {Najarro}, {de Koter}, {Sana}, {Evans}, {Crowther}, {H{\'e}nault-Brunet},
  {Herrero}, {Langer}, {Schneider}, {Sim{\'o}n-D{\'{\i}}az}, {Taylor}, \&
  {Walborn}}]{Bestenlehner:2014}
{Bestenlehner}, J.~M., {Gr{\"a}fener}, G., {Vink}, J.~S., {et~al.} 2014, \aap,
  570, A38

\bibitem[{{Bromm} \& {Larson}(2004)}]{Bromm:2004}
{Bromm}, V. \& {Larson}, R.~B. 2004, \araa, 42, 79

\bibitem[{{Brott} {et~al.}(2011){Brott}, {de Mink}, {Cantiello}, {Langer}, {de
  Koter}, {Evans}, {Hunter}, {Trundle}, \& {Vink}}]{Brott:2011a}
{Brott}, I., {de Mink}, S.~E., {Cantiello}, M., {et~al.} 2011, \aap, 530, A115

\bibitem[{Böhm-Vitense(1958)}]{Boehm:1958}
Böhm-Vitense, E. 1958, ZAp, 46, 108

\bibitem[{{Cantiello} {et~al.}(2007){Cantiello}, {Yoon}, {Langer}, \&
  {Livio}}]{Cantiello:2007}
{Cantiello}, M., {Yoon}, S.-C., {Langer}, N., \& {Livio}, M. 2007, \aap, 465,
  L29

\bibitem[{Caretta(2010)}]{Caretta:2010}
Caretta. 2010, A\&A, 516, A55

\bibitem[{Caretta {et~al.}(2005)}]{Caretta:2005}
Caretta, E. {et~al.} 2005, A\&A, 433, 597

\bibitem[{{Castro} {et~al.}(2014){Castro}, {Fossati}, {Langer},
  {Sim{\'o}n-D{\'{\i}}az}, {Schneider}, \& {Izzard}}]{Castro:2014}
{Castro}, N., {Fossati}, L., {Langer}, N., {et~al.} 2014, \aap, 570, L13

\bibitem[{{Chabrier} {et~al.}(2014){Chabrier}, {Hennebelle}, \&
  {Charlot}}]{Chabrier:2014}
{Chabrier}, G., {Hennebelle}, P., \& {Charlot}, S. 2014, \apj, 796, 75

\bibitem[{{Chiappini} {et~al.}(2011){Chiappini}, {Frischknecht}, {Meynet},
  {Hirschi}, {Barbuy}, {Pignatari}, {Decressin}, \& {Maeder}}]{Chiappini:2011}
{Chiappini}, C., {Frischknecht}, U., {Meynet}, G., {et~al.} 2011, \nat, 472,
  454

\bibitem[{{Chini} {et~al.}(2012){Chini}, {Hoffmeister}, {Nasseri}, {Stahl}, \&
  {Zinnecker}}]{Chini:2012}
{Chini}, R., {Hoffmeister}, V.~H., {Nasseri}, A., {Stahl}, O., \& {Zinnecker},
  H. 2012, \mnras, 424, 1925

\bibitem[{{Ciardi} {et~al.}(2003){Ciardi}, {Ferrara}, \& {White}}]{Ciardi:2003}
{Ciardi}, B., {Ferrara}, A., \& {White}, S.~D.~M. 2003, \mnras, 344, L7

\bibitem[{{Crowther} \& {Hadfield}(2006)}]{Crowther:2006}
{Crowther}, P.~A. \& {Hadfield}, L.~J. 2006, \aap, 449, 711

\bibitem[{{Crowther} {et~al.}(2010){Crowther}, {Schnurr}, {Hirschi}, {Yusof},
  {Parker}, {Goodwin}, \& {Kassim}}]{Crowther:2010}
{Crowther}, P.~A., {Schnurr}, O., {Hirschi}, R., {et~al.} 2010, \mnras, 408,
  731

\bibitem[{{Dabringhausen} {et~al.}(2009){Dabringhausen}, {Kroupa}, \&
  {Baumgardt}}]{Dabringhausen:2009}
{Dabringhausen}, J., {Kroupa}, P., \& {Baumgardt}, H. 2009, \mnras, 394, 1529

\bibitem[{D'Antona \& Ventura(2010)}]{DAntona:2010}
D'Antona, F. \& Ventura, P. 2010, Proceedings IAU Symposium, No.268

\bibitem[{{de Jager} {et~al.}(1988){de Jager}, {Nieuwenhuijzen}, \& {van der
  Hucht}}]{deJager:1988}
{de Jager}, C., {Nieuwenhuijzen}, H., \& {van der Hucht}, K.~A. 1988, \aaps,
  72, 259

\bibitem[{de~Mink {et~al.}(2009)}]{deMink:2009}
de~Mink, S. {et~al.} 2009, A\&A, 507, L1

\bibitem[{{de Mink} {et~al.}(2013){de Mink}, {Langer}, {Izzard}, {Sana}, \& {de
  Koter}}]{deMink:2013}
{de Mink}, S.~E., {Langer}, N., {Izzard}, R.~G., {Sana}, H., \& {de Koter}, A.
  2013, \apj, 764, 166

\bibitem[{{de Mink} {et~al.}(2014){de Mink}, {Sana}, {Langer}, {Izzard}, \&
  {Schneider}}]{deMink:2014}
{de Mink}, S.~E., {Sana}, H., {Langer}, N., {Izzard}, R.~G., \& {Schneider},
  F.~R.~N. 2014, \apj, 782, 7

\bibitem[{{Decressin} {et~al.}(2007){Decressin}, {Meynet}, {Charbonnel},
  {Prantzos}, \& {Ekstr{\"o}m}}]{Decressin:2007}
{Decressin}, T., {Meynet}, G., {Charbonnel}, C., {Prantzos}, N., \&
  {Ekstr{\"o}m}, S. 2007, \aap, 464, 1029

\bibitem[{Denissenkov \& Hartwick(2014)}]{Denissenkov:2014}
Denissenkov, P. \& Hartwick, F. 2014, MNRAS, 437

\bibitem[{{Dufton} {et~al.}(2013){Dufton}, {Langer}, {Dunstall}, {Evans},
  {Brott}, {de Mink}, {Howarth}, {Kennedy}, {McEvoy}, {Potter},
  {Ram{\'{\i}}rez-Agudelo}, {Sana}, {Sim{\'o}n-D{\'{\i}}az}, {Taylor}, \&
  {Vink}}]{Dufton:2013}
{Dufton}, P.~L., {Langer}, N., {Dunstall}, P.~R., {et~al.} 2013, \aap, 550,
  A109

\bibitem[{{Ekstr{\"o}m} {et~al.}(2011){Ekstr{\"o}m}, {Georgy}, {Meynet},
  {Maeder}, \& {Granada}}]{Ekstroem:2011}
{Ekstr{\"o}m}, S., {Georgy}, C., {Meynet}, G., {Maeder}, A., \& {Granada}, A.
  2011, in IAU Symposium, Vol. 272, IAU Symposium, ed. C.~{Neiner}, G.~{Wade},
  G.~{Meynet}, \& G.~{Peters}, 62--72

\bibitem[{{Ekstr{\"o}m} {et~al.}(2008){Ekstr{\"o}m}, {Meynet}, {Maeder}, \&
  {Barblan}}]{Ekstroem:2008}
{Ekstr{\"o}m}, S., {Meynet}, G., {Maeder}, A., \& {Barblan}, F. 2008, \aap,
  478, 467

\bibitem[{{Eldridge} {et~al.}(2008){Eldridge}, {Izzard}, \&
  {Tout}}]{Eldridge:2008}
{Eldridge}, J.~J., {Izzard}, R.~G., \& {Tout}, C.~A. 2008, \mnras, 384, 1109

\bibitem[{{Eldridge} {et~al.}(2011){Eldridge}, {Langer}, \&
  {Tout}}]{Eldridge:2011}
{Eldridge}, J.~J., {Langer}, N., \& {Tout}, C.~A. 2011, \mnras, 414, 3501

\bibitem[{{Eldridge} \& {Stanway}(2012)}]{Eldridge:2012}
{Eldridge}, J.~J. \& {Stanway}, E.~R. 2012, \mnras, 419, 479

\bibitem[{{Espinosa Lara} \& {Rieutord}(2013)}]{Espinosa:2013}
{Espinosa Lara}, F. \& {Rieutord}, M. 2013, \aap, 552, A35

\bibitem[{{Frebel} {et~al.}(2005){Frebel}, {Aoki}, {Christlieb}, {Ando},
  {Asplund}, {Barklem}, {Beers}, {Eriksson}, {Fechner}, {Fujimoto}, {Honda},
  {Kajino}, {Minezaki}, {Nomoto}, {Norris}, {Ryan}, {Takada-Hidai},
  {Tsangarides}, \& {Yoshii}}]{Frebel:2005}
{Frebel}, A., {Aoki}, W., {Christlieb}, N., {et~al.} 2005, \nat, 434, 871

\bibitem[{{Georgy} {et~al.}(2013){Georgy}, {Ekstr{\"o}m}, {Eggenberger},
  {Meynet}, {Haemmerl{\'e}}, {Maeder}, {Granada}, {Groh}, {Hirschi}, {Mowlavi},
  {Yusof}, {Charbonnel}, {Decressin}, \& {Barblan}}]{Georgy:2013}
{Georgy}, C., {Ekstr{\"o}m}, S., {Eggenberger}, P., {et~al.} 2013, \aap, 558,
  A103

\bibitem[{{Georgy} {et~al.}(2012){Georgy}, {Ekstr{\"o}m}, {Meynet}, {Massey},
  {Levesque}, {Hirschi}, {Eggenberger}, \& {Maeder}}]{Georgy:2012}
{Georgy}, C., {Ekstr{\"o}m}, S., {Meynet}, G., {et~al.} 2012, \aap, 542, A29

\bibitem[{{Georgy} {et~al.}(2009){Georgy}, {Meynet}, {Walder}, {Folini}, \&
  {Maeder}}]{Georgy:2009}
{Georgy}, C., {Meynet}, G., {Walder}, R., {Folini}, D., \& {Maeder}, A. 2009,
  \aap, 502, 611

\bibitem[{{Gr{\"a}fener} \& {Hamann}(2008)}]{Graefener:2008}
{Gr{\"a}fener}, G. \& {Hamann}, W.-R. 2008, \aap, 482, 945

\bibitem[{{Gr{\"a}fener} {et~al.}(2011){Gr{\"a}fener}, {Vink}, {de Koter}, \&
  {Langer}}]{Grafener:2011}
{Gr{\"a}fener}, G., {Vink}, J.~S., {de Koter}, A., \& {Langer}, N. 2011, \aap,
  535, A56

\bibitem[{{Graham} \& {Fruchter}(2013)}]{Graham:2013}
{Graham}, J.~F. \& {Fruchter}, A.~S. 2013, \apj, 774, 119

\bibitem[{{Gratton} {et~al.}(2001){Gratton}, {Bonifacio}, {Bragaglia},
  {Carretta}, {Castellani}, {Centurion}, {Chieffi}, {Claudi}, {Clementini},
  {D'Antona}, {Desidera}, {Fran{\c c}ois}, {Grundahl}, {Lucatello}, {Molaro},
  {Pasquini}, {Sneden}, {Spite}, \& {Straniero}}]{Gratton:2001}
{Gratton}, R.~G., {Bonifacio}, P., {Bragaglia}, A., {et~al.} 2001, \aap, 369,
  87

\bibitem[{{Greif} {et~al.}(2010){Greif}, {Glover}, {Bromm}, \&
  {Klessen}}]{Greif:2010}
{Greif}, T.~H., {Glover}, S.~C.~O., {Bromm}, V., \& {Klessen}, R.~S. 2010,
  \apj, 716, 510

\bibitem[{{Grevesse} {et~al.}(1996){Grevesse}, {Noels}, \&
  {Sauval}}]{Grevesse:1996}
{Grevesse}, N., {Noels}, A., \& {Sauval}, A.~J. 1996, in Astronomical Society
  of the Pacific Conference Series, Vol.~99, Cosmic Abundances, ed. S.~S.
  {Holt} \& G.~{Sonneborn}, 117

\bibitem[{{Groenewegen} {et~al.}(2009){Groenewegen}, {Sloan}, {Soszy{\'n}ski},
  \& {Petersen}}]{Groenewegen:2009}
{Groenewegen}, M.~A.~T., {Sloan}, G.~C., {Soszy{\'n}ski}, I., \& {Petersen},
  E.~A. 2009, \aap, 506, 1277

\bibitem[{Hamann {et~al.}(1995)Hamann, Koesterke, \& Wessolowski}]{Hamann:1995}
Hamann, W.-R., Koesterke, L., \& Wessolowski, U. 1995, A\&A, 299, 151

\bibitem[{{Heap} {et~al.}(2015){Heap}, {Bouret}, \& {Hubeny}}]{Heap:2015}
{Heap}, S., {Bouret}, J.-C., \& {Hubeny}, I. 2015, ArXiv e-print 1504.02742

\bibitem[{Heger \& Langer(2000)}]{HegerII:2000}
Heger, A. \& Langer, N. 2000, ApJ, 544, 1016

\bibitem[{{Heger} {et~al.}(2000){Heger}, {Langer}, \& {Woosley}}]{HegerI:2000}
{Heger}, A., {Langer}, N., \& {Woosley}, S.~E. 2000, \apj, 528, 368

\bibitem[{Heger \& Woosley(2010)}]{Heger:2010}
Heger, A. \& Woosley, S. 2010, ApJ, 724, 341

\bibitem[{{Heger} {et~al.}(2005){Heger}, {Woosley}, \& {Spruit}}]{Heger:2005}
{Heger}, A., {Woosley}, S.~E., \& {Spruit}, H.~C. 2005, \apj, 626, 350

\bibitem[{{Hirschi} {et~al.}(2005){Hirschi}, {Meynet}, \&
  {Maeder}}]{Hirschi:2005}
{Hirschi}, R., {Meynet}, G., \& {Maeder}, A. 2005, \aap, 443, 581

\bibitem[{{Horv{\'a}th} {et~al.}(2014){Horv{\'a}th}, {Hakkila}, \&
  {Bagoly}}]{Horvath:2014}
{Horv{\'a}th}, I., {Hakkila}, J., \& {Bagoly}, Z. 2014, \aap, 561, L12

\bibitem[{{Hosokawa} {et~al.}(2012){Hosokawa}, {Yoshida}, {Omukai}, \&
  {Yorke}}]{Hosokawa:2012}
{Hosokawa}, T., {Yoshida}, N., {Omukai}, K., \& {Yorke}, H.~W. 2012, \apjl,
  760, L37

\bibitem[{{Huang} {et~al.}(2010){Huang}, {Gies}, \& {McSwain}}]{Huang:2010}
{Huang}, W., {Gies}, D.~R., \& {McSwain}, M.~V. 2010, \apj, 722, 605

\bibitem[{Hunter \& Thronson(1995)}]{Hunter:1995}
Hunter, D. \& Thronson, H. 1995, ApJ, 452, 238

\bibitem[{{Hunter} {et~al.}(2007){Hunter}, {Dufton}, {Smartt}, {Ryans},
  {Evans}, {Lennon}, {Trundle}, {Hubeny}, \& {Lanz}}]{Hunter:2007}
{Hunter}, I., {Dufton}, P.~L., {Smartt}, S.~J., {et~al.} 2007, \aap, 466, 277

\bibitem[{{Hunter} {et~al.}(2008){Hunter}, {Lennon}, {Dufton}, {Trundle},
  {Sim{\'o}n-D{\'{\i}}az}, {Smartt}, {Ryans}, \& {Evans}}]{Hunter:2008}
{Hunter}, I., {Lennon}, D.~J., {Dufton}, P.~L., {et~al.} 2008, \aap, 479, 541

\bibitem[{{Iglesias} \& {Rogers}(1996)}]{Iglesias:1996}
{Iglesias}, C.~A. \& {Rogers}, F.~J. 1996, \apj, 464, 943

\bibitem[{{Inserra} {et~al.}(2013){Inserra}, {Smartt}, {Jerkstrand}, {Valenti},
  {Fraser}, {Wright}, {Smith}, {Chen}, {Kotak}, {Pastorello}, {Nicholl},
  {Bresolin}, {Kudritzki}, {Benetti}, {Botticella}, {Burgett}, {Chambers},
  {Ergon}, {Flewelling}, {Fynbo}, {Geier}, {Hodapp}, {Howell}, {Huber},
  {Kaiser}, {Leloudas}, {Magill}, {Magnier}, {McCrum}, {Metcalfe}, {Price},
  {Rest}, {Sollerman}, {Sweeney}, {Taddia}, {Taubenberger}, {Tonry},
  {Wainscoat}, {Waters}, \& {Young}}]{Inserra:2013}
{Inserra}, C., {Smartt}, S.~J., {Jerkstrand}, A., {et~al.} 2013, \apj, 770, 128

\bibitem[{Izotov {et~al.}(1999)Izotov, Papaderos, Thuan,
  {et~al.}}]{Izotov:1999}
Izotov, Y., Papaderos, P., Thuan, T., {et~al.} 1999, unpublished

\bibitem[{Izotov \& Thuan(2002)}]{Izotov:2002}
Izotov, Y. \& Thuan, T. 2002, ApJ, 567

\bibitem[{Izotov \& Thuan(2004)}]{Izotov:2004}
Izotov, Y. \& Thuan, T. 2004, ApJ, 616, 768

\bibitem[{{Kehrig} {et~al.}(2013){Kehrig}, {P{\'e}rez-Montero},
  {V{\'{\i}}lchez}, {Brinchmann}, {Kunth}, {Garc{\'{\i}}a-Benito}, {Crowther},
  {Hern{\'a}ndez-Fern{\'a}ndez}, {Durret}, {Contini},
  {Fern{\'a}ndez-Mart{\'{\i}}n}, \& {James}}]{Kehrig:2013}
{Kehrig}, C., {P{\'e}rez-Montero}, E., {V{\'{\i}}lchez}, J.~M., {et~al.} 2013,
  \mnras, 432, 2731

\bibitem[{{Kehrig} {et~al.}(2015){Kehrig}, {V{\'{\i}}lchez},
  {P{\'e}rez-Montero}, {Iglesias-P{\'a}ramo}, {Brinchmann}, {Kunth}, {Durret},
  \& {Bayo}}]{Kehrig:2015}
{Kehrig}, C., {V{\'{\i}}lchez}, J.~M., {P{\'e}rez-Montero}, E., {et~al.} 2015,
  \apjl, 801, L28

\bibitem[{{Keller} {et~al.}(2014){Keller}, {Bessell}, {Frebel}, {Casey},
  {Asplund}, {Jacobson}, {Lind}, {Norris}, {Yong}, {Heger}, {Magic}, {da
  Costa}, {Schmidt}, \& {Tisserand}}]{Keller:2014}
{Keller}, S.~C., {Bessell}, M.~S., {Frebel}, A., {et~al.} 2014, \nat, 506, 463

\bibitem[{{K{\"o}hler} {et~al.}(2015){K{\"o}hler}, {Langer}, {de Koter}, {de
  Mink}, {Crowther}, {Evans}, {Gr{\"a}fener}, {Sana}, {Sanyal}, {Schneider}, \&
  {Vink}}]{Koehler:2015}
{K{\"o}hler}, K., {Langer}, N., {de Koter}, A., {et~al.} 2015, \aap, 573, A71

\bibitem[{{Kozyreva} {et~al.}(2014){Kozyreva}, {Blinnikov}, {Langer}, \&
  {Yoon}}]{Kozyreva:2014}
{Kozyreva}, A., {Blinnikov}, S., {Langer}, N., \& {Yoon}, S.-C. 2014, \aap,
  565, A70

\bibitem[{{Kroupa}(2001)}]{Kroupa:2001}
{Kroupa}, P. 2001, \mnras, 322, 231

\bibitem[{{Krti{\v c}ka} {et~al.}(2011){Krti{\v c}ka}, {Owocki}, \&
  {Meynet}}]{Krticka:2011}
{Krti{\v c}ka}, J., {Owocki}, S.~P., \& {Meynet}, G. 2011, \aap, 527, A84

\bibitem[{{Kub{\'a}t}(2012)}]{Kubat:2012}
{Kub{\'a}t}, J. 2012, \apjs, 203, 20

\bibitem[{{Kudritzki}(2002)}]{Kudritzki:2002}
{Kudritzki}, R.~P. 2002, \apj, 577, 389

\bibitem[{{Kudritzki} {et~al.}(1987){Kudritzki}, {Pauldrach}, \&
  {Puls}}]{Kudritzki:1987}
{Kudritzki}, R.~P., {Pauldrach}, A., \& {Puls}, J. 1987, \aap, 173, 293

\bibitem[{Langer(1989)}]{Langer:1989a}
Langer, N. 1989, A\&A, 210, 93

\bibitem[{Langer(1991)}]{Langer:1991}
Langer, N. 1991, A\&A, 252, 669

\bibitem[{Langer(1997)}]{Langer:1997}
Langer, N. 1997, Luminous Blue Variables: Massive Stars in Transition. ASP
  Conference Series. Ed. A.Nota, H.Lamers., 120, 83

\bibitem[{{Langer}(1998)}]{Langer:1998}
{Langer}, N. 1998, \aap, 329, 551

\bibitem[{{Langer}(2012)}]{Langer:2012}
{Langer}, N. 2012, \araa, 50, 107

\bibitem[{Langer {et~al.}(1983)Langer, Fricke, \& Sugimoto}]{Langer:1983}
Langer, N., Fricke, K.~J., \& Sugimoto, D. 1983, A\&A, 126, 207

\bibitem[{Langer \& Maeder(1995)}]{Langer:1995}
Langer, N. \& Maeder, A. 1995, A\&A, 295, 685

\bibitem[{{Langer} \& {Norman}(2006)}]{Langer:2006}
{Langer}, N. \& {Norman}, C.~A. 2006, \apjl, 638, L63

\bibitem[{{Langer} {et~al.}(2007){Langer}, {Norman}, {de Koter}, {Vink},
  {Cantiello}, \& {Yoon}}]{Langer:2007}
{Langer}, N., {Norman}, C.~A., {de Koter}, A., {et~al.} 2007, \aap, 475, L19

\bibitem[{{Lebouteiller} {et~al.}(2013){Lebouteiller}, {Heap}, {Hubeny}, \&
  {Kunth}}]{Lebouteiller:2013}
{Lebouteiller}, V., {Heap}, S., {Hubeny}, I., \& {Kunth}, D. 2013, \aap, 553,
  A16

\bibitem[{{Lecavelier des Etangs} {et~al.}(2004){Lecavelier des Etangs},
  {D{\'e}sert}, {Kunth}, {Vidal-Madjar}, {Callejo}, {Ferlet}, {H{\'e}brard}, \&
  {Lebouteiller}}]{LecavelierdesEtangs:2004}
{Lecavelier des Etangs}, A., {D{\'e}sert}, J.-M., {Kunth}, D., {et~al.} 2004,
  \aap, 413, 131

\bibitem[{Lee {et~al.}(2014)Lee, Suda, Beers, \& Stancliffe}]{Lee:2014}
Lee, Y.~S., Suda, T., Beers, T.~C., \& Stancliffe, R.~J. 2014, ApJ, 788, 131

\bibitem[{{Legrand} {et~al.}(1997){Legrand}, {Kunth}, {Roy}, {Mas-Hesse}, \&
  {Walsh}}]{Legrand:1997}
{Legrand}, F., {Kunth}, D., {Roy}, J.-R., {Mas-Hesse}, J.~M., \& {Walsh}, J.~R.
  1997, \aap, 326, L17

\bibitem[{{Leloudas} {et~al.}(2015){Leloudas}, {Schulze}, {Kr{\"u}hler},
  {Gorosabel}, {Christensen}, {Mehner}, {de Ugarte Postigo}, {Amor{\'{\i}}n},
  {Th{\"o}ne}, {Anderson}, {Bauer}, {Gallazzi}, {He{\l}miniak}, {Hjorth},
  {Ibar}, {Malesani}, {Morell}, {Vinko}, \& {Wheeler}}]{Leloudas:2015}
{Leloudas}, G., {Schulze}, S., {Kr{\"u}hler}, T., {et~al.} 2015, \mnras, 449,
  917

\bibitem[{{Levesque} {et~al.}(2010){Levesque}, {Kewley}, {Berger}, \&
  {Zahid}}]{Levesque:2010}
{Levesque}, E.~M., {Kewley}, L.~J., {Berger}, E., \& {Zahid}, H.~J. 2010, \aj,
  140, 1557

\bibitem[{{Longmore} {et~al.}(2014){Longmore}, {Kruijssen}, {Bastian}, {Bally},
  {Rathborne}, {Testi}, {Stolte}, {Dale}, {Bressert}, \&
  {Alves}}]{Longmore:2014}
{Longmore}, S.~N., {Kruijssen}, J.~M.~D., {Bastian}, N., {et~al.} 2014,
  Protostars and Planets VI, 291

\bibitem[{{Lunnan} {et~al.}(2013){Lunnan}, {Chornock}, {Berger},
  {Milisavljevic}, {Drout}, {Sanders}, {Challis}, {Czekala}, {Foley}, {Fong},
  {Huber}, {Kirshner}, {Leibler}, {Marion}, {McCrum}, {Narayan}, {Rest},
  {Roth}, {Scolnic}, {Smartt}, {Smith}, {Soderberg}, {Stubbs}, {Tonry},
  {Burgett}, {Chambers}, {Kudritzki}, {Magnier}, \& {Price}}]{Lunnan:2013}
{Lunnan}, R., {Chornock}, R., {Berger}, E., {et~al.} 2013, \apj, 771, 97

\bibitem[{MacFadyen \& Woosley(1999)}]{MacFadyen:1999}
MacFadyen, A. \& Woosley, S. 1999, ApJ, 524, 262

\bibitem[{{Mackey} {et~al.}(2014){Mackey}, {Mohamed}, {Gvaramadze}, {Kotak},
  {Langer}, {Meyer}, {Moriya}, \& {Neilson}}]{Mackey:2014}
{Mackey}, J., {Mohamed}, S., {Gvaramadze}, V.~V., {et~al.} 2014, \nat, 512, 282

\bibitem[{Maeder(1987)}]{Maeder:1987}
Maeder, A. 1987, A\&A, 178

\bibitem[{{Maeder} \& {Meynet}(2000)}]{Maeder:2000}
{Maeder}, A. \& {Meynet}, G. 2000, \araa, 38, 143

\bibitem[{{Marigo} {et~al.}(2003){Marigo}, {Chiosi}, \&
  {Kudritzki}}]{Marigo:2003}
{Marigo}, P., {Chiosi}, C., \& {Kudritzki}, R.-P. 2003, \aap, 399, 617

\bibitem[{{Martins} {et~al.}(2013){Martins}, {Depagne}, {Russeil}, \&
  {Mahy}}]{Martins:2013}
{Martins}, F., {Depagne}, E., {Russeil}, D., \& {Mahy}, L. 2013, \aap, 554, A23

\bibitem[{{Mathews} {et~al.}(2005){Mathews}, {Kajino}, \&
  {Shima}}]{Mathews:2005}
{Mathews}, G.~J., {Kajino}, T., \& {Shima}, T. 2005, \prd, 71

\bibitem[{Mauron \& Josselin(2011)}]{Mauron:2011}
Mauron, N. \& Josselin, E. 2011, A\&A, 526

\bibitem[{{McEvoy} {et~al.}(2015){McEvoy}, {Dufton}, {Evans}, {Kalari},
  {Markova}, {Sim{\'o}n-D{\'{\i}}az}, {Vink}, {Walborn}, {Crowther}, {de
  Koter}, {de Mink}, {Dunstall}, {H{\'e}nault-Brunet}, {Herrero}, {Langer},
  {Lennon}, {Ma{\'{\i}}z Apell{\'a}niz}, {Najarro}, {Puls}, {Sana},
  {Schneider}, \& {Taylor}}]{McEvoy:2015}
{McEvoy}, C.~M., {Dufton}, P.~L., {Evans}, C.~J., {et~al.} 2015, \aap, 575, A70

\bibitem[{{Meynet} \& {Maeder}(2000)}]{Meynet:2000}
{Meynet}, G. \& {Maeder}, A. 2000, \aap, 361, 101

\bibitem[{{Meynet} \& {Maeder}(2002)}]{Meynet:2002}
{Meynet}, G. \& {Maeder}, A. 2002, \aap, 390, 561

\bibitem[{{Meynet} \& {Maeder}(2005)}]{Meynet:2005}
{Meynet}, G. \& {Maeder}, A. 2005, \aap, 429, 581

\bibitem[{{Meynet} \& {Maeder}(2007)}]{Meynet:2007}
{Meynet}, G. \& {Maeder}, A. 2007, \aap, 464, L11

\bibitem[{{Modjaz} {et~al.}(2011){Modjaz}, {Kewley}, {Bloom}, {Filippenko},
  {Perley}, \& {Silverman}}]{Modjaz:2011}
{Modjaz}, M., {Kewley}, L., {Bloom}, J.~S., {et~al.} 2011, \apjl, 731, L4

\bibitem[{{Mokiem} {et~al.}(2006){Mokiem}, {de Koter}, {Evans}, {Puls},
  {Smartt}, {Crowther}, {Herrero}, {Langer}, {Lennon}, {Najarro}, {Villamariz},
  \& {Yoon}}]{Mokiem:2006}
{Mokiem}, M.~R., {de Koter}, A., {Evans}, C.~J., {et~al.} 2006, \aap, 456, 1131

\bibitem[{{Mokiem} {et~al.}(2007){Mokiem}, {de Koter}, {Vink}, {Puls}, {Evans},
  {Smartt}, {Crowther}, {Herrero}, {Langer}, {Lennon}, {Najarro}, \&
  {Villamariz}}]{Mokiem:2007}
{Mokiem}, M.~R., {de Koter}, A., {Vink}, J.~S., {et~al.} 2007, \aap, 473, 603

\bibitem[{{Moriya} {et~al.}(2013){Moriya}, {Blinnikov}, {Tominaga}, {Yoshida},
  {Tanaka}, {Maeda}, \& {Nomoto}}]{Moriya:2013}
{Moriya}, T.~J., {Blinnikov}, S.~I., {Tominaga}, N., {et~al.} 2013, \mnras,
  428, 1020

\bibitem[{{Moriya} \& {Langer}(2015)}]{Moriya:2015}
{Moriya}, T.~J. \& {Langer}, N. 2015, \aap, 573, A18

\bibitem[{Muijres {et~al.}(2012)}]{Muijres:2012}
Muijres, L. {et~al.} 2012, A\&A, 537

\bibitem[{Müller \& Vink(2014)}]{Mueller:2014}
Müller, P. \& Vink, J. 2014, A\&A, 564

\bibitem[{{Nicholls} {et~al.}(2014){Nicholls}, {Dopita}, {Sutherland},
  {Jerjen}, {Kewley}, \& {Basurah}}]{Nicholls:2014}
{Nicholls}, D.~C., {Dopita}, M.~A., {Sutherland}, R.~S., {et~al.} 2014, \apj,
  786, 155

\bibitem[{Nieuwenhuijzen \& de~Jager(1990)}]{Nieuwenhuijzen:1990}
Nieuwenhuijzen, H. \& de~Jager, C. 1990, A\&A, 231, 134

\bibitem[{{Niino}(2011)}]{Niino:2011}
{Niino}, Y. 2011, \mnras, 417, 567

\bibitem[{{Nugis} \& {Lamers}(2000)}]{Nugis:2000}
{Nugis}, T. \& {Lamers}, H.~J.~G.~L.~M. 2000, \aap, 360, 227

\bibitem[{{Peimbert} {et~al.}(2007){Peimbert}, {Luridiana}, \&
  {Peimbert}}]{Peimbert:2007}
{Peimbert}, M., {Luridiana}, V., \& {Peimbert}, A. 2007, \apj, 666, 636

\bibitem[{{Penny} \& {Gies}(2009)}]{Penny:2009}
{Penny}, L.~R. \& {Gies}, D.~R. 2009, \apj, 700, 844

\bibitem[{{Peters} {et~al.}(2010){Peters}, {Banerjee}, {Klessen}, {Mac Low},
  {Galv{\'a}n-Madrid}, \& {Keto}}]{Peters:2010}
{Peters}, T., {Banerjee}, R., {Klessen}, R.~S., {et~al.} 2010, \apj, 711, 1017

\bibitem[{{Podsiadlowski} {et~al.}(2004){Podsiadlowski}, {Mazzali}, {Nomoto},
  {Lazzati}, \& {Cappellaro}}]{Podsiadlowski:2004}
{Podsiadlowski}, P., {Mazzali}, P.~A., {Nomoto}, K., {Lazzati}, D., \&
  {Cappellaro}, E. 2004, \apjl, 607, L17

\bibitem[{{Portegies Zwart} {et~al.}(2010){Portegies Zwart}, {McMillan}, \&
  {Gieles}}]{PortegiesZwart:2010}
{Portegies Zwart}, S.~F., {McMillan}, S.~L.~W., \& {Gieles}, M. 2010, \araa,
  48, 431

\bibitem[{{Puls} {et~al.}(2008){Puls}, {Vink}, \& {Najarro}}]{Puls:2008}
{Puls}, J., {Vink}, J.~S., \& {Najarro}, F. 2008, \aapr, 16, 209

\bibitem[{{Quimby} {et~al.}(2011){Quimby}, {Kulkarni}, {Kasliwal}, {Gal-Yam},
  {Arcavi}, {Sullivan}, {Nugent}, {Thomas}, {Howell}, {Nakar}, {Bildsten},
  {Theissen}, {Law}, {Dekany}, {Rahmer}, {Hale}, {Smith}, {Ofek}, {Zolkower},
  {Velur}, {Walters}, {Henning}, {Bui}, {McKenna}, {Poznanski}, {Cenko}, \&
  {Levitan}}]{Quimby:2011}
{Quimby}, R.~M., {Kulkarni}, S.~R., {Kasliwal}, M.~M., {et~al.} 2011, \nat,
  474, 487

\bibitem[{{Quimby} {et~al.}(2013){Quimby}, {Yuan}, {Akerlof}, \&
  {Wheeler}}]{Quimby:2013}
{Quimby}, R.~M., {Yuan}, F., {Akerlof}, C., \& {Wheeler}, J.~C. 2013, \mnras,
  431, 912

\bibitem[{{Ram{\'{\i}}rez-Agudelo} {et~al.}(2013){Ram{\'{\i}}rez-Agudelo},
  {Sim{\'o}n-D{\'{\i}}az}, {Sana}, {de Koter}, {Sab{\'{\i}}n-Sanjul{\'{\i}}an},
  {de Mink}, {Dufton}, {Gr{\"a}fener}, {Evans}, {Herrero}, {Langer}, {Lennon},
  {Ma{\'{\i}}z Apell{\'a}niz}, {Markova}, {Najarro}, {Puls}, {Taylor}, \&
  {Vink}}]{RamirezAgudelo:2013}
{Ram{\'{\i}}rez-Agudelo}, O.~H., {Sim{\'o}n-D{\'{\i}}az}, S., {Sana}, H.,
  {et~al.} 2013, \aap, 560, A29

\bibitem[{Salpeter(1955)}]{Salpeter:1955}
Salpeter, E. 1955, ApJ, 121, 161

\bibitem[{{Sana} {et~al.}(2012){Sana}, {de Mink}, {de Koter}, {Langer},
  {Evans}, {Gieles}, {Gosset}, {Izzard}, {Le Bouquin}, \&
  {Schneider}}]{Sana:2012}
{Sana}, H., {de Mink}, S.~E., {de Koter}, A., {et~al.} 2012, Science, 337, 444

\bibitem[{Sanders {et~al.}(2012)Sanders, Soderberg, Valenti, Foley, Chornock,
  Chomiuk, Berger, Smartt, Hurley, Barthelmy, Levesque, Narayan, Botticella,
  Briggs, Connaughton, Terada, Gehrels, Golenetskii, Mazets, Cline, von
  Kienlin, Boynton, Chambers, Grav, Heasley, Hodapp, Jedicke, Kaiser, Kirshner,
  Kudritzki, Luppino, Lupton, Magnier, Monet, Morgan, Onaka, Price, Stubbs,
  Tonry, Wainscoat, \& Waterson}]{Sanders:2012}
Sanders, N.~E., Soderberg, A.~M., Valenti, S., {et~al.} 2012, ApJ, 756, 184

\bibitem[{Sanyal {et~al.}(2015)}]{Sanyal:2015}
Sanyal, D. {et~al.} 2015, A\&A, accepted

\bibitem[{{Schaerer} {et~al.}(1999{\natexlab{a}}){Schaerer}, {Contini}, \&
  {Kunth}}]{Schaerer:1999a}
{Schaerer}, D., {Contini}, T., \& {Kunth}, D. 1999{\natexlab{a}}, \aap, 341,
  399

\bibitem[{{Schaerer} {et~al.}(1999{\natexlab{b}}){Schaerer}, {Contini}, \&
  {Pindao}}]{Schaerer:1999b}
{Schaerer}, D., {Contini}, T., \& {Pindao}, M. 1999{\natexlab{b}}, \aaps, 136,
  35

\bibitem[{{Schneider} {et~al.}(2014){Schneider}, {Izzard}, {de Mink}, {Langer},
  {Stolte}, {de Koter}, {Gvaramadze}, {Hu{\ss}mann}, {Liermann}, \&
  {Sana}}]{Schneider:2014}
{Schneider}, F.~R.~N., {Izzard}, R.~G., {de Mink}, S.~E., {et~al.} 2014, \apj,
  780, 117

\bibitem[{Searle \& Sargent(1972)}]{Searle:1972}
Searle, L. \& Sargent, W. 1972, ApJ, 173, 25

\bibitem[{{Shirazi} \& {Brinchmann}(2012)}]{Shirazi:2012}
{Shirazi}, M. \& {Brinchmann}, J. 2012, \mnras, 421, 1043

\bibitem[{{Sobral} {et~al.}(2015){Sobral}, {Matthee}, {Darvish}, {Schaerer},
  {Mobasher}, {R{\"o}ttgering}, {Santos}, \& {Hemmati}}]{Sobral:2015}
{Sobral}, D., {Matthee}, J., {Darvish}, B., {et~al.} 2015, ArXiv e-prints
  1504.01734

\bibitem[{{Sonnenfeld} {et~al.}(2012){Sonnenfeld}, {Treu}, {Gavazzi},
  {Marshall}, {Auger}, {Suyu}, {Koopmans}, \& {Bolton}}]{Sonnenfeld:2012}
{Sonnenfeld}, A., {Treu}, T., {Gavazzi}, R., {et~al.} 2012, \apj, 752, 163

\bibitem[{Spruit(2002)}]{Spruit:2002}
Spruit, H. 2002, A\&A, 381, 923

\bibitem[{Spruit(2006)}]{Spruit:2006}
Spruit, H. 2006, arXiv:astro-ph/0607164

\bibitem[{{Suijs} {et~al.}(2008){Suijs}, {Langer}, {Poelarends}, {Yoon},
  {Heger}, \& {Herwig}}]{Suijs:2008}
{Suijs}, M.~P.~L., {Langer}, N., {Poelarends}, A.-J., {et~al.} 2008, \aap, 481,
  L87

\bibitem[{Thompson {et~al.}(2004)Thompson, Chang, \& Quataert}]{Thompson:2004}
Thompson, T., Chang, P., \& Quataert, E. 2004, ApJ, 611, 380

\bibitem[{{Tolstoy} {et~al.}(2009){Tolstoy}, {Hill}, \& {Tosi}}]{Tolstoy:2009}
{Tolstoy}, E., {Hill}, V., \& {Tosi}, M. 2009, \araa, 47, 371

\bibitem[{{Tramper} {et~al.}(2011){Tramper}, {Sana}, {de Koter}, \&
  {Kaper}}]{Tramper:2011}
{Tramper}, F., {Sana}, H., {de Koter}, A., \& {Kaper}, L. 2011, \apjl, 741, L8

\bibitem[{{Treu} {et~al.}(2010){Treu}, {Auger}, {Koopmans}, {Gavazzi},
  {Marshall}, \& {Bolton}}]{Treu:2010}
{Treu}, T., {Auger}, M.~W., {Koopmans}, L.~V.~E., {et~al.} 2010, \apj, 709,
  1195

\bibitem[{{Vaduvescu} {et~al.}(2007){Vaduvescu}, {McCall}, \&
  {Richer}}]{Vaduvescu:2007}
{Vaduvescu}, O., {McCall}, M.~L., \& {Richer}, M.~G. 2007, \aj, 134, 604

\bibitem[{Vink {et~al.}(2000)Vink, de~Koter, \& Lamers}]{Vink:2000}
Vink, J., de~Koter, A., \& Lamers, H. 2000, A\&A, 362, 295

\bibitem[{Vink {et~al.}(2001)Vink, de~Koter, \& Lamers}]{Vink:2001}
Vink, J., de~Koter, A., \& Lamers, H. 2001, A\&A, 369, 574

\bibitem[{{Vink} {et~al.}(2010){Vink}, {Brott}, {Gr{\"a}fener}, {Langer}, {de
  Koter}, \& {Lennon}}]{Vink:2010}
{Vink}, J.~S., {Brott}, I., {Gr{\"a}fener}, G., {et~al.} 2010, \aap, 512, L7

\bibitem[{{Vink} {et~al.}(2011){Vink}, {Muijres}, {Anthonisse}, {de Koter},
  {Gr{\"a}fener}, \& {Langer}}]{Vink:2011}
{Vink}, J.~S., {Muijres}, L.~E., {Anthonisse}, B., {et~al.} 2011, \aap, 531,
  A132

\bibitem[{{Walborn} {et~al.}(2004){Walborn}, {Morrell}, {Howarth}, {Crowther},
  {Lennon}, {Massey}, \& {Arias}}]{Walborn:2004}
{Walborn}, N.~R., {Morrell}, N.~I., {Howarth}, I.~D., {et~al.} 2004, \apj, 608,
  1028

\bibitem[{{Weisz} {et~al.}(2014){Weisz}, {Dolphin}, {Skillman}, {Holtzman},
  {Gilbert}, {Dalcanton}, \& {Williams}}]{Weisz:2014}
{Weisz}, D.~R., {Dolphin}, A.~E., {Skillman}, E.~D., {et~al.} 2014, \apj, 789,
  147

\bibitem[{Woosley \& Heger(2006)}]{Woosley:2006}
Woosley, S. \& Heger, A. 2006, ApJ, 637, 914

\bibitem[{{Woosley}(2010)}]{Woosley:2010}
{Woosley}, S.~E. 2010, \apjl, 719, L204

\bibitem[{{Yong} {et~al.}(2003){Yong}, {Grundahl}, {Lambert}, {Nissen}, \&
  {Shetrone}}]{Yong:2003}
{Yong}, D., {Grundahl}, F., {Lambert}, D.~L., {Nissen}, P.~E., \& {Shetrone},
  M.~D. 2003, \aap, 402, 985

\bibitem[{{Yoon}(2015)}]{Yoon:2015}
{Yoon}, S.-C. 2015, \pasa, 32, 15

\bibitem[{{Yoon} {et~al.}(2012){Yoon}, {Dierks}, \& {Langer}}]{Yoon:2012}
{Yoon}, S.-C., {Dierks}, A., \& {Langer}, N. 2012, \aap, 542, A113

\bibitem[{Yoon \& Langer(2005)}]{Yoon:2005}
Yoon, S.-C. \& Langer, N. 2005, A\&A, 443, 643

\bibitem[{{Yoon} {et~al.}(2006){Yoon}, {Langer}, \& {Norman}}]{Yoon:2006}
{Yoon}, S.-C., {Langer}, N., \& {Norman}, C. 2006, \aap, 460, 199

\bibitem[{{Yoshida} {et~al.}(2007){Yoshida}, {Oh}, {Kitayama}, \&
  {Hernquist}}]{Yoshida:2007}
{Yoshida}, N., {Oh}, S.~P., {Kitayama}, T., \& {Hernquist}, L. 2007, \apj, 663,
  687

\bibitem[{{Yusof} {et~al.}(2013){Yusof}, {Hirschi}, {Meynet}, {Crowther},
  {Ekstr{\"o}m}, {Frischknecht}, {Georgy}, {Abu Kassim}, \&
  {Schnurr}}]{Yusof:2013}
{Yusof}, N., {Hirschi}, R., {Meynet}, G., {et~al.} 2013, \mnras, 433, 1114

\bibitem[{Zhao {et~al.}(2013)Zhao, Gao, \& Gu}]{Zhao:2013}
Zhao, Y., Gao, Y., \& Gu, Q. 2013, ApJ, 764, 44

\end{thebibliography}

%---------------------------------------------------------------------------
% Online material -- sumtab

\onecolumn
\appendix 
\section{Key quantities of our model sequences (Online material)}

\LTcapwidth=0.9\textwidth
\begin{longtable}{ccccccccccclcc}

	%first header
\caption{Important characteristics of our models. 
M$^i$: initial mass. 
v$_{\rm rot}^i$: initial rotational velocity. 
v$_{\rm rot}^{Y_{C}=0.28}$: rotational velocity at the zero-age main-sequence (defined as Y$_{\rm C}$=0.28). 
$\tau_{\rm MS}$: Main-sequence lifetime. M$^f$: mass at the end of the calculation. 
v$_{\rm rot}^f$: rotational velocity at the end of the calculation. 
Y$_C^f$: central mass fraction of helium at the end of the calculation. 
Y$_S^f$: surface mass fraction of helium at the end of the calculation; bold face indicates if the last model is chemically-homogeneous (with Y$_S^f\geq0.7$), i.e. a TWUIN star. 
$A$(C)$_S^f$=log[C/H]$_S^f$+12: surface abundance of carbon ($^{12}$C) at the end of the calculation. 
$A$(N)$_S^f$=log[N/H]$_S^f$+12: surface abundance of nitrogen ($^{14}$N) at the end of the calculation. 
$\log$L$^f$/L$_{\odot}$: luminosity at the end of the calculation. 
T$_{\rm eff}^f$: effective temperature at the end of the calculation. 
R$^f$: radius at the end of the calculation; bold face indicates if the last model is a cool supergiant (with T$_{\rm eff}^f<12\mathrm{~kK}$). 
Type: evolutionary class as defined in Sect.~\ref{sec:evol}.
}\\ \hline \hline
M$^i$ & v$_{\rm rot}^i$ & v$_{\rm rot}^{Y_{C}=0.28}$ & $\tau_{\rm MS}$ & M$^f$ & v$_{\rm rot}^f$ & Y$_C^f$ & Y$_S^f$ & $A$(C)$_S^f$ & $A$(N)$_S^f$ & $\log$L$^f$/L$_{\odot}$ & T$_{\rm eff}^f$ & R$^f$ & Type  \\ 

[M$_{\odot}$] & [km~s$^{-1}$] & [km~s$^{-1}$] & [Myr] & [M$_{\odot}$] & [km~s$^{-1}$] &  & &  &  &  & [K] & [R$_{\odot}$] & \\ \hline \hline
\endfirsthead

	%header
\caption{Continued.}\\ \hline \hline
M$^i$ & v$_{\rm rot}^i$ & v$_{\rm rot}^{Y_{C}=0.28}$ & $\tau_{\rm MS}$ & M$^f$ & v$_{\rm rot}^f$ & Y$_C^f$ & Y$_S^f$ & $A$(C)$_S^f$ & $A$(N)$_S^f$ & $\log$L$^f$/L$_{\odot}$ & T$_{\rm eff}^f$ & R$^f$ & Type \\ 

[M$_{\odot}$] & [km~s$^{-1}$] & [km~s$^{-1}$] & [Myr] & [M$_{\odot}$] & [km~s$^{-1}$] &  & &  &  &  & [K] & [R$_{\odot}$] & \\ \hline \hline
\endhead

	%footer
\hline \hline
\endfoot

	%last footer
\endlastfoot

	%table content

9 & 0 & 0 & 23.1 & 9.0 & 0 & 0.98 & {0.25} & 6.37 & 5.50 & 4.14 & 23828 & {6.9} & NE \\
9 & 100 & 136 & 23.3 & 9.0 &  116 & 0.98 & {0.25} & 6.30 & 5.87 & 4.15 & 23532 & {7.1} & NE \\
9 & 200 & 276 & 23.8 & 9.0 &  246 & 0.98 & {0.25} & 5.83 & 6.46 & 4.14 & 22567 & {7.7} & NE \\
9 & 300 & 435 & 24.8 & 9.0 &  412 & 0.98 & {0.25} & 5.28 & 6.70 & 4.14 & 20948 & {8.9} & NE \\
9 & 375 & 572 & 26.3 & 8.9 &  440 & 0.98 & {0.26} & 5.03 & 6.85 & 4.14 & 20778 & {9.1} & NE \\
9 & 400 & 618 & 32.7 & 8.8 &  474 & 0.98 & {0.40} & 5.17 & 7.13 & 4.30 & 24777 & {7.7} & TE \\
9 & 425 & 663 & 31.4 & 8.8 &  459 & 0.98 & {0.35} & 5.13 & 7.10 & 4.24 & 23067 & {8.2} & TE \\
9 & 450 & 709 & 41.4 & 8.9 &  615 & 0.98 & \textbf{0.95} & 6.68 & 8.29 & 4.89 & 56521 & {2.9} & CHE \\
\end{longtable}

%---------------------------------------------------------------------------
% Online material -- fluxtab

\section{Photoionizing flux (Online material)}

\begin{longtable}{cc|cccccccccccc}

	%first header
\caption{Time averaged ionizing fluxes ($L$), and time averaged ($Q$) and peak photon emission rates ($Q^{\rm max}$)  of the ionizing photons for hydrogen (H), neutral helium (He) and singly ionised helium (He$^+$) during the main-sequence evolution for each stellar sequence. We emphasize that the given quantities rely on the black body approximation, and refer to Sect.~\ref{sec:valid} for a discussion of its validity.
}\\ \hline \hline
M$^i$ & v$_{\rm rot}^i$ & L($\gamma_{H}$) & Q($\gamma_{H}$) & Q$^{\rm max}$($\gamma_{H}$) & L($\gamma_{He}$) & Q($\gamma_{He}$) & Q$^{\rm max}$($\gamma_{He}$) & L($\gamma_{He^+}$) & Q($\gamma_{He^+}$) & Q$^{\rm max}$($\gamma_{He^+}$) \\

[M$_{\odot}$] & [km~s$^{-1}$] & [erg/s] & [s$^{-1}$] & [s$^{-1}$] & [erg/s] & [s$^{-1}$] & [s$^{-1}$] & [erg/s] &  [s$^{-1}$] & [s$^{-1}$] \\ \hline \hline

\endfirsthead

	%header
\caption{Continued.}\\ \hline \hline
M$^i$ & v$_{\rm rot}^i$ & L($\gamma_{H}$) & Q($\gamma_{H}$) & Q$^{\rm max}$($\gamma_{H}$) & L($\gamma_{He}$) & Q($\gamma_{He}$) & Q$^{\rm max}$($\gamma_{He}$) & L($\gamma_{He^+}$) & Q($\gamma_{He^+}$) & Q$^{\rm max}$($\gamma_{He^+}$) \\ 

[M$_{\odot}$] & [km~s$^{-1}$] & [erg/s] & [s$^{-1}$] & [s$^{-1}$] & [erg/s] & [s$^{-1}$] & [s$^{-1}$] & [erg/s] &  [s$^{-1}$] & [s$^{-1}$] \\ \hline \hline
\endhead

	%footer
\hline \hline
\endfoot

	%last footer
\endlastfoot

	%table content

9 & 0 & 4.94E+36 & 1.83E+47 & 2.22E+47 & 2.60E+35 & 5.89E+45 & 6.84E+45 & 1.53E+31 & 1.66E+41 & 2.28E+41  \\
9 & 100 & 4.84E+36 & 1.80E+47 & 2.19E+47 & 2.50E+35 & 5.67E+45 & 6.61E+45 & 1.40E+31 & 1.52E+41 & 2.10E+41  \\
9 & 200 & 4.40E+36 & 1.65E+47 & 1.97E+47 & 2.09E+35 & 4.77E+45 & 5.58E+45 & 9.53E+30 & 1.04E+41 & 1.49E+41  \\
9 & 300 & 3.67E+36 & 1.37E+47 & 1.61E+47 & 1.46E+35 & 3.35E+45 & 3.99E+45 & 4.33E+30 & 4.73E+40 & 7.59E+40  \\
9 & 375 & 3.20E+36 & 1.21E+47 & 1.43E+47 & 1.11E+35 & 2.55E+45 & 3.08E+45 & 2.19E+30 & 2.39E+40 & 3.75E+40  \\
9 & 400 & 4.95E+36 & 1.84E+47 & 3.33E+47 & 2.38E+35 & 5.40E+45 & 1.00E+46 & 1.00E+31 & 1.09E+41 & 3.02E+41  \\
9 & 425 & 4.09E+36 & 1.53E+47 & 2.34E+47 & 1.73E+35 & 3.95E+45 & 5.81E+45 & 5.21E+30 & 5.69E+40 & 1.17E+41  \\
9 & 450 & 1.59E+37 & 5.30E+47 & 5.61E+48 & 2.99E+36 & 6.29E+46 & 1.39E+48 & 1.73E+34 & 1.81E+44 & 1.12E+46  \\
\end{longtable}
%---------------------------------------------------------------------------
% Online material -- isochrones

\section{Isochrones (Online material)}

\begin{figure}[h!]
\resizebox{0.45\hsize}{!}{\includegraphics[angle=270,width=1.\columnwidth,page=1]{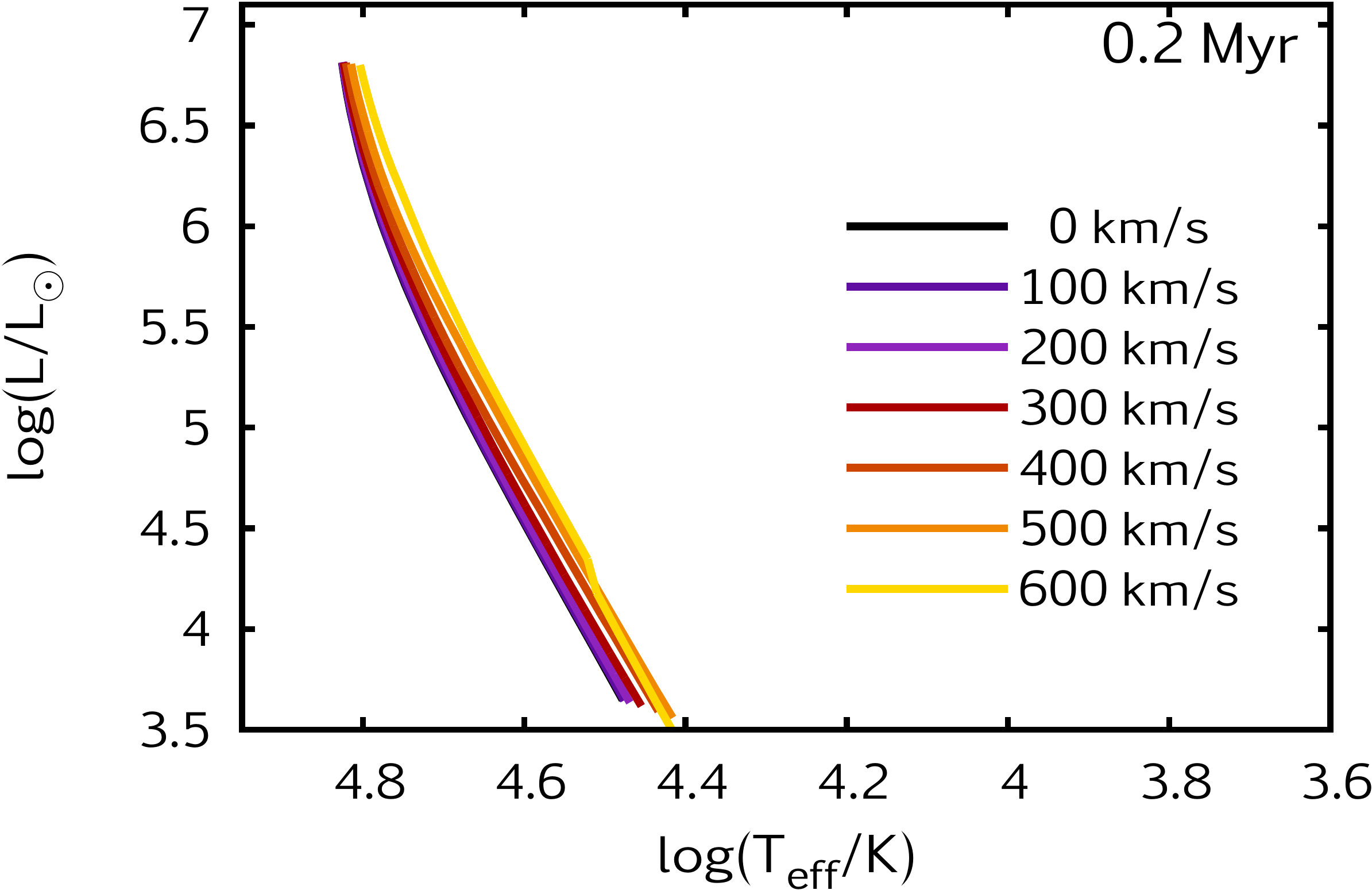}} \hspace{15pt}
\resizebox{0.45\hsize}{!}{\includegraphics[angle=270,width=1.\columnwidth,page=9]{img-crop}}\\ 

\vspace{10pt} 

\resizebox{0.45\hsize}{!}{\includegraphics[angle=270,width=1.\columnwidth,page=3]{img-crop}} \hspace{15pt}
\resizebox{0.45\hsize}{!}{\includegraphics[angle=270,width=1.\columnwidth,page=11]{img-crop}}\\ 

\vspace{10pt}

\resizebox{0.45\hsize}{!}{\includegraphics[angle=270,width=1.\columnwidth,page=5]{img-crop}} \hspace{15pt}
\resizebox{0.45\hsize}{!}{\includegraphics[angle=270,width=1.\columnwidth,page=13]{img-crop}}\\ 

\vspace{10pt}

\resizebox{0.45\hsize}{!}{\includegraphics[angle=270,width=1.\columnwidth,page=7]{img-crop}} \hspace{15pt}
\resizebox{0.45\hsize}{!}{\includegraphics[angle=270,width=1.\columnwidth,page=15]{img-crop}}
\caption{Isochrones of different ages of rotating stellar evolutionary models are shown in the HR~diagram. The initial surface rotational velocity is chosen in steps of 100~km~s$^{-1}$ from non-rotating to 600~km~s$^{-1}$
}
\label{fig:iso}
\end{figure}

Fig.~\ref{fig:iso} shows isochrones of our stellar evolutionary calculations for several ages and rotational rates. 

%---------------------------------------------------------------------------

\twocolumn

\end{document}